\documentclass[%
superscriptaddress,
twocolumn,
nofootinbib,
amsmath,amssymb,
aps,
prd,
floatfix,
]{revtex4}

\usepackage{graphicx}
\usepackage{dcolumn}
\usepackage{bm}
\usepackage{xcolor}
\usepackage[colorlinks=true, linkcolor=blue, citecolor=blue, urlcolor=blue]{hyperref}
\usepackage{cleveref}
\usepackage{diagbox}
\usepackage{multirow}
\usepackage{physics}
\usepackage{braket}

\DeclareRobustCommand{\Eq}[1]{Eq.~\eqref{eq:#1}}

\DeclareRobustCommand{\fig}[1]{Fig.~\ref{fig:#1}}

\DeclareRobustCommand{\refcite}[1]{Ref.~\cite{#1}}

\crefname{figure}{fig.}{figs.}
\Crefname{figure}{Fig.}{Figs.}
\crefname{table}{tab.}{tabs.}
\Crefname{table}{Tab.}{Tabs.}
\crefname{equation}{eq.}{eqns.}
\Crefname{equation}{Eq.}{Eqns.}
\crefname{listing}{lst.}{lsts.}
\Crefname{listing}{Lst.}{Lsts.}
\crefname{section}{sec.}{secs.}
\Crefname{section}{Sec.}{Secs.}
\crefname{appendix}{app.}{apps.}
\Crefname{appendix}{App.}{Apps.}

\newcommand{\pvec}{\vec{p}}
\newcommand{\qvec}{\vec{q}}
\newcommand{\xvec}{\vec{x}}
\newcommand{\yvec}{\vec{y}}
\newcommand{\zvec}{\vec{z}}
\newcommand{\ts}{t_{\rm sep}}
\newcommand{\ti}{t_{\rm ins}}

\newcommand{\qcdcutoff}{\Lambda_{\rm QCD}}

\begin{document}


\title{Transversity PDFs of the proton from lattice QCD with physical quark masses}

\author{Xiang Gao}
\email{gaox@anl.gov}
\affiliation{Physics Division, Argonne National Laboratory, Lemont, IL 60439, USA}

\author{Andrew D. Hanlon}
\email{ahanlon@bnl.gov}
\affiliation{Physics Department, Brookhaven National Laboratory, Bldg. 510A, Upton, New York 11973, USA}

\author{Swagato Mukherjee}
\affiliation{Physics Department, Brookhaven National Laboratory, Bldg. 510A, Upton, New York 11973, USA}

\author{Peter Petreczky}
\affiliation{Physics Department, Brookhaven National Laboratory, Bldg. 510A, Upton, New York 11973, USA}

\author{Qi Shi}
\affiliation{Physics Department, Brookhaven National Laboratory, Bldg. 510A, Upton, New York 11973, USA}
\affiliation{Key Laboratory of Quark \& Lepton Physics (MOE) and Institute of Particle Physics, Central China Normal University, Wuhan 430079, China}

\author{Sergey Syritsyn}
\affiliation{RIKEN-BNL Research Center, Brookhaven National Laboratory, Upton, New York 11973}
\affiliation{Department of Physics and Astronomy, Stony Brook University, Stony Brook, New York 11790}

\author{Yong Zhao}
\affiliation{Physics Division, Argonne National Laboratory, Lemont, IL 60439, USA}


\date{\today}

\begin{abstract}
We present a lattice QCD calculation of the transversity isovector- and isoscalar-quark parton distribution functions (PDFs) of the proton utilizing a perturbative matching at next-to-leading-order (NLO) accuracy.
Additionally, we determine the isovector and isoscalar tensor charges for the proton.
In both calculations, the disconnected contributions to the isoscalar matrix elements have been ignored.
The calculations are performed using a single ensemble of $N_f = 2 +1$ highly-improved staggered quarks simulated with physical-mass quarks and a lattice spacing of $a = 0.076$ fm.
The Wilson-clover action, with physical quark masses and smeared gauge links obtained from one iteration of hypercubic smearing, is used in the valence sector.
Using the NLO operator product expansion, we extract the lowest four to six Mellin moments and the PDFs via a neural network from the matrix elements in the pseudo-PDF approach.
In addition, we calculate the PDFs in the quasi-PDF approach with hybrid-scheme renormalization and the recently developed leading-renormalon resummation technique, at NLO with the resummation of leading small-$x$ logarithms.
\end{abstract}


\maketitle


\section{Introduction}
\label{sec:intro}

A significant goal of hadron physics is the determination of the full structure of nucleons.
There has been much progress towards this end, both experimentally and theoretically.
Experimentally, the leading-twist parton distributions functions (PDFs) for both the unpolarized and longitudinally polarized proton have been determined with high precision through global analyses~\cite{Alekhin:2017kpj,Hou:2019efy,Bailey:2020ooq,NNPDF:2021njg} of experimental data collected for example at HERA, the Tevatron, the LHC, etc.
In addition, to obtain the full collinear structure at leading twist requires the transversity PDF, which gives the difference in the probability to find a parton aligned and anti-aligned with the transversely polarized hadron.
However, the transversity PDF is less well constrained experimentally, but measurements of the single transverse-spin asymmetries from semi-inclusive deep-inelastic scattering (SIDIS) processes by COMPASS~\cite{COMPASS:2014bze} and HERMES~\cite{HERMES:2020ifk}, as well as dihadron production in SIDIS by COMPASS~\cite{COMPASS:2012bfl,COMPASS:2014ysd} and HERMES~\cite{HERMES:2008mcr}, and $pp$ collisions by RHIC~\cite{Yang:2009zzr,Fatemi:2012ry,STAR:2015jkc}, have led to a series of extractions of the transversity PDF~\cite{Bacchetta:2011ip,Bacchetta:2012ty,Anselmino:2013vqa,Kang:2014zza,Martin:2014wua,Radici:2015mwa,Anselmino:2015sxa,Kang:2015msa,Radici:2016lam,Lin:2017stx,Radici:2018iag,Benel:2019mcq,Anselmino:2020nrk,DAlesio:2020vtw,Cammarota:2020qcw,Gamberg:2022kdb,Cocuzza:2023oam,Cocuzza:2023vqs}.
However, the uncertainties can still be as large as 40\% or more in the valence region~\cite{Cocuzza:2023vqs}. One of the major goals of the JLab 12 GeV upgrade and the upcoming Electron-Ion Collider (EIC) is to gain more information on the spin structure of the nucleon, including the transversity PDF~\cite{Dudek:2012vr,Burkert:2022hjz}.

On the theoretical side, 
there has been significant development in the first-principles calculations of PDFs through lattice QCD 
(see reviews in Refs.~\cite{Cichy:2018mum,Zhao:2018fyu,Radyushkin:2019mye,Ji:2020ect,Constantinou:2020pek,Constantinou:2020hdm,Cichy:2021lih}).
Among them, the two most widely used approaches utilize either the quasi-PDF within the framework of large-momentum effective theory (LaMET)~\cite{Ji:2013dva,Ji:2014gla,Ji:2020ect} or the pseudo-PDF~\cite{Radyushkin:2017cyf,Orginos:2017kos}. Both the quasi-PDF and pseudo-PDF are defined from the matrix elements of a gauge-invariant equal-time bilinear operator in a boosted hadron state~\cite{Ji:2013dva}, which can be directly simulated on the lattice. In the LaMET approach, the  PDF can be calculated from the quasi-PDF through a power expansion and effective theory matching at large hadron momentum, with controlled precision for a range of moderate $x$. On the other hand, the pseudo-PDF method relies on a short-distance factorization in coordinate space~\cite{Ji:2017rah,Radyushkin:2017lvu,Izubuchi:2018srq}, which allows for a model-independent extraction of the lowest Mellin moments~\cite{Izubuchi:2018srq} or a model-dependent extraction of the $x$-dependent PDF. Both methods require larger hadron momenta to extract more information on the PDF, and can complement each other in practical calculations~\cite{Ji:2022ezo,Holligan:2023rex}.

Over the past decade, there have been a few calculations of the transversity PDF from lattice QCD using both the quasi-~\cite{Chen:2016utp,Alexandrou:2018eet,Liu:2018hxv,Alexandrou:2019lfo,Alexandrou:2021oih,LatticeParton:2022xsd} and pseudo-PDF~\cite{HadStruc:2021qdf} approaches, which were all carried out with a next-to-leading-order (NLO) perturbative matching correction. Among them, the first physical pion mass calculations~\cite{Alexandrou:2018eet,Liu:2018hxv,Alexandrou:2019lfo} were accomplished with the regularization-independent momentum subtraction (RI/MOM) scheme~\cite{Constantinou:2017sej,Alexandrou:2017huk,Chen:2017mzz,Stewart:2017tvs} for the lattice renormalization, which is flawed by the introduction of non-perturbative effects at large quark-bilinear separation. To overcome this problem, the hybrid scheme~\cite{Ji:2020brr} was proposed to subtract the Wilson line mass with a matching to the $\overline{\rm MS}$ scheme at large quark-bilinear separation, which was used in the recent calculation of Ref.~\cite{LatticeParton:2022xsd} with continuum and physical pion mass extrapolations. More recently, a systematic way to remove the renormalon ambiguity in the Wilson-line mass matching, called leading-renormalon resummation (LRR), was proposed in Refs.~\cite{Holligan:2023rex,Zhang:2023bxs}.

In this work we carry out a lattice QCD calculation of the proton isovector and isoscalar quark transversity PDFs at physical quark masses, where the latter have been calculated without the inclusion of disconnected diagrams. This is an extension of our previous calculation of the proton isovector unpolarized PDF~\cite{Gao:2022uhg}.
Here we utilize both methods for calculation, which can help to understand the significance of the different systematics within them. In particular, for the quasi-PDF method, we adopt the hybrid scheme with LRR and work at NLO with leading-logarithmic (LL) resummation that accounts for PDF evolution, which gives us a reliable estimate of the sysmtematic uncertainty in the small-$x$ region~\cite{Zhang:2023bxs}.

The rest of the paper is organized as follows.
First, in \Cref{sec:lattice}, we review the setup of our lattice calculation.
Then in \Cref{sec:analysis} we describe our analysis strategy to extract the ground-state matrix elements, which includes an estimate for the tensor charge.
In \Cref{sec:leading_twist_OPE} we use the ground-state matrix elements to extract the lowest few Mellin moments from the leading-twist operator product expansion (OPE).
We then move on to the determination of the transversity PDF with the pseudo-PDF method in \Cref{sec:dnn} and the quasi-PDF method in \Cref{sec:xspace}.
And, finally, we conclude in \Cref{sec:conc}.
\section{Lattice Details}
\label{sec:lattice}

Our setup is nearly identical to that used in our previous work on the unpolarized proton PDF~\cite{Gao:2022uhg},
and is also similar to our work on the pion valence PDF~\cite{Gao:2021dbh,Gao:2022iex}.
There are only two differences here: i)
the specific correlators needed for the transversity PDF, which were, in fact, computed at the same time as the correlators needed for the unpolarized PDF;
and ii) an increase in statistics for the $P_z = \frac{2\pi 6}{L}$, $t_{\rm sep} = 12a$ data.
Therefore, we only repeat the most pertinent details here.

The calculations are performed on a $64^3 \times 64$ ensemble of $N_f = 2 + 1$ highly-improved staggered quarks (HISQ)~\cite{Follana:2006rc} with masses tuned to their physical values and a lattice spacing of $a = 0.076$ fm,
which was generated by the HotQCD collaboration~\cite{Bazavov:2019www}.
For the valence quarks the tree-level tadpole-improved Wilson-clover action is used
on the smeared gauge field background. We use one iteration of hyper-cubic (HYP)
smearing \cite{Hasenfratz:2001hp} on the original gauge fields. 
The valence quark masses are set to their physical value. 
Because of the use of a HYP smeared background, we do not see any exceptional
configurations. This feature of the Wilson-clover action with HYP smeared
gauge background holds even on coarser lattices, as was pointed out some time
ago~\cite{Bhattacharya:2015wna}.
Note that, in principle, the mixed action setup leads to violations of unitarity at non-zero lattice spacing.
However, the effects of this appear to be very mild (e.g. see Ref.~\cite{Mondal:2020cmt} which makes consistent comparisons for nucleon structure observables between different mixed and unmixed setups).

In order to build a nucleon operator with good overlap onto a highly-boosted nucleon state, the quark fields are smeared using Coulomb-gauge momentum smearing~\cite{Bali:2016lva} as described in App. A of Ref.~\cite{Izubuchi:2019lyk}.
Within this method, for a given desired momentum $P_z \equiv \frac{2 \pi n_z}{L}$ of the nucleon,
the momentum smearing assumes a quark boost of $\frac{2 \pi k_z}{L}$,
where $n_z, k_z \in \mathbb{Z}$.
For an optimal signal, $k_z$ should be about half of $n_z$.

We use the Qlua software suite~\cite{qlua} for calculating the quark propagators and subsequently constructing the final correlators.
The needed inversions are performed using the multigrid solver in QUDA~\cite{Clark:2009wm,Babich:2011np} and utilize all-mode averaging (AMA)~\cite{Shintani:2014vja} to reduce the total computational cost.
The residual used in our solver is $10^{-10}$ and $10^{-4}$ for exact and sloppy solves, respectively.

Some of the more important details, including the total statistics achieved, can be found in \Cref{tab:setup}.

\begin{table}
\centering
\begin{tabular}{c|c|c|c|c|c|c}
\hline
\hline
Ensembles & $m_\pi$ & $N_{\rm cfg}$ & $n_z$ & $k_z$ & $\ts/a$ & (\#ex,\#sl) \\
$a,L_t \times L_s^3$ & (GeV) & & & \\
\hline
$a=0.076$ fm     & 0.14 & 350 & 0 & 0 & 6         & (1, 16) \\
$64 \times 64^3$ &      &     & 0 & 0 & 8,10      & (1, 32) \\
                 &      &     & 0 & 0 & 12        & (2, 64) \\
                 &      &     & 1 & 0 & 6,8,10,12 & (1, 32) \\
                 &      &     & 4 & 2 & 6         & (1, 32) \\
                 &      &     & 4 & 2 & 8,10,12   & (4, 128) \\
                 &      &     & 6 & 3 & 6         & (1, 20) \\
                 &      &     & 6 & 3 & 8         & (4, 100) \\
                 &      &     & 6 & 3 & 10        & (5, 140) \\
                 &      &     & 6 & 3 & 12        & (13, 416)* \\
\hline
\hline
\end{tabular}
\caption{The more important details on the ensemble and the statistics gathered for our calculation.
         The integer momentum $n_z$ of the nucleon and the corresponding integer boost momentum $k_z$ of the quarks are given.
         The sink-source separations used are given by $t_{\rm sep}$.
         And, finally, the number of samples used for the exact and sloppy solves is given by \#ex and \#sl, respectively.
         The asterisk indicates where extra samples were generated as compared to our previous work in Ref.~\cite{Gao:2022uhg}.}
\label{tab:setup}
\end{table}

\subsection{Correlation functions}
\label{sec:correlators}

We use the standard interpolating operator for the nucleon, given by
\begin{equation}
    N_\alpha (x, t) = \varepsilon_{abc} u_{a \alpha} (x, t) (u_b (x, t)^T C \gamma_5 d_c (x, t)) ,
\end{equation}
where $C = \gamma_t \gamma_y$ is the charge-conjugation matrix.
These are then used to construct the two-point correlation functions as follows
\begin{equation}
\begin{split}
    C^{\rm 2pt}_{\mathcal{P}} & (\pvec, \ts; \xvec, t_0) = \\
    &\sum_{\yvec} e^{-i \pvec \cdot (\yvec - \xvec)} \mathcal{P}^{\rm 2pt}_{\alpha \beta} \braket{N^{(s)}_\alpha (\yvec, \ts + t_0) \overline{N}^{(s^\prime)}_\beta (\xvec, t_0)} ,
\end{split}
\end{equation}
where the superscripts on the nucleon operators specify whether the quarks are smeared ($s = S$) or not ($s = P$),
and $\mathcal{P}^{\rm 2pt}$ is a projection operator.
As described in Ref.~\cite{Gao:2022uhg}, we always use smeared quarks at the source time, but consider both smeared and unsmeared quarks at the sink time, which helps to more reliably extract the spectrum by looking for agreement between the independent analysis of both correlators.

The three-point correlators computed are given by
\begin{equation}
\begin{split}
    C^{\rm 3pt}_{\mathcal{P}, \Gamma, f} & (\pvec , \qvec , \ts, \ti, z ; \xvec, t_0) = \\
    & \sum_{\yvec, \zvec_0} e^{-i \pvec \cdot (\yvec - \xvec)} e^{-i \qvec \cdot (\xvec - \zvec_0)} \mathcal{P}^{\rm 3pt}_{\alpha \beta} \\
    &\times \braket{N_\alpha (\yvec, \ts + t_0) \mathcal{O}^{f}_\Gamma (\zvec_0 + z \hat{z}, \ti + t_0) \overline{N}_\beta (\xvec, t_0)} ,
\end{split}
\end{equation}
where $\mathcal{P}^{\rm 3pt}$ is a projection operator, $\pvec$ is the momentum of the sink nucleon, $\qvec$ is the momentum transfer, and the inserted operator is
\begin{equation}\label{eq.3pt}
\begin{split}
    \mathcal{O}&_\Gamma^{f} (\zvec_0 + z \hat{z}, \ti + t_0) = \overline{\psi}^{f} (\zvec_0, \ti + t_0) \Gamma \\
    &\times W(\zvec_0, \ti + t_0 ; \zvec_0 + z\hat{z} , \ti + t_0) \psi^{f}(\zvec_0 + z\hat{z}, \ti + t_0) ,
\end{split}
\end{equation}
where $\Gamma$ is a product of gamma matrices, $\psi^{f}(\zvec, t)$ is a quark field of flavor $f$,
and $W$ is a straight Wilson line of length $z$ connecting the quark fields.
For the three-point functions, we only consider nucleon operators built from smeared quarks.
The Wilson line is formed from products of the HYP-smeared gauge links and is needed to construct a gauge-invariant operator.
In this work, we consider the light quark flavors $f = u,d$ separately, allowing us to access the isovector ($u-d$) and isoscalar ($u+d$) combinations.
Note, however, that all disconnected contributions are ignored, leading to uncontrolled errors due to their neglect in the isoscalar combination.
This approximation is expected to be reasonable given that estimates from PNDME for the disconnected contributions to the tensor charge have indicated they are smaller than the statistical error on the connected contributions~\cite{Bhattacharya:2015esa,Bhattacharya:2015wna}.
In what follows, we only consider zero momentum transfer $\qvec = 0$ and the sink momenta are always in the $z$-direction $\pvec = \frac{2 \pi n_z}{L} \hat{z} \equiv P_z \hat{z}$.
We use four different values for the sink momenta $n_z \in \left\{0, 1, 4, 6 \right\}$ which in physical units corresponds to $P_z = \left\{0, 0.25, 1.02, 1.53 \right\}$ GeV.
The statistics gathered and quark boosts used for each $n_z$ are given in \Cref{tab:setup}.

In this work, we are interested in the tensor charge and the transversity PDF, which can be accessed with $\Gamma \propto \sigma^{z j}$ (with $j$ being either $x$ or $y$) and
\begin{equation}
    \mathcal{P}^{\rm 3pt} = \frac{1}{2} (1 + \gamma_t)(1 - i \gamma_5 \hat{s}\cdot\vec{\gamma})
\end{equation}
which projects the nucleon to positive parity and its spin to be aligned along the direction given by $\hat{s}$.
Here we use $\Gamma = -i\sigma^{zy} = -i \gamma_z \gamma_y$ and $\hat{s} = \hat{x}$.
Throughout the remainder of the text, we use $\delta$ to denote the specific operator and polarization used, which is motivated by the standard usage of $\delta q (x)$ in the literature for the transversity PDF.

In order to guarantee the cancellation of amplitudes that appear in the spectral decompositions of the three- and two-point functions, we set $\mathcal{P}^{\rm 2pt} = \mathcal{P}^{\rm 3pt} \equiv \mathcal{P}$, and we will denote this in the two-point functions by $C^{\rm 2pt}_{S_x}$.
\section{Ground-state matrix elements}
\label{sec:analysis}

In this section, we extract the ground-state bare matrix elements from the three-point correlation functions.
Our analysis strategy is nearly identical to that used in our previous work of Ref.~\cite{Gao:2022uhg}, and we repeat the most important details here for convenience.
The only difference in the strategy is the choice in our preferred fit ranges.
In this work, the quality of our data has increased, giving us more confidence in our fits, and therefore, we end up excluding less time insertions from our final fits.

Our approach first extracts the spectrum and ratios of amplitudes from the two-point correlation functions in order to use these as priors on the parameters shared in our fits to the ratio of three-point to two-point functions.
Although the two-point correlation functions differ slightly from those used in our previous work in Ref.~\cite{Gao:2022uhg}, because $\mathcal{P}^{\rm 2pt}$ is different, we do not include any discussions here, as the analysis strategy is identical and the results only change by a slight increase in the error.
The increase in error can be understood from the fact that the change in $\mathcal{P}^{\rm 2pt}$ amounts to only using a single spin polarization, as opposed to averaging over both spin polarizations as done previously.

\subsection{Analysis strategy}
\label{subsec:analysis_strat}

\begin{figure*}
    \centering
    \includegraphics[width=0.32\textwidth]{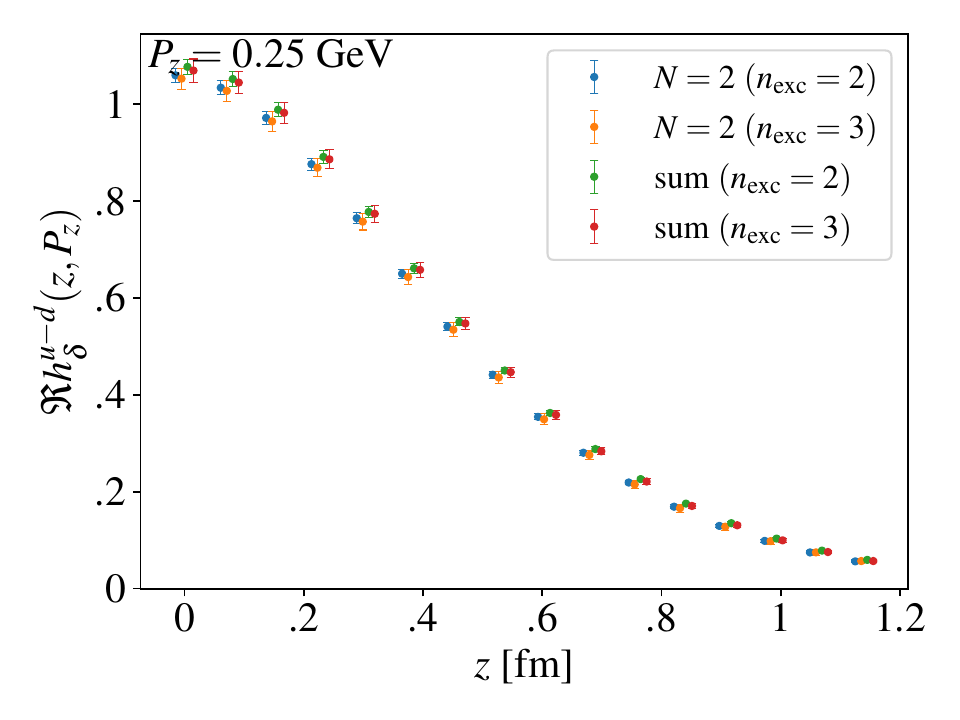}
    \includegraphics[width=0.32\textwidth]{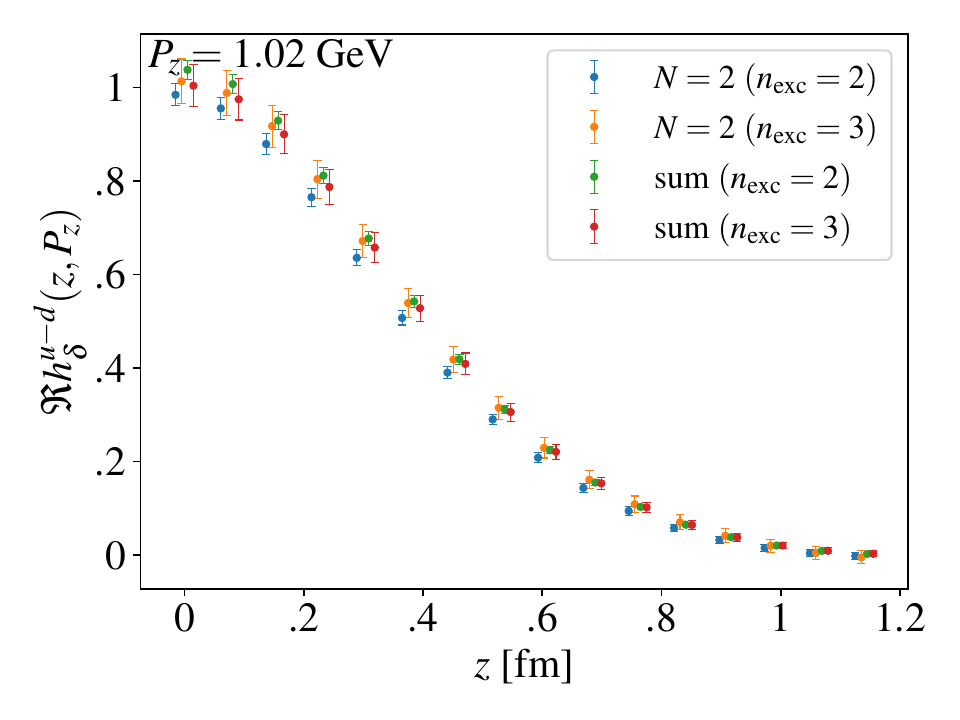}
    \includegraphics[width=0.32\textwidth]{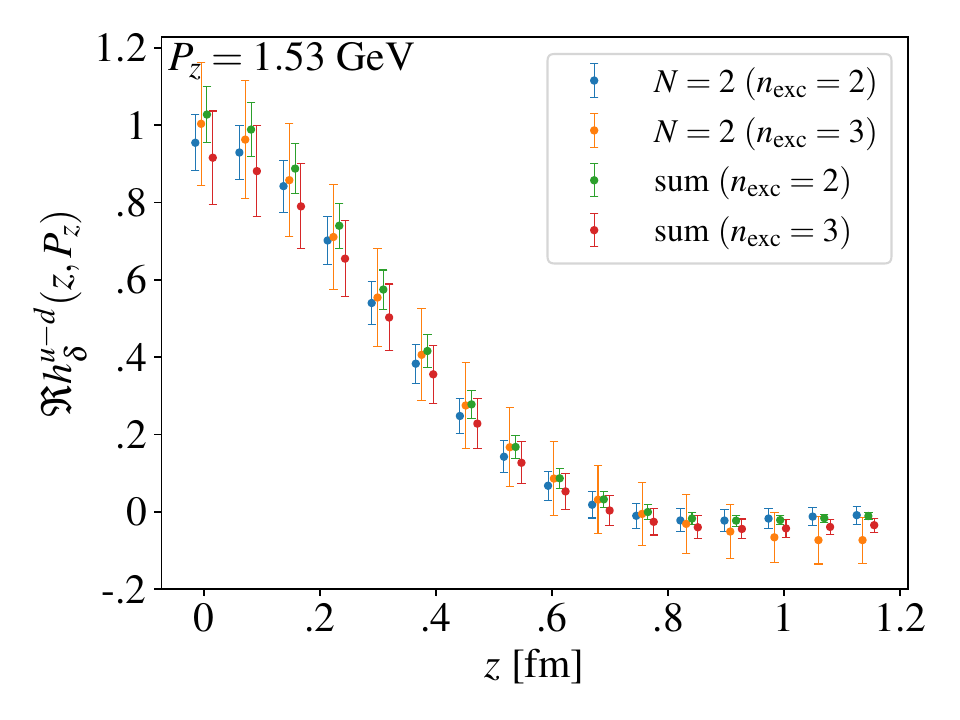}
    \includegraphics[width=0.32\textwidth]{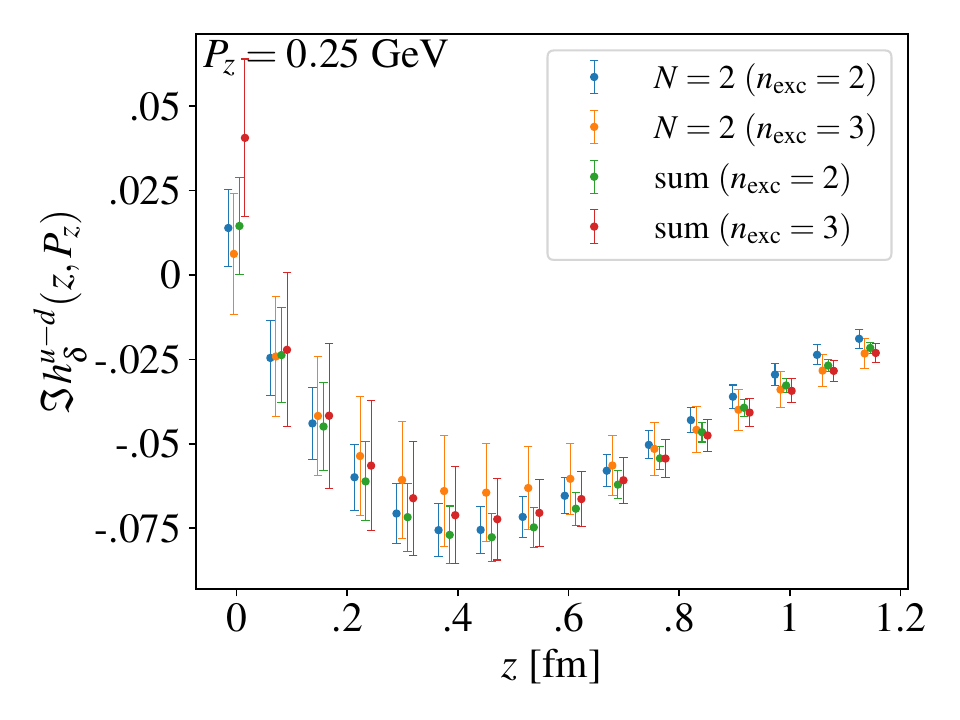}
    \includegraphics[width=0.32\textwidth]{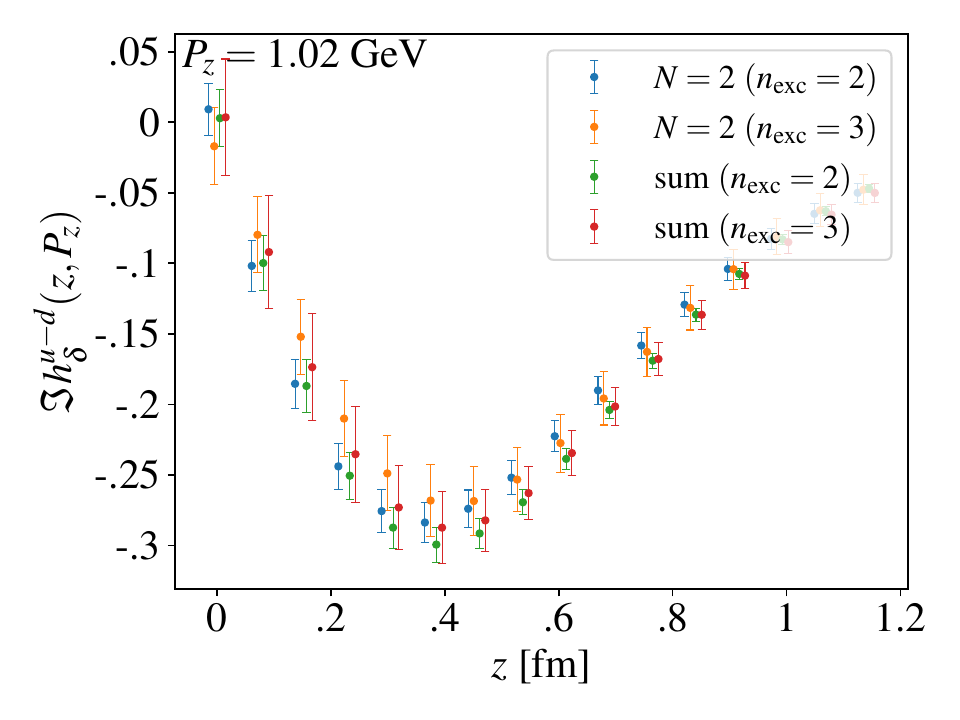}
    \includegraphics[width=0.32\textwidth]{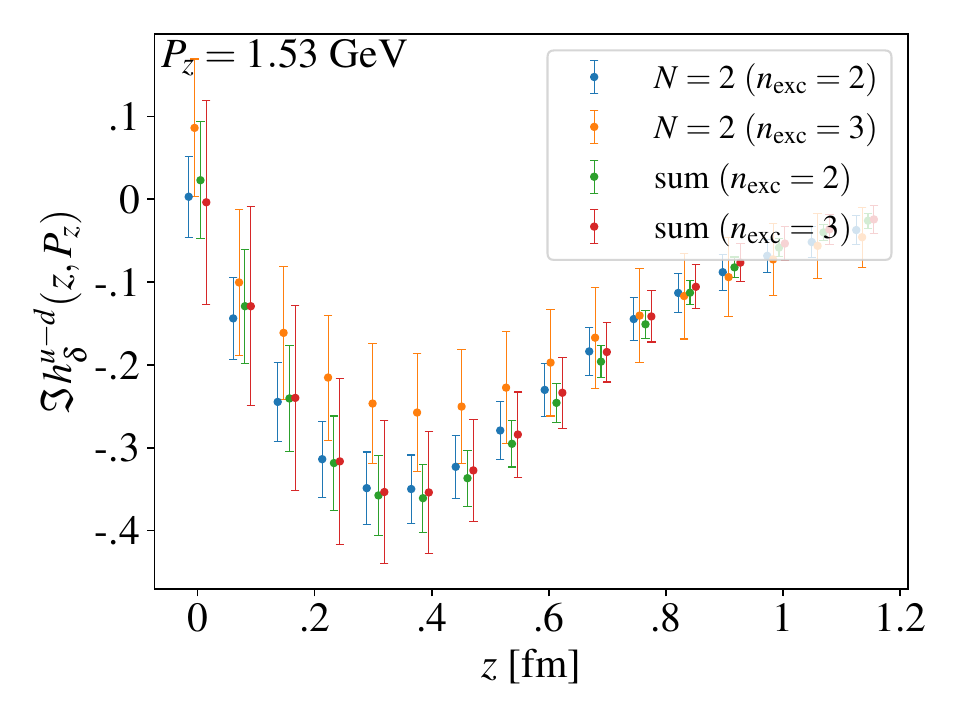}    
    \caption{The Wilson-line length dependence of the (upper) real and (lower) imaginary parts of the
             isovector ground-state bare matrix elements from two-state and summation fits with $n_{\rm exc} = 2,3$
             for the three nonzero values of momentum considered (one for each column).
             The results shown are averaged with the negative $z$ fits.}
    \label{fig:u-d_bare_mat_comp}
\end{figure*}

We follow the standard approach for extracting the bare matrix elements, which begins by forming an appropriate ratio of the three-point to two-point correlation functions given by
\begin{equation}
    R^{f}_{\delta}(P_z, \ts, \ti, z) \equiv \frac{C^{\rm 3pt}_{\delta, f}(\pvec=P_z \hat{z}, \qvec=0, \ts, \ti, z)}{C^{\rm 2pt}_{S_x}(\pvec, \ts)} .
    \label{eq:ratio}
\end{equation}
The main reason for this choice is that it can be shown that
\begin{equation}
    \lim_{\ti,\ts \to \infty} R^{f}_{\delta} (P_z, \ts, \ti, z) = h_{\delta;0,0}^{f} (z, P_z) ,
\end{equation}
where $h_{\delta;0,0}^{f} (z, P_z)$ is the desired bare ground-state matrix element.
Since the values of $\ts$ considered here are not likely in the asymptotic region,
we include the effects from the lowest $N$ states by substituting the spectral decompositions of the three- and two-point functions truncated at the $N$th state.
After some algebra, we find
\begin{equation}
    \begin{split}
    &R^{f}_{\delta} (P_z, \ts, \ti, z; N) = \\
     &\frac{\sum_{m,n=0}^{N-1} h^{\prime f}_{\delta;m,n} \prod_{l,k,r=1}^{m} e^{-\Delta_{l,l-1}\ts} e^{(\Delta_{k,k-1} - \Delta_{r,r-1}) \ti}}{1 + \sum_{i=1}^{N-1} r_i \prod_{j=1}^{i} e^{-\Delta_{j,j-1}\ts}} ,
    \end{split}
\end{equation}
where $\Delta_{i,j} \equiv E_i - E_j$, $r_i \equiv |A_{\alpha}^{(i)}(P_z)|^2/|A_{\alpha}^{(0)}(P_z)|^2$, $A_{\alpha}^{(n)} (P_z) \equiv \bra{\Omega} N_{\beta} \mathcal{P}_{\beta \alpha} \ket{n, P_z}$ ($\ket{\Omega}$ is the vacuum state and $\ket{n, P_z}$ is the $n$-th nucleon state with momentum $P_z$), and
\begin{equation}
    h^{\prime f}_{\delta;m,n} \equiv \frac{A^{(m)}_\alpha (P_z) A^{(n)}_\alpha (P_z)^\ast h^{f}_{\delta; m,n} (z, P_z)}{A^{(0)}_\alpha (P_z) A^{(0)}_\alpha (P_z)^\ast} .
\end{equation}
The parameters $h^{\prime f}_{\delta;m,n}$ depend on $z$ and $P_z$, but this dependence is suppressed to save space.
For convenience, we typically suppress the indices on the matrix elements when referring to the ground state matrix element (i.e. $h^{f}_{\delta} (z, P_z) \equiv h^{f}_{\delta;0,0} (z, P_z)$).
As the excited-state matrix elements are never used, this should not cause any confusion.
We fit the ratio of data in \Cref{eq:ratio} to $R^f_\delta(P_z, \ts, \ti, z; N)$,
where $h^{\prime f}_{\delta;m,n}$, $\Delta_{i,j}$, and $r_i$ are the fit parameters.
The parameters $\Delta_{i,j}$ and $r_i$ are priored using the fit results from the two-point functions (see Ref.~\cite{Gao:2022uhg} for details).
In this work, we only consider $N=1,2$, as our limited data tend to lead to unreliable fits when $N > 2$.

In order to reduce the effects from unaccounted for excited-states as much as possible, we remove some of the data points nearest the sink and source times.
We do this in a symmetric way, i.e. for each $\ti$ not included in the fit, we also do not include $\ts - \ti - 1$.
We define $n_{\rm exc}$ to be the number of insertion times removed on each side of the middle point for each $\ts$.
Therefore, for each $\ts$, the insertion times included in the fit are $\ti \in [n_{\rm exc}+1, \ts - n_{\rm exc} - 1]$.
However, making $n_{\rm exc}$ too large can leave too little data left, and therefore we only consider $n_{\rm exc} \le 3$.
As described in our previous work of Ref.~\cite{Gao:2022uhg}, the two-point function fits show contributions from three states for $\ts \le n_{\rm exc} + 1$ requiring the use of an effective value for the prior on the gap $\Delta_{1,0}$ that takes into account effects from higher states.
The specific value used for the prior on the gap comes from the two-state fit to the two-point function with the lower fit range $t_{\rm min} = n_{\rm exc} + 1$.

\begin{figure*}
    \centering
    \includegraphics[width=0.32\textwidth]{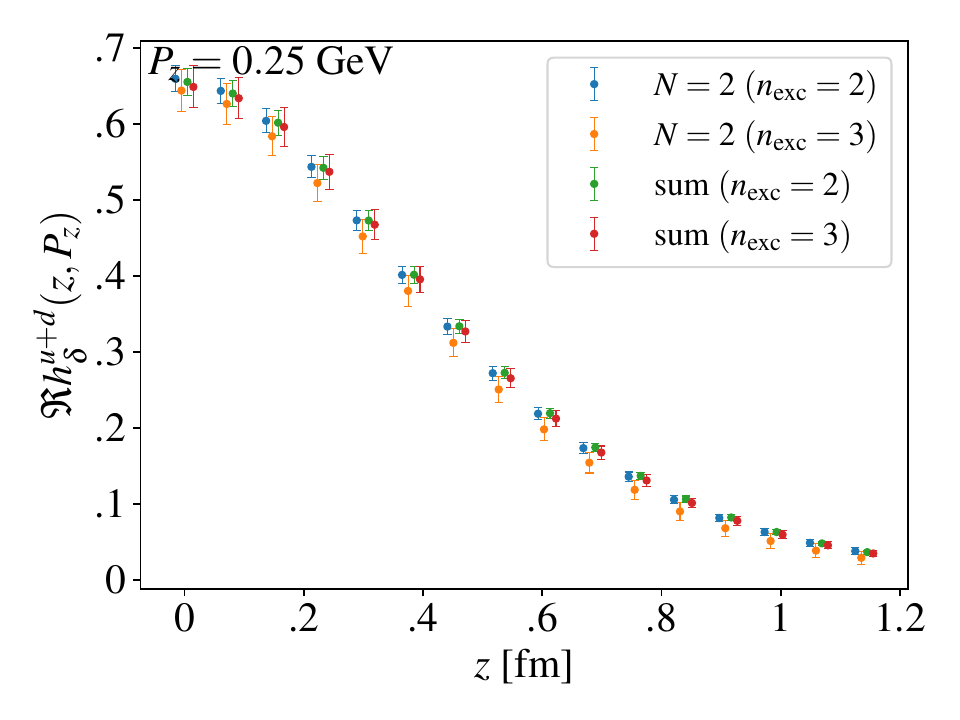}
    \includegraphics[width=0.32\textwidth]{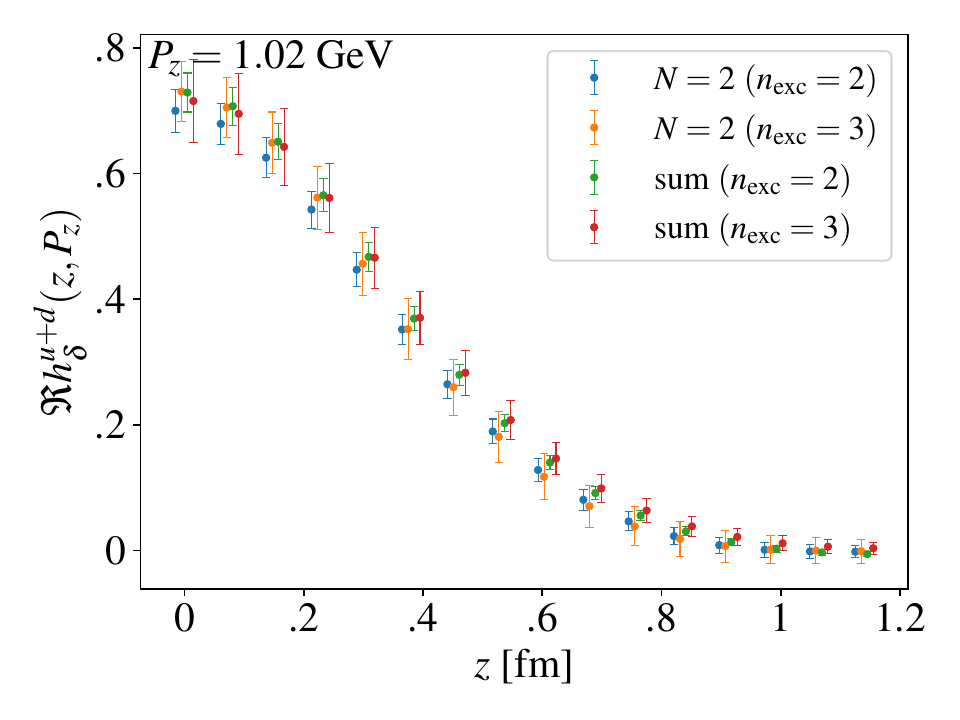}
    \includegraphics[width=0.32\textwidth]{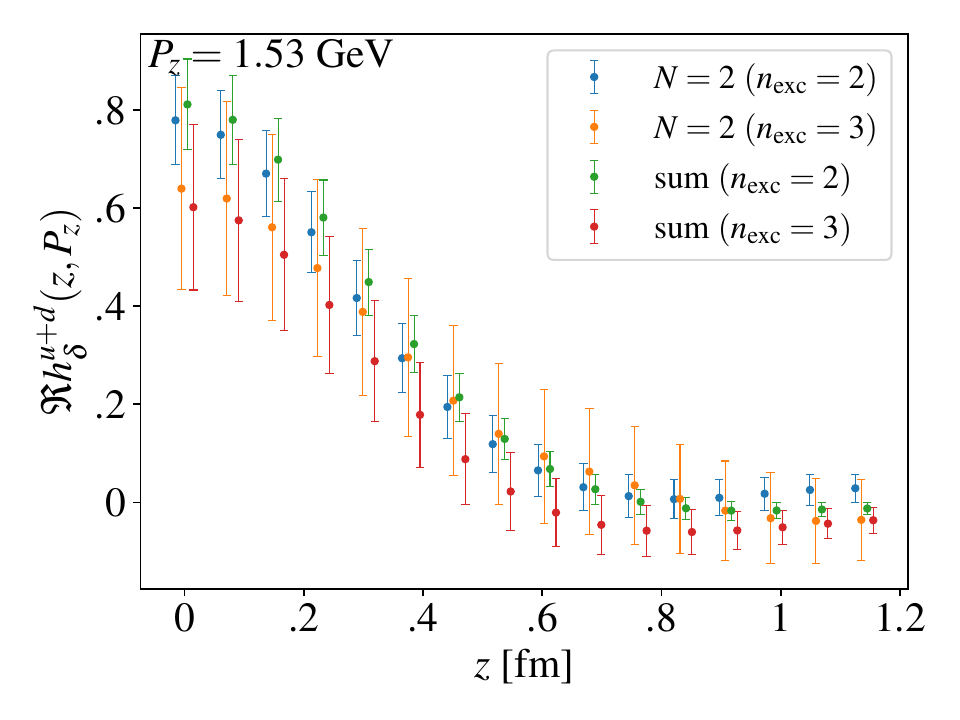}
    \includegraphics[width=0.32\textwidth]{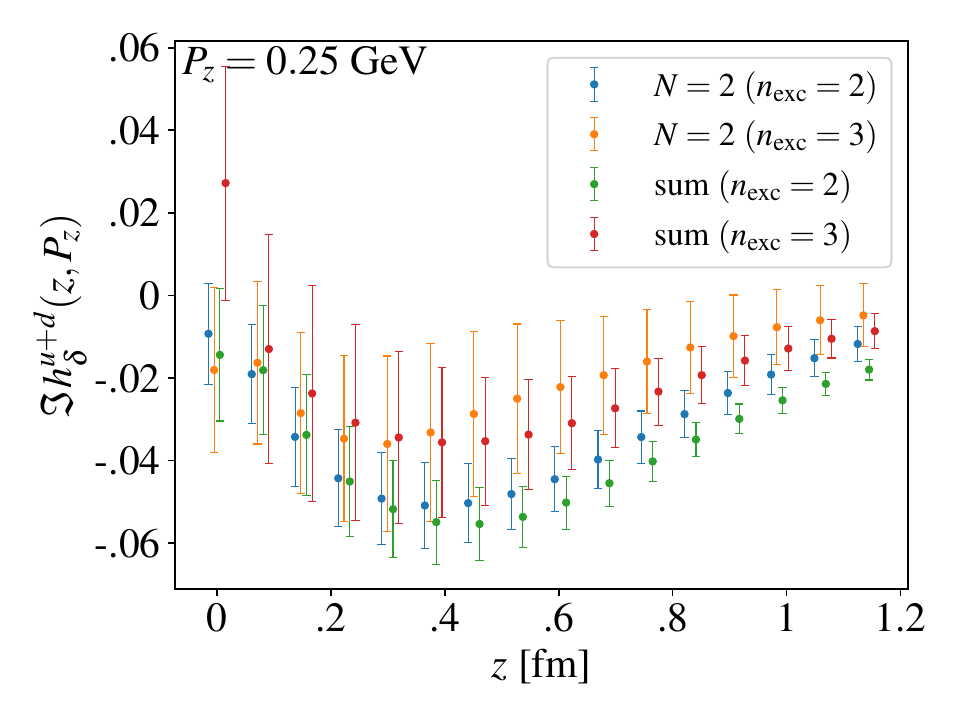}
    \includegraphics[width=0.32\textwidth]{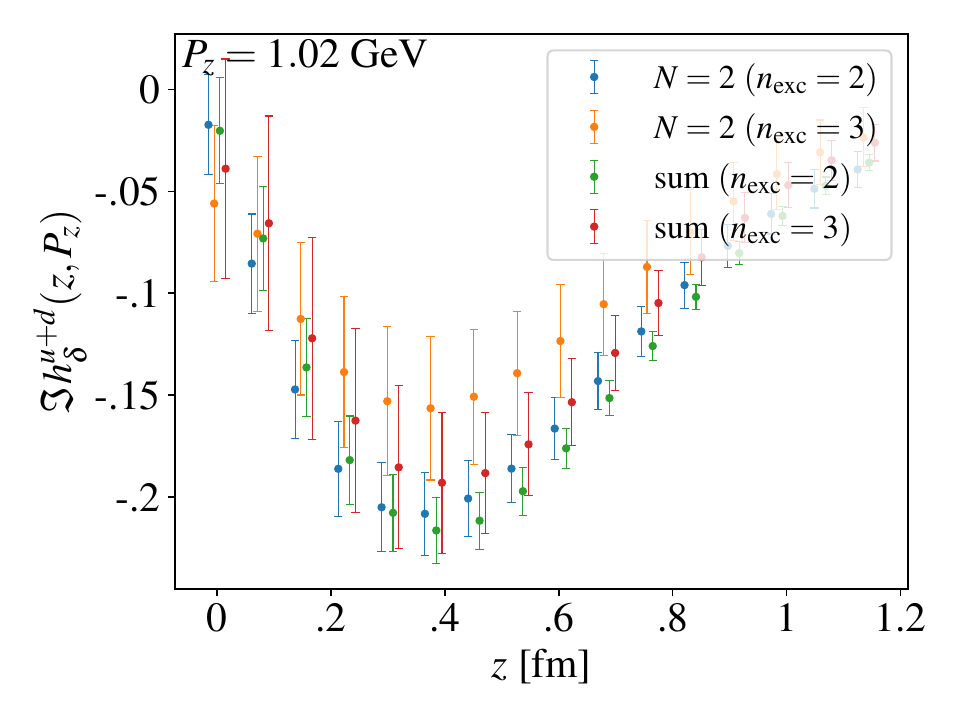}
    \includegraphics[width=0.32\textwidth]{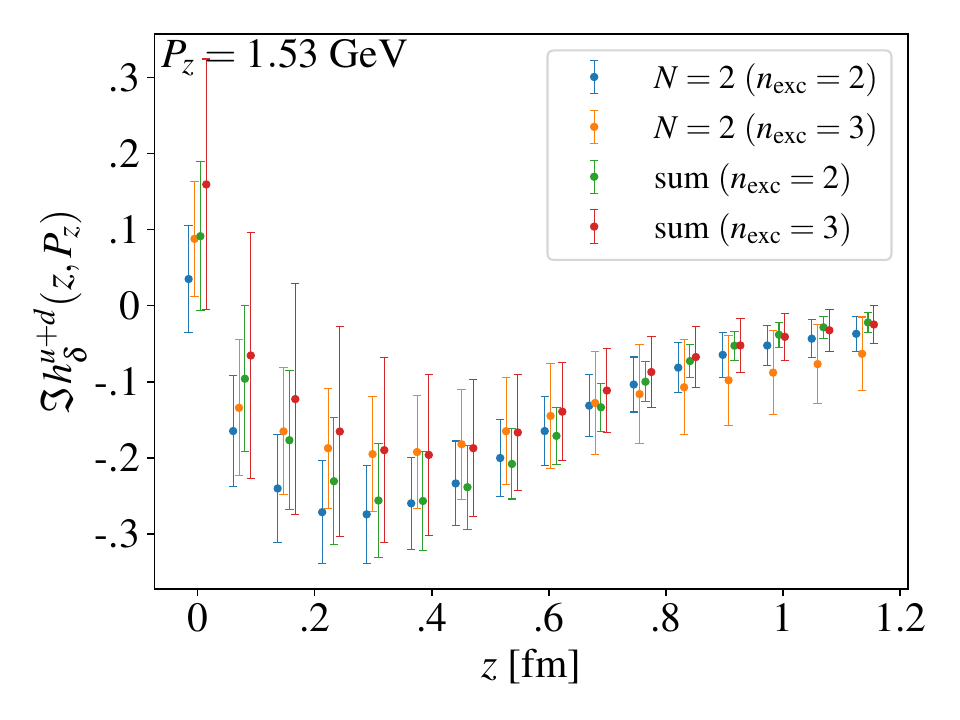}    
    \caption{The same as \Cref{fig:u-d_bare_mat_comp}, but for the isoscalar
             matrix elements.}
    \label{fig:u+d_bare_mat_comp}
\end{figure*}

As an additional consistency check on our fit results, we also make use of the summation method, which involves first summing $R^{f}_{\delta}(P_z, \ts, \ti, z)$ over the subset $\ti \in [n_{\rm exc}+1, \ts - n_{\rm exc} - 1]$ for each $\ts$
\begin{equation}
    S^{f}_{\delta} (P_z, \ts, z; n_{\rm exc}) \equiv \sum_{\ti = n_{\rm exc}+1}^{\ts - n_{\rm exc} - 1} R^{f}_{\delta} (P_z, \ts, \ti, z) ,
\end{equation}
which reduces the leading contamination from excited states.
The bare ground-state matrix element can then be extracted from a linear fit to the sum as
\begin{equation}
    S^{f}_{\delta} (P_z, \ts, z; n_{\rm exc}) = B_0 + \ts h^{f}_{\delta} (z, P_z) .
\end{equation}

In \Cref{fig:u-d_bare_mat_comp,fig:u+d_bare_mat_comp}, we show comparisons of the two-state and summation fit results for the isovector and isoscalar combinations, respectively, as a function of the Wilson line length for both $n_{\rm exc} = 2$ and $3$.
We see generally good agreement across these different fits, and, with the better data quality as compared to the unpolarized case,
we choose our preferred fit as the two-state fit to the ratio \Cref{eq:ratio} with $n_{\rm exc} = 2$.

Several representative fits, all using our preferred fit strategy, are shown in \Cref{app:three_point_fits}.
There we include the fits to the zero-momentum local matrix elements, relevant for the tensor charge, in \Cref{fig:c3pt_loc_u-d_and_u+d_repr_fits_re} for the isovector and isoscalar combinations.
We also include various fits to the non-local matrix elements, relevant for information on the PDFs, in \Cref{fig:c3pt_u-d_repr_fits_re,fig:c3pt_u-d_repr_fits_im} for the isovector combination and in  \Cref{fig:c3pt_u+d_repr_fits_re,fig:c3pt_u+d_repr_fits_im} for the isoscalar combination.

\subsection{Tensor charge \texorpdfstring{$g_T$}{gT}}

Although the focus of this work is based on the non-local matrix elements,
we first turn our attention to the local ones which give us access to the nucleon tensor charge $g_T$.
The bare matrix elements must be renormalized in a standard scheme (like $\overline{\text{MS}}$) in order to make comparisons with phenomenological results and other lattice determinations.
The matrix elements are multiplicatively renormalizable,
and we first determine the ratio of renormalization constants $Z_T/Z_V$ in the RI-MOM scheme which is then converted to $\overline{\text{MS}}$ at four-loop accuracy~\cite{Gracey:2022vqr} and subsequently evolved to the scale $\mu = 2$ GeV using the two loop evolution function~\cite{Chetyrkin:1997dh} (see \Cref{app:ZT} for details).
Then, using the ratio of bare charges $g_T^{\rm bare} / g_V^{\rm bare}$ along with the expectation of $Z_V g_V^{\rm bare} = 1$,
the renormalized tensor charge $g_T = Z_T g_T^{\rm bare}$ can be determined.
Using our estimate for $Z_T/Z_V$, we find
\begin{equation}
\begin{split}
    g_T^{u-d} = 1.01(3) , \;\; \overline{\text{MS}} (\mu = 2 \; \text{GeV}) ,\\
    g_T^{u+d} = 0.61(2) , \;\; \overline{\text{MS}} (\mu = 2 \; \text{GeV}) .
\end{split}
\end{equation}
In \Cref{tab:tensor_charge}, we show a comparison of our results to the other $N_f = 2 + 1$ results given in the FLAG review 2021~\cite{FlavourLatticeAveragingGroupFLAG:2021npn}.

\begin{table}
\centering
\begin{tabular}{cccc}
\hline
\hline
& $g_T^{u-d}$ & $g_T^u$ & $g_T^d$ \\
\hline
This work & 1.01(3) & 0.81(2) & -0.20(1) \\
NME~\cite{Park:2021ypf} & 0.95(5)(2) & & \\
RBC/UKQCD~\cite{Abramczyk:2019fnf} & 1.04(5) & & \\
Mainz~\cite{Harris:2019bih,Djukanovic:2019gvi} & $0.965(38)(^{13}_{41})$ & 0.77(4)(6) & -0.19(4)(6) \\
LHPC~\cite{Hasan:2019noy} & 0.972(41) & & \\
JLQCD~\cite{Yamanaka:2018uud} & 1.08(3)(3)(9) & 0.85(3)(2)(7) & -0.24(2)(0)(2) \\
LHPC~\cite{Green:2012ej} & 1.038(11)(12) & & \\
RBC/UKQCD~\cite{Aoki:2010xg} & 0.9(2) & & \\
\hline
\end{tabular}
\caption{Comparison of our extracted tensor charges with those in the FLAG review 2021~\cite{FlavourLatticeAveragingGroupFLAG:2021npn} with $N_f = 2 + 1$.
The results are ordered by year.}
\label{tab:tensor_charge}
\end{table}

\section{Mellin Moments from the Leading-twist OPE}
\label{sec:leading_twist_OPE}

We now move on to the extraction of the lowest few Mellin moments using the leading-twist OPE approximation.
Here we avoid the need for the renormalization factors, which depend on the Wilson-line length $z$ and the lattice spacing $a$, by forming the renormalization-group invariant ratio~\cite{Orginos:2017kos,Fan:2020nzz}
\begin{equation}
    \mathcal{M}^{f/f^\prime}_{\delta} (\lambda, z^2; P_z^0) = \frac{h^{f}_{\delta} (z, P_z)}{h^{f^\prime}_{\delta} (z, P_z^0)} \big/ \frac{h^{f}_{\delta} (0, P_z)}{h^{f^\prime}_{\delta} (0, P_z^0)} ,
    \label{eq:pITD}
\end{equation}
where $\lambda \equiv z P_z$ is known as the Ioffe time.
In the literature, this ratio is referred to as the Ioffe time pseudo-distribution (pseudo-ITD).
In order to cancel the renormalization factors, the $z=0$ matrix elements are not necessary,
but this choice is favorable in that it enforces a normalization and cancels correlations.
In this work, we only consider the case with $P_z^0 = 0$, commonly referred to as the reduced pseudo-ITD~\cite{Orginos:2017kos,Joo:2019jct,Joo:2020spy,Bhat:2020ktg,Karpie:2021pap,Egerer:2021ymv,Bhat:2022zrw}.
Additionally, since there are no gluons involved in the case of the transversity distributions,
the leading-twist OPE expansion of the pseudo-ITD does not depend on the flavor combination $f^\prime$, even if $f \neq f^\prime$, and we, therefore, opt to omit the $f^\prime$ from our notation in order to not be overly cumbersome.
In what follows, when extracting the isovector flavor combination, $f = f^\prime = u-d$, and for the isoscalar flavor combination, $f = u+d$ and $f^\prime=u-d$.

Then, using the leading-twist OPE approximation, we can write down the reduced pseudo-ITD as an expansion in Mellin moments~\cite{Karpie:2018zaz}
\begin{equation}
\begin{split}
    \mathcal{M}^{f}_{\delta}(\lambda, z^2, P_z^0=0) = \sum_{n=0} \frac{C^\delta_n(\mu^2 z^2)}{C^\delta_0(\mu^2 z^2)} & \frac{(-i \lambda)^n}{n!} \frac{\braket{x^n}^{f}_\delta (\mu)}{\braket{x^0}^{f}_\delta (\mu)} \\
    &+ \mathcal{O}(\Lambda^2_{\rm QCD} z^2) ,
\end{split}
\label{eq:ope}
\end{equation}
where $C^\delta_n (\mu^2 z^2)$ are the Wilson coefficients for the transversity computed in the ratio scheme up to NLO in the strong coupling $\alpha_s(\mu)$,\footnote{We note that the anomalous dimension of the transversity matrix elements are known to N3LO~\cite{Blumlein:2021enk}.} which at fixed order are given by~\cite{Chen:2016utp,HadStruc:2021qdf}
\begin{equation}
\begin{split}
    C_{n,{\rm NLO}}^\delta (\mu^2 z^2) = 1 + \frac{\alpha_s(\mu) C_F}{2 \pi} & \Bigg[2\ln (\frac{\mu^2 z^2 e^{2 \gamma_E + 1}}{4}) \sum_{j = 2}^{n+1} \frac{1}{j} \\
     &- 2 \Bigg(\sum_{j=1}^n \frac{1}{j}\Bigg)^2 - 2 \sum_{j=1}^n \frac{1}{j^2}\Bigg] ,
\end{split}
\end{equation}
$C_F = 4/3$,
and $\braket{x^n}^{f}_\delta(\mu)$ is the $n$th Mellin moment of the transversity PDF of flavor $f$ defined at the factorization scale $\mu$, i.e.
\begin{equation}
    \braket{x^n}^{f}_\delta (\mu) = \int_{-1}^{1} \mathrm{d}x \, x^n \delta q^f (x, \mu) ,
\end{equation}
where $\delta q^{f} (x, \mu)$ is the transversity PDF of a quark with flavor $f$ for $x \geq 0$ and of its antiquark for $x < 0$.
Estimates for the strong coupling itself are determined from Ref.~\cite{Petreczky:2020tky}, and we exclusively work at the scale $\mu = 2$ GeV, resulting in $\alpha_s (\mu = 2$ GeV$) = 0.2930$.
Further, we also consider the effects from target mass corrections (TMCs), which can be incorporated with the following substitution
\begin{equation}
    \braket{x^n}^f_\delta \to \braket{x^n}^f_\delta \sum_{k=0}^{\lfloor n/2 \rfloor} \frac{(n-k)!}{k! (n-2k)!} \bigg(\frac{m^2_N}{4 P^2_z} \bigg)^k .
\end{equation}

As the Wilson coefficients are all real, it is clear from \Cref{eq:ope} that the real and imaginary parts of the reduced pseudo-ITD, $\mathcal{M}_\delta^f(\lambda, z^2, 0)$, can be written solely in terms of the even and odd moments, respectively.
Therefore, we choose to separately fit the real and imaginary parts of the reduced pseudo-ITD to 
\begin{equation}
\begin{split}
    \Re \mathcal{M}^{f}_{\delta} & (\lambda, z^2, P_z^0=0) = \\
    &\sum_{n=0}^{\lfloor N_{\rm max}/2 \rfloor} \frac{C^\delta_{2n} (\mu^2 z^2)}{C^\delta_0 (\mu^2 z^2)} \frac{(-i \lambda)^{2n}}{(2n)!} \braket{x^{2n}}^{\prime f}_\delta , \\
    \Im \mathcal{M}^{f}_{\delta} & (\lambda, z^2, P_z^0=0) = \\
    &\sum_{n=1}^{\lceil N_{\rm max}/2 \rceil} \frac{C^\delta_{2n-1} (\mu^2 z^2)}{C^\delta_0 (\mu^2 z^2)} \frac{(-i \lambda)^{2n-1}}{(2n-1)!} \braket{x^{2n-1}}^{\prime f}_\delta ,
\end{split}
\label{eq:ope_fits}
\end{equation}
respectively, where the reduced moments $\braket{x^n}^{\prime f}_\delta \equiv \braket{x^n}^f_\delta / \braket{x^0}^f_\delta$ with $n>0$ are the fit parameters.
The $n=0$ reduced moment is identically one, which is enforced explicitly in the fit.
Additionally, $g_T^f \equiv \braket{x^0}^f_\delta$, which implies that the reduced moments are the original moments in units of the tensor charge, and we express all results as such.

We start the analysis as before in Ref.~\cite{Gao:2022uhg} by first assessing the validity of the leading-twist approximation (i.e. how important are the $\mathcal{O}(\Lambda^2_{\rm QCD} z^2)$ corrections which are ignored).
To this end, we perform fits to \Cref{eq:ope_fits} at only a single value for $z^2$ (referred to as a fixed-$z^2$ analysis) and look for any dependence of the extracted moments on the specific value of $z^2$.
Observing little or no dependence on $z^2$ would suggest that the higher-twist contributions are negligible within our statistics and that the leading-twist approximation is valid.

As $z$ increases, the higher-moment terms in \Cref{eq:ope} begin to become important.
We can determine when these higher-moment terms are expected to be non-negligible by using the leading-twist OPE with the moments extracted from the global analysis of JAM3D-22~\cite{Gamberg:2022kdb}.
We found that including two $n \neq 0$ moments in the OPE for both the real and imaginary parts is necessary for $z > 4a \sim 0.304$ fm, and that a third $n \neq 0$ moment becomes necessary for $z > 8a \sim 0.608$ fm.
However, using only one value of $z^2$ allows for up to two moments to be fit to both the real and imaginary data, as the number of non-zero $P_z$ considered is three.
Therefore, for the fixed-$z^2$ analysis, the largest value of $z$ used is $z = 8a$ and we use two moments in the fits when $z > 4a$.

\begin{figure}
    \centering
    \includegraphics[width=\columnwidth]{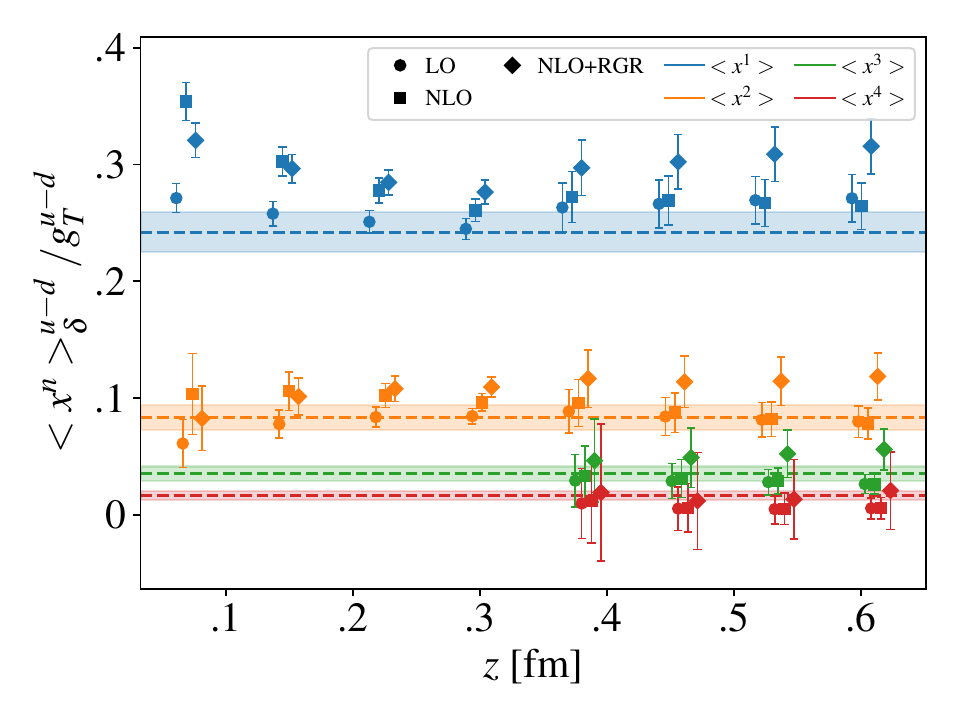}
    \includegraphics[width=\columnwidth]{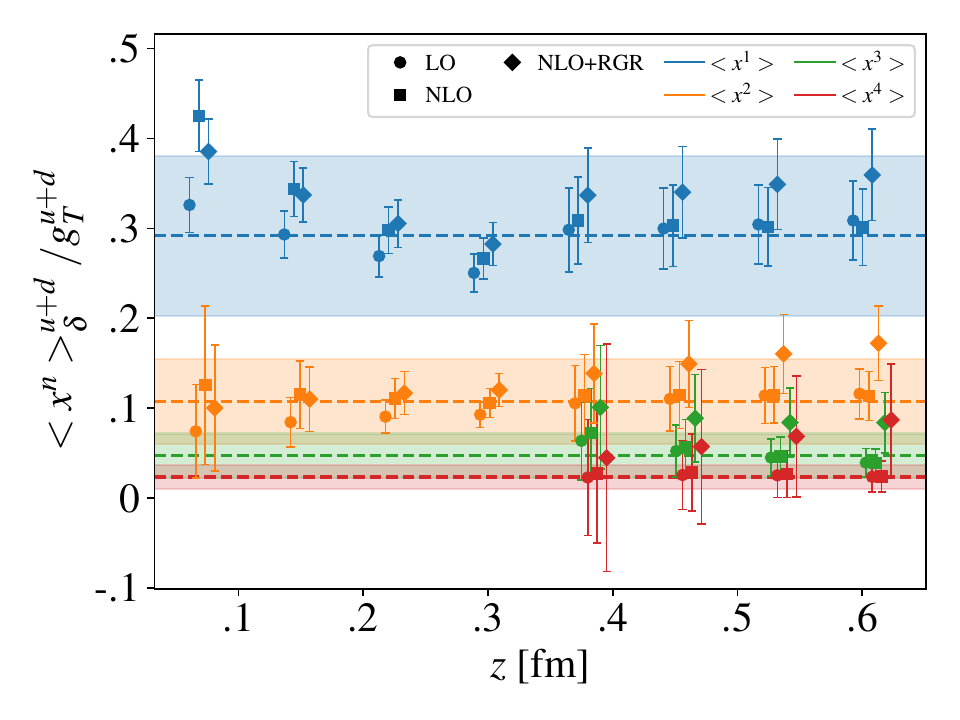}
    \caption{Results for the lowest four Mellin moments of the (upper) isovector and (lower) isoscalar PDF as a function of $z$
             from fits of the
             reduced pseudo-ITD at fixed $z$ with $n_z \in [1,4,6]$ to
             the leading-twist OPE using LO, fixed-order NLO, and NLO+RGR
             Wilson coefficients evaluated at $\mu = 2$ GeV, all of which include TMCs.
             Only the first two
             moments are extracted for $z \leq 4a$.
             The horizontal dashed lines and bands correspond to the central values and errors, respectively,
             of the moments extracted from the global analysis of JAM3D-22~\cite{Gamberg:2022kdb} defined at the scale $Q = 2$ GeV.}
    \label{fig:fixed_z}
\end{figure}

\begin{figure}
    \centering
    \includegraphics[width=\columnwidth]{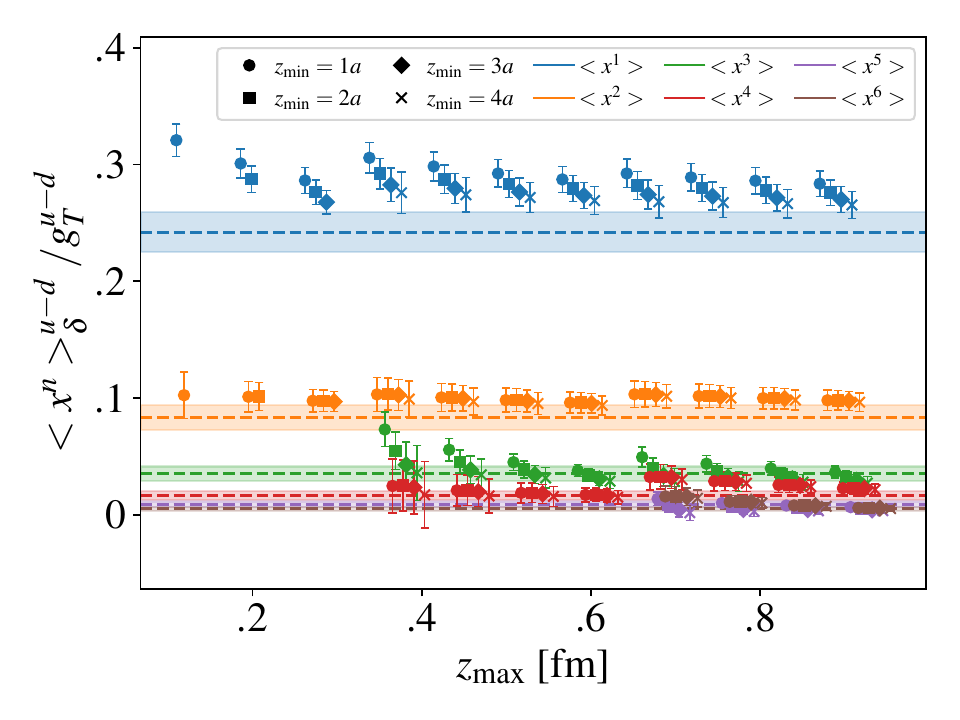}
    \includegraphics[width=\columnwidth]{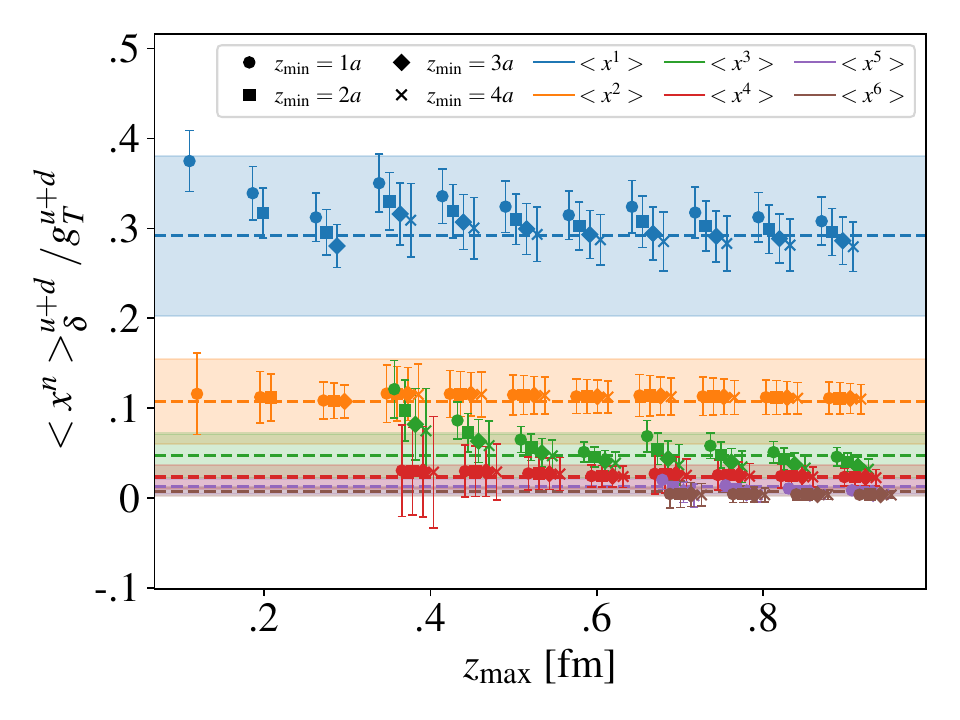}
    \caption{Results for the lowest four Mellin moments of the (upper) isovector and
             (lower) isoscalar PDF from uncorrelated fits of the reduced pseudo-ITD to
             the leading-twist OPE as a function of $z_{\rm max}$, with
             $z \in [z_{\rm min}, z_{\rm max}]$ and $n_z \in [1, 4, 6]$.
             The results use the fixed-order NLO Wilson coefficients evaluated at $\mu = 2$ GeV
             and include TMCs.
             Only the first two moments are considered for $z_{\rm max} \leq 4a$.
             The next two moments, $\braket{x^3}$ and $\braket{x^4}$ are included for $4a < z_{\rm max} \leq 8a$.
             And, two more moments, $\braket{x^5}$ and $\braket{x^6}$ are included for $z_{\rm max} > 8a$.
             The horizontal dashed lines are the same as in \Cref{fig:fixed_z}.}
    \label{fig:global_z}
\end{figure}

\begin{figure*}
    \centering
    \includegraphics[width=\columnwidth]{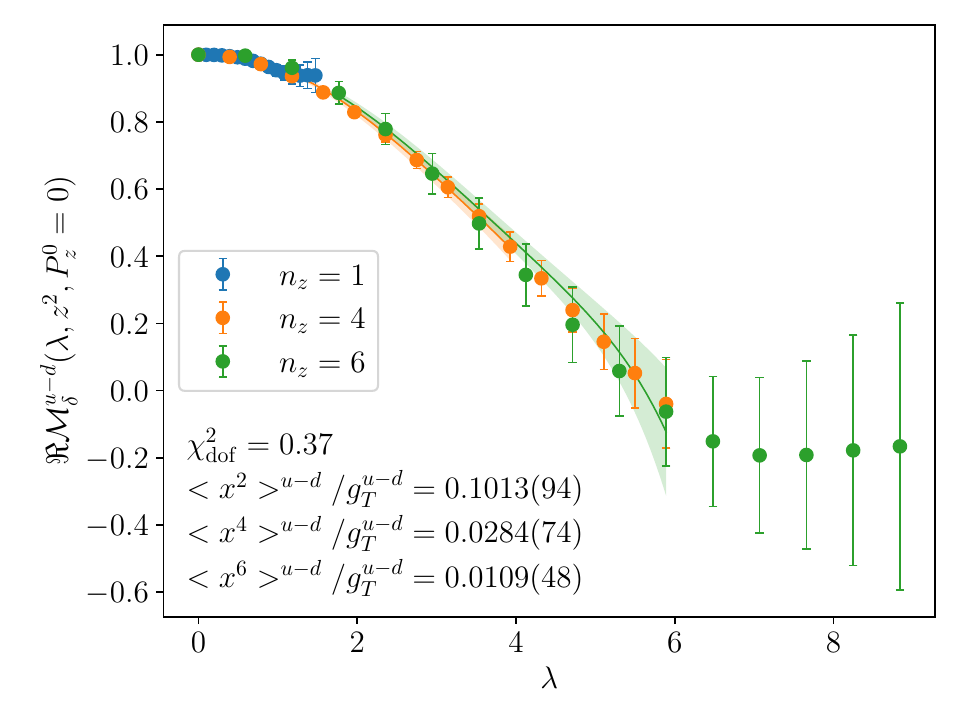}
    \includegraphics[width=\columnwidth]{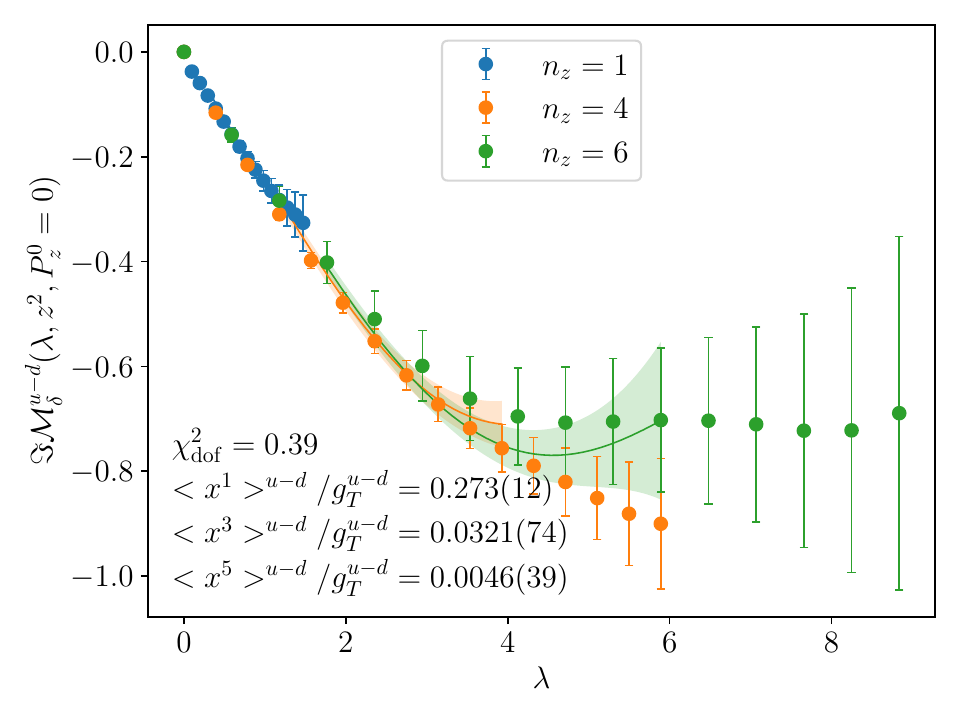}
    \includegraphics[width=\columnwidth]{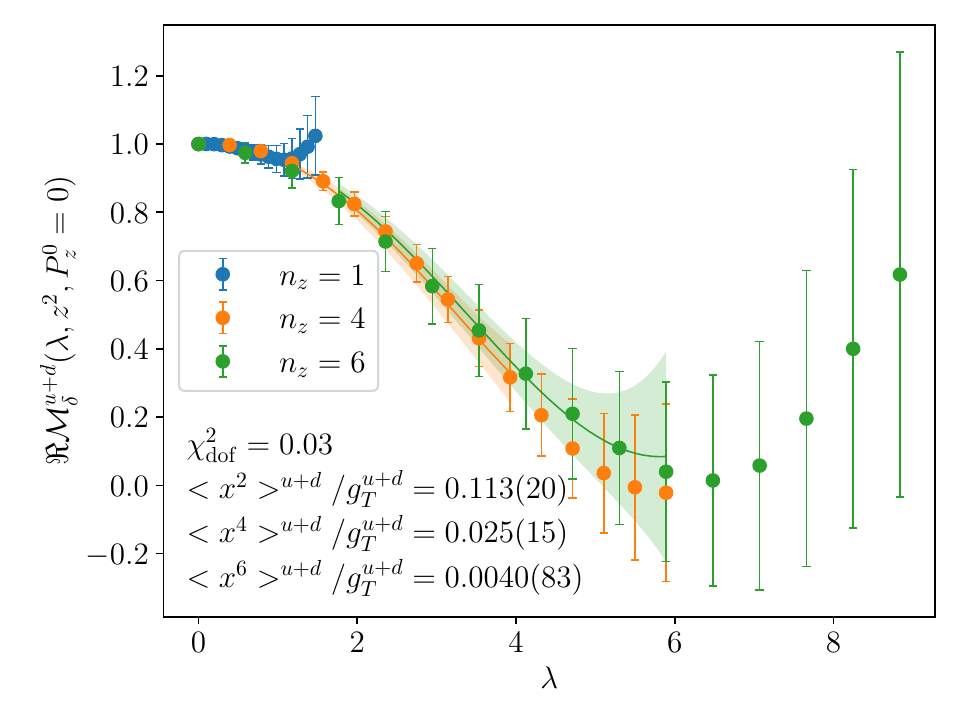}
    \includegraphics[width=\columnwidth]{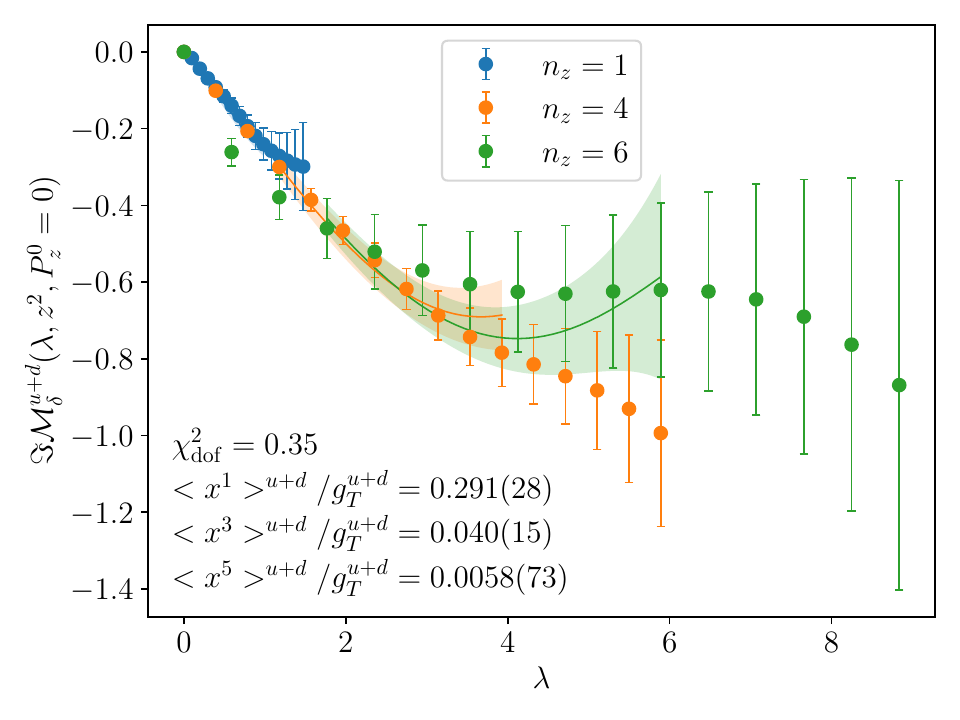}
    \caption{The (left) real and (right) imaginary parts of the (upper) isovector and (lower) isoscalar reduced pseudo-ITD
             for the three momentum used in this work.
             The data come from our preferred fit strategy
             described in \Cref{subsec:analysis_strat}.
             The fit ranges used are $z \in [3a, 10a]$.
             The shaded bands correspond to the fits using the leading-twist
             OPE with fixed-order NLO Wilson coefficients and TMCs evaluated at $\mu = 2$ GeV and
             including three moments in both the real and imaginary parts.}
    \label{fig:fit_ratio_matrix_elements}
\end{figure*}

Our results for the fixed-$z^2$ analysis for both the isovector and isoscalar combinations are shown in \Cref{fig:fixed_z}.
All results shown always include TMCs.
Initially, we only considered the LO and fixed-order NLO Wilson coefficients in the analysis.
However, the fixed-order NLO results show significant $z$ dependence for $\braket{x}$ at small values of $z$.
This is not completely unexpected, as discretization errors~\cite{Gao:2020ito} and large logs can be significant for small values of $z$, see Appendix B in Ref.~\cite{Gao:2022iex}.
Note, however, in that work, the analysis was done for the pion PDF, where the large logs become important at somewhat smaller $z$ compared to the range of $z$ where we see strong dependence here.
To better understand the effects of large logs for the transversity PDF of the proton, we also use the NLO Wilson coefficients combined with renormalization group resummation (RGR) at next-to-leading logarithm (NLL) accuracy, given by
\begin{equation}
\begin{split}
    & C^\delta_{n, {\rm NLO+RGR}} (\mu^2 z^2) = C^\delta_{n, {\rm NLO}} (\mu_0^2 z^2) \\
    & \;\;\;\;\; \times e^{-\frac{\gamma_n^{(1)}\textup{ln}\frac{a_s(\mu)}{a_s(\mu_0)}}{\beta_0}-\frac{(-\beta_1\gamma_n^{(1)}+\beta_0\gamma_n^{(2)})\textup{ln}\frac{\beta_0+\beta_1a_s(\mu)}{\beta_0+\beta_1a_s(\mu_0)}}{\beta_0\beta_1}} ,
\end{split}
\end{equation}
where $a_s=\alpha_s/(2\pi)$, $\beta_n$ is the $n$th order coefficient of the $\beta$ function,
and $\gamma_n^{(1)}$ and $\gamma_n^{(2)}$ are the anomalous dimensions of the $n$th moments~\cite{Vogelsang:1997ak, Ji:2022thb}. The RGR evolve the running coupling $\alpha_s$ from the physical scale $\mu_0 = 2 e^{-\gamma_E}/z$ to the factorization scale $\mu$. As can be seen from \Cref{fig:fixed_z}, the use of the NLO+RGR Wilson coefficients produces results mostly consistent with the NLO case at small $z$.
This suggests the significant $z$ dependence at small $z$ is mainly a discretization effect rather than due to large logs.

\begin{figure*}
    \centering
    \includegraphics[width=\columnwidth]{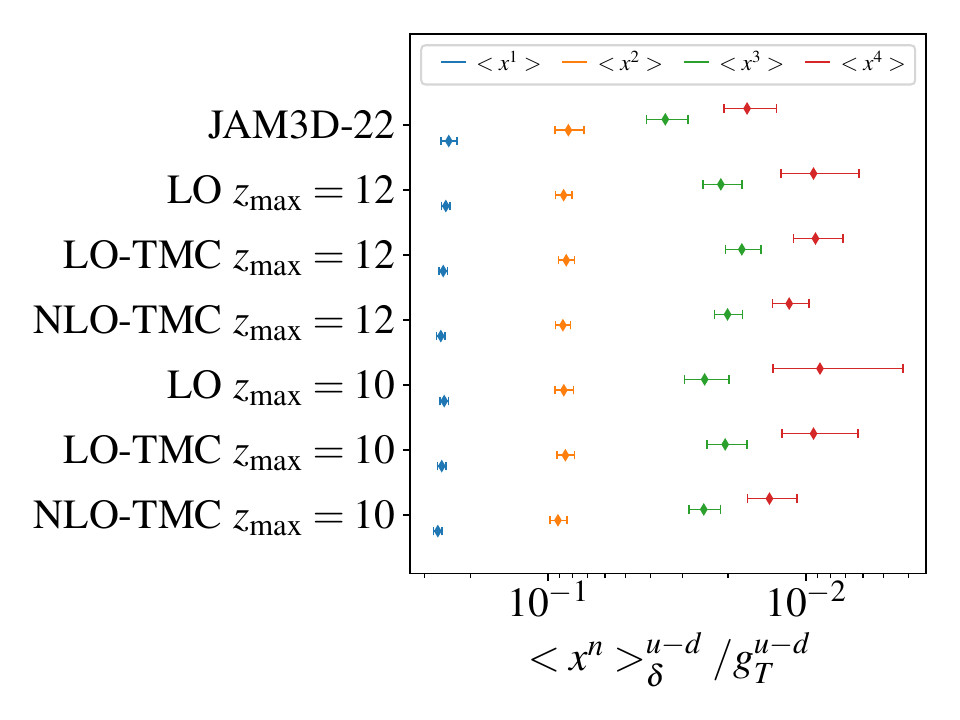}
    \includegraphics[width=\columnwidth]{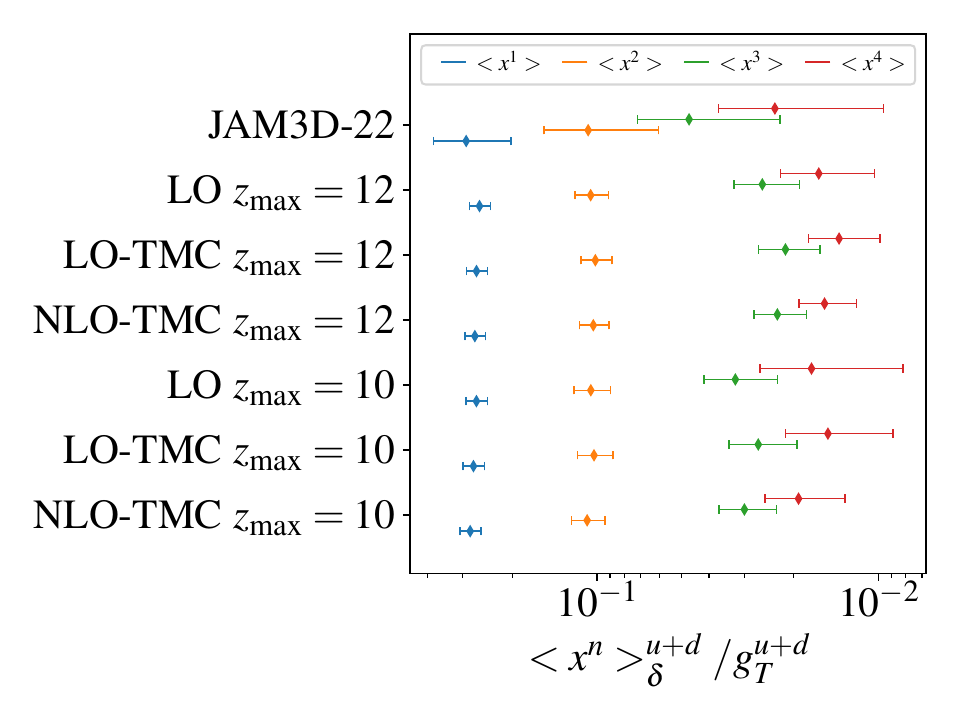}
    \caption{Summary plots of our extracted (left) isovector and (right) isoscalar moments from the
             leading-twist OPE approximation evaluated at $\mu = 2$ GeV for various fitting strategies compared to the results from JAM3D-22~\cite{Gamberg:2022kdb} defined at the scale $Q = 2$ GeV.
             Two $z_{\rm max}$ are considered, as well as Wilson coefficients at LO and fixed-order NLO.}
    \label{fig:moments_summary}
\end{figure*}

Next, we move on to including a range of values of $z$ in our fits, considering various ranges $z \in [z_{\rm min}, z_{\rm max}]$.
With the extra data, we can include an extra moment in the fits for $z > 8a$.
The results for both the isovector and isoscalar moments are shown in \Cref{fig:global_z}.
Given the small effect from the RGR which also becomes unstable when $a_s(\mu_0)$ runs close to the Landau pole, we opt to use the fixed-order NLO Wilson coefficients, and also always include TMCs for the final results.
There is some dependence on the choice of $z_{\rm min}$, as expected from the fixed-$z^2$ analysis results, however it is rather mild.
Our preferred fit range is $z \in [3a, 10a]$, which removes most of the effects from discretization errors and large logs at small $z$ and keeps $z_{\rm max}$ small enough to likely keep higher-twist contributions negligible.
The results of these preferred fits are shown in \Cref{fig:fit_ratio_matrix_elements}.
Finally, we show a summary of the results from different strategies and their comparison to JAM3D-22~\cite{Gamberg:2022kdb} in \Cref{fig:moments_summary}.

It is interesting to note the rather good agreement with the global analysis from JAM3D-22, especially for the lowest two moments, whereas we found tension for the lowest non-trivial moment in the unpolarized case~\cite{Gao:2022uhg}.
However, comparing the matrix elements presented here versus those from the unpolarized ones, there is some hint of smaller excited-state contamination in the matrix elements of this work, which may be responsible for the better agreement.

\section{PDF from leading-twist OPE: DNN reconstruction}
\label{sec:dnn}

\subsection{Method}\label{subsec:dnnMethod}

\begin{figure*}
    \centering
    \includegraphics[width=0.45\textwidth]{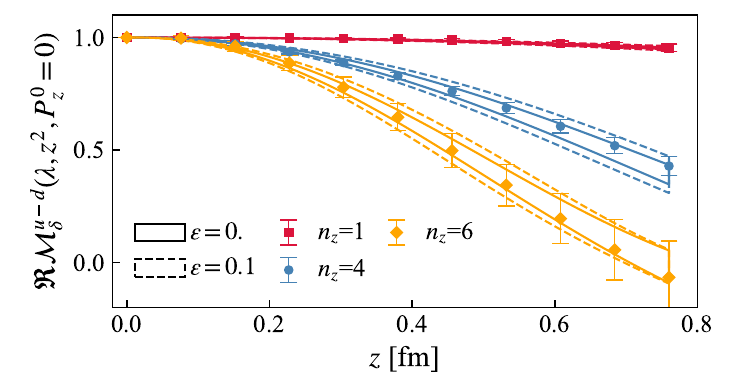}
    \includegraphics[width=0.45\textwidth]{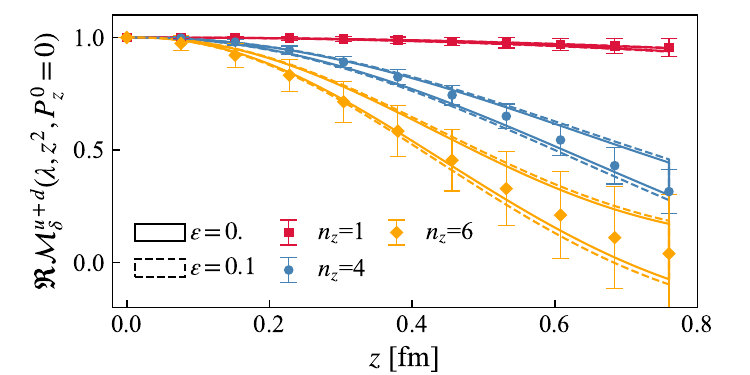}
    \includegraphics[width=0.45\textwidth]{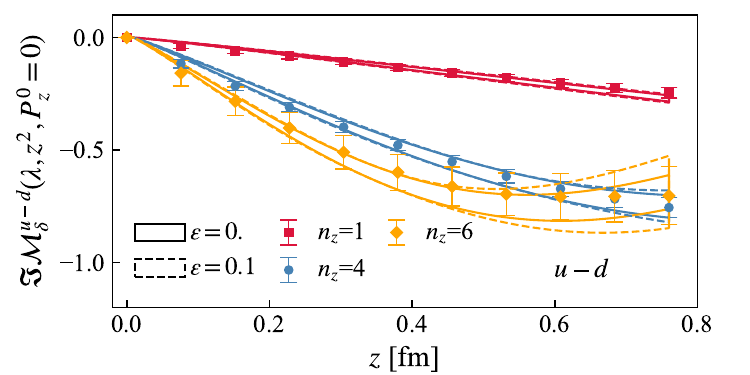}
    \includegraphics[width=0.45\textwidth]{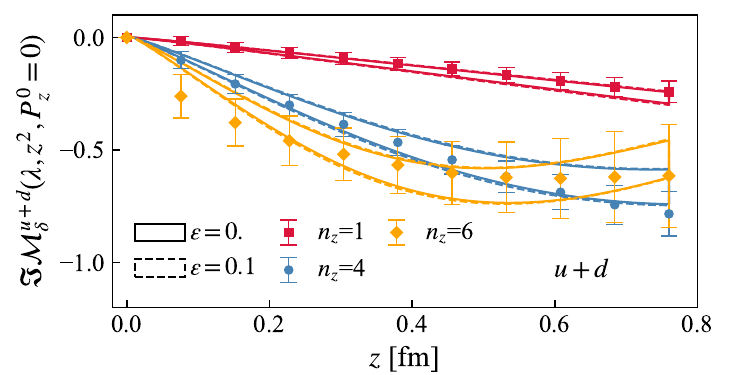}
    \caption{The DNN training results using the (left) isovector and (right) isoscalar reduced pseudo-ITD matrix elements in the range $z\in[2a, 10a]$ for the (upper) real part and (lower) imaginary part (lower panel) are shown. The results using $\epsilon=0$ and $\epsilon=0.1$ are shown as the solid and dotted curves, respectively.}
    \label{fig:DNN_zmax76_rITD_isoV_S}
\end{figure*}

It has been shown that we can extract the Mellin moments of transversity PDFs by applying the OPE formula to the ratio-scheme renormalized matrix elements, model independently.
Limited by the finite $\lambda=zP_z$, the lattice data is only sensitive to the first few moments while the higher ones are factorially suppressed.
As a result, to predict the $x$ dependence of the PDFs, one needs to introduce additional prior knowledge or a reasonable choice of model.
Commonly used models are usually of the form
\begin{align}
    q(x)=Ax^\alpha(1-x)^\beta(1+\textup{sub-leading~terms}),
\end{align}
which is inspired by the end-point behavior of the PDFs.
However, the sub-leading terms may play an important role, particularly in moderate regions of $x$, and one may find reasonable models for the sub-leading terms that give acceptable fits to the data.
But, unless the data is precise, the model could introduce an uncontrolled bias.
The use of a deep neural network (DNN) is a flexible way to maximally avoid any model bias (see e.g. Refs.~\cite{Karpie:2019eiq,DelDebbio:2020rgv} which also employed a neural network for this purpose) --- but cannot remove the bias entirely, as a nueral network is still a model ---
which is capable of approximating any functional form given a complicated enough network structure.
As proposed in \refcite{Gao:2022uhg}, we parametrize the PDFs by,
\begin{equation}\label{eq:DNNform}
	q(x;\alpha,\beta,\boldsymbol{\theta})\equiv A x^{\alpha}(1-x)^{\beta} [1+\epsilon(x)\sin({f_{\rm DNN}(x,\boldsymbol{\theta})})] ,
\end{equation}
where $f_\textup{DNN}(x, \boldsymbol{\theta})$ is a DNN multistep iterative function, constructed layer by layer. The initial layer consists of a single node, denoted as $a_1^1$, which represents the input variable $x$. Subsequently, in the hidden layers, a linear transformation is performed using the equation:
\begin{align}
z_{i}^{(l)} = b_{i}^{(l)}+\sum_jW_{ij}^{(l)}a_j^{(l-1)} .
\end{align}
Here, $z_{i}^{(l)}$ is the intermediate result obtained by adding the bias term $b_{i}^{(l)}$ to the sum of the weighted inputs from the previous layer, represented by $W_{ij}^{(l)}a_j^{(l-1)}$. Following this linear transformation, a nonlinear activation function $\sigma^{(l)}(z_{i}^{(l)})$ is applied, and the resulting output serves as the input to the next layer, represented by $a_i^{(l)}$. We specifically employed the exponential linear unit activation function $\sigma_{\textsf{elu}}(z)=\theta(-z)(e^z-1)+\theta(z)z$. Lastly, the final layer generates the output $f_\text{DNN}(x,\boldsymbol{\theta})$, which is subsequently utilized to evaluate $q(x;\alpha,\beta,\mathbf{\theta})$. The lower indices $i = 1,...,n^{(l)}$ are used to identify specific nodes within the $l$th layer, where $n^{(l)}$ denotes the number of nodes in the $l$th layer. The upper indices, enclosed in parentheses, $l=1,...,N$, are employed to indicate the individual layers, where $N$ corresponds to the number of layers, representing the depth of the DNN. The parameters of the DNN, namely the biases $b_{i}^{(l)}$ and weights $W_{ij}^{(l)}$, represented by $\boldsymbol{\theta}$, are subject to optimization (training) by minimizing the loss function defined as
\begin{align}
\begin{split}
J(\boldsymbol{\theta})
\equiv
\frac{\eta}{2} \boldsymbol{\theta}\cdot\boldsymbol{\theta}+\frac{1}{2} \chi^2(\boldsymbol{\theta},\alpha,\beta,...) .
\end{split}
\end{align}
The first term in the loss function serves the purpose of preventing overfitting and ensuring that the function represented by the DNN remains well-behaved and smooth. The details of the $\chi^2$ function can be found in the appendix of \refcite{Gao:2022uhg}. Due to the limited statistics, a simple network structure such as $\{1,16,16,1\}$ (indicating the number of nodes in each layer) is sufficient to provide a smooth approximation of the sub-leading contribution. In practice, we experimented with different values of $\eta$ ranging from $10^{0}$ to $10^{-2}$ and considered network structures of sizes $\{1,16,16,1\}$, $\{1,16,16,16,1\}$, and $\{1,32,32,1\}$. However, the results remained consistent across these variations. Therefore, we opted for $\eta=0.1$ and selected a DNN structure with four layers, including the input and output layers, specified as $\{1,16,16,1\}$.

To balance the model bias and data precision, the contribution of the DNN is limited by $|\epsilon(x)\sin(f_{\rm DNN})|\lesssim \epsilon(x)$, which can be fully removed by setting $\epsilon(x)=0$. It is also possible to control the size of the DNN parametrized sub-leading contribution at each specific $x$. However, in this work, given the limited statistics, we simply fix $\epsilon(x)$ to be a small constant, e.g. $0.1$.

\subsection{DNN represented PDF}

To train the PDFs, we re-write the short distance factorization as,
\begin{equation}
    \tilde{h}^f_\delta(z,P_z,\mu) =\int_{-1}^1d\alpha\, \mathcal{C}^\delta(\alpha, \mu^2z^2)\, \int_{-1}^1 dy\, e^{-iy \alpha\lambda} \delta q^f(y,\mu),
\end{equation}
where the renormalized matrix elements $\tilde{h}^f_\delta (z, P_z, \mu)$ are directly connected to the $x$-dependent PDFs $\delta q^f (x, \mu)$, and $\mathcal{C}^\delta(\alpha, \mu^2 z^2)$ can be determined from the Wilson coefficients $C_n^\delta (\mu^2 z^2)$~\cite{Izubuchi:2018srq, Ji:2022thb}.
In this section we use the NLO fixed-order Wilson coefficients.
In our case, the real and imaginary parts of the reduced pseudo-ITD $\mathcal{M}^{\frac{f}{f^\prime}}_\delta (\lambda, z^2, P_z^0=0)$ are related to $\delta q^{f,-}(x)$ and $\delta q^{f,+}(x)$, defined as
\begin{align}\label{eq:isoVrelation}
\begin{split}
	&\delta q^{f,-}(x)\equiv \delta q^f(x) - \delta q^{\bar f}(x) ,\\
	&\delta q^{f,+}(x)\equiv \delta q^f(x) + \delta q^{\bar f}(x),
\end{split}
\end{align}
in the region $x \in [0, 1]$ and 
where $\delta q^f(x)$ and $\delta q^{\bar f}(x)$ are the quark and anti-quark transversity distributions of flavor $f$, respectively. However, as observed in the literature~\cite{HadStruc:2021qdf, LatticeParton:2022xsd}, with the current lattice accuracy, the anti-quark distributions are mostly consistent with zero. We therefore ignore the anti-quark contribution and fit the real and imaginary parts together to $\delta q^f(x;\alpha,\beta,\boldsymbol{\theta})=\delta q^{f,-}(x)=\delta q^{f,+}(x)$. 

\begin{figure}
    \centering
    \includegraphics[width=0.45\textwidth]{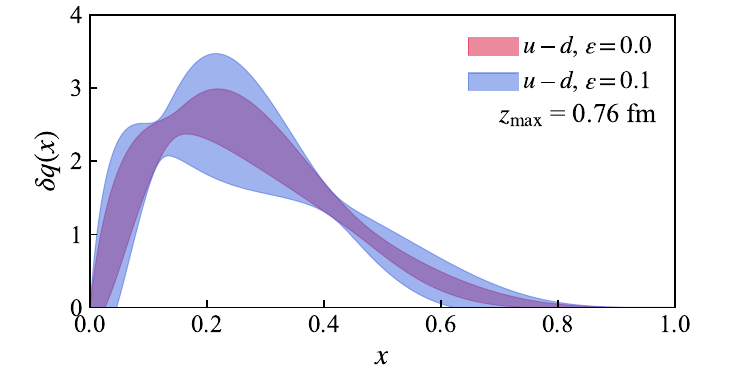}
    \includegraphics[width=0.45\textwidth]{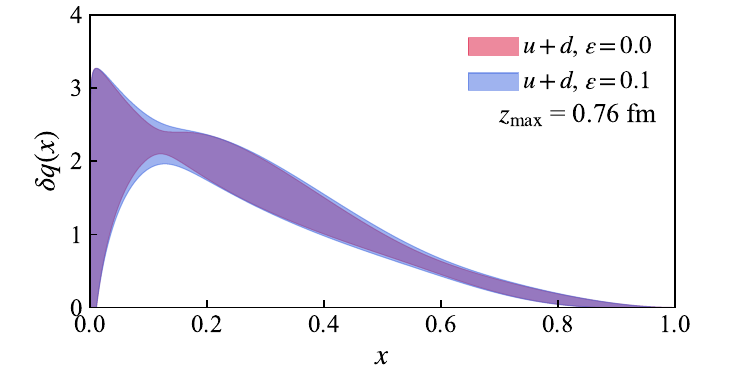}
    \caption{The DNN represented PDFs using the matrix elements in the range $z\in[2a, 10a]$ for the (upper) isovector and (lower) isoscalar cases are shown. The results with $\epsilon=0$ and $\epsilon=0.1$ are shown as the red and blue bands, respectively. }
    \label{fig:DNN_zmax76_PDF}
\end{figure}

\begin{figure}
    \centering
    \includegraphics[width=0.45\textwidth]{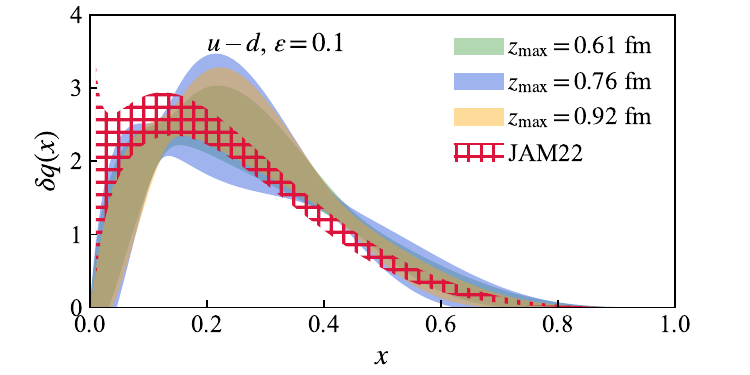}
    \includegraphics[width=0.45\textwidth]{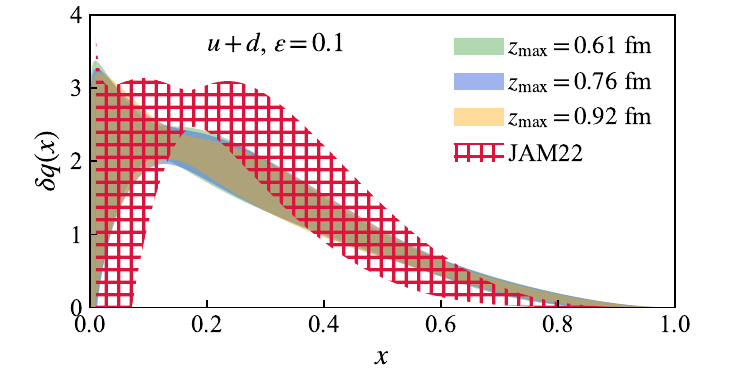}
    \caption{The DNN represented PDFs using the matrix elements in the range $z\in[2a, z_{\rm{max}}]$ for the (upper) isovector and (lower) isoscalar cases are shown. For comparison, we also show the global analysis results from JAM3D-22~\cite{Gamberg:2022kdb}.}
    \label{fig:DNN_zmax_PDF}
\end{figure}

We use the matrix elements in the range $z\in[2a, z_{\rm max}]$ for the parameter training, skipping $z=a$ in order to avoid the most serious discretization effects. In \Cref{fig:DNN_zmax76_rITD_isoV_S}, we show the fit results for $z_{\rm max} = 10a$ with $\epsilon = 0$ and $\epsilon = 0.1$ which both lead to a good description of the data.
The corresponding PDFs are shown in \Cref{fig:DNN_zmax76_PDF}, and the results from $\epsilon=0.1$ exhibit slightly larger errors but mostly overlap with the $\epsilon=0$ case. It is evident that the effects of the DNN were minimal which is likely a result of the limited statistics. We anticipate the DNN playing a more significant role when more precise data becomes available.
In what follows, we use the results with $\epsilon=0.1$.

The short distance factorization could suffer from power corrections at large values of $z^2$. To check this, we vary the $z_{\rm max}$ used to train the PDFs to investigate such systematic errors. As shown in \Cref{fig:DNN_zmax_PDF}, by slightly increasing $z_{\rm max}$, the results do not change significantly within the large errors, suggesting that higher-twist effects are less important compared to the statistics of our data. For comparison, we also show the most recent global analysis results from JAM3D-22~\cite{Gamberg:2022kdb}, and overall agreement is observed.

\section{\texorpdfstring{$x$}{x}-space Matching}
\label{sec:xspace}

We now move on to our final method for extracting information on the transversity PDF.
This method utilizes LaMET to match the quasi-PDF --- determined from the Fourier transform of hybrid-renormalized matrix elements --- to the light-cone PDF.

\subsection{Hybrid renormalization}

It is well known that the bare matrix elements can be multiplicatively renormalized by removing the linear divergence originating from the Wilson line self energy and the overall logarithmic divergence
\begin{equation}
    h^f_\delta(z,P_z) = Z_T(a)e^{-\delta m(a)z}e^{-\bar{m}_0z}\tilde{h}_\delta^{f}(z,P_z), 
\end{equation}
where $\tilde{h}_\delta^f$ is the renormalized matrix element, $\delta m(a)$ contains the Wilson-line self-energy linear UV divergences,
$Z_T(a)$ contains the logarithmic UV divergences, and $\bar{m}_0$ is used to fix the scheme dependence present in $\delta m(a)$.
The Wilson-line self-energy divergence term $\delta m(a)$ can be extracted from physical matrix elements, like those involving Wilson loops.
Here we use the value $a \delta m(a) = 0.1597(16)$ determined from the static quark-antiquark potential taken from Refs~\cite{HotQCD:2014kol,Bazavov:2016uvm,Bazavov:2017dsy,Bazavov:2018wmo,Petreczky:2021mef}.
The scheme dependence in $\delta m(a)$ can be attributed to a renormalon ambiguity, but can be fixed to a particular scheme by appropriate determination of $\bar{m}_0$~\cite{Gao:2021dbh, Zhang:2023bxs}, and here we choose the $\overline{\rm MS}$ scheme.
Our strategy for determining $\bar{m}_0$ is to compare the $P_z=0$ bare matrix elements $h_\delta^f(z,P_z=0)$ to the Wilson coefficient $C_0^\delta(\mu^2 z^2)$ computed in the $\overline{\rm MS}$ scheme
\begin{equation}
    h_\delta^f(z,P_z=0) = Z_T(a)e^{-\delta m(a)z}e^{-\bar{m}_0z}C_0^\delta(\mu^2z^2) .
    \label{eq:mult_renorm}
\end{equation}
In order to remove $Z_T(a)$ and hopefully cancel some of the discretization effects, we next divide \eqref{eq:mult_renorm} by itself with $z$ shifted by one unit of the lattice spacing.
Then, after rearranging, we arrive at
\begin{equation}\label{eq:Effm0}
    e^{a \delta m(a)}\frac{h^f_\delta(z,P_z=0,a)}{h^f_\delta(z-a,P_z=0,a)}=e^{-a\bar{m}_0}\frac{C_0^\delta(\mu^2z^2)}{C_0^\delta(\mu^2(z-a)^2)} .
\end{equation}
Before proceeding, we must first discuss the specifics of the Wilson coefficients used.

\begin{figure}
    \centering
    \includegraphics[width=\columnwidth]{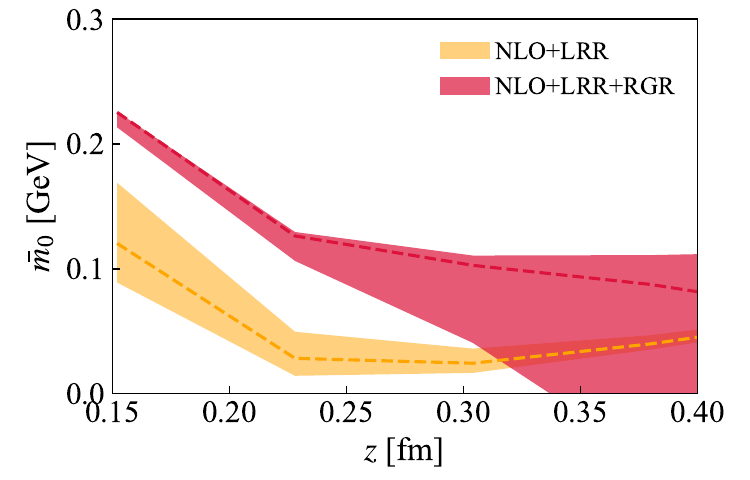}
    \caption{The $\bar{m}_0$ determined using NLO+LRR and NLO+LRR+RGR Wilson coefficients are shown. The bands come from the scale variation.}
    \label{fig:m0Fit}
\end{figure}

\begin{figure}
    \centering
    \includegraphics[width=\columnwidth]{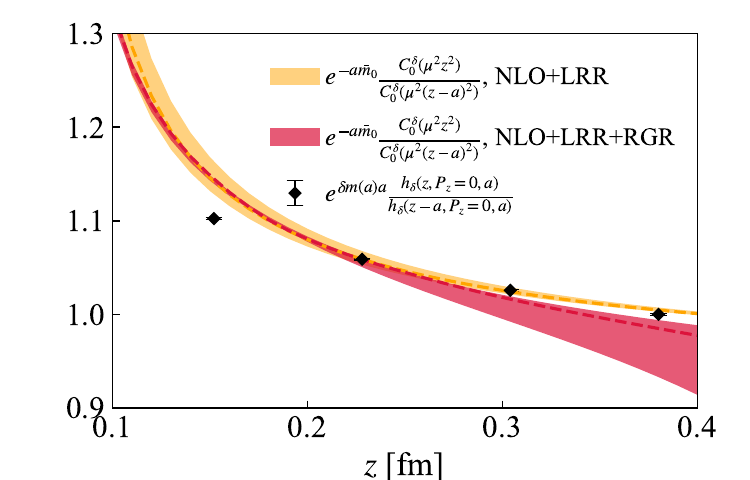}
    \caption{The ratio of $P_z=0$ matrix elements (black points) defined in \Eq{Effm0} are shown. The bands are inferred from the NLO+LRR and NLO+LRR+RGR Wilson coefficients respectively with scale variation.}
    \label{fig:m0Ratio}
\end{figure}

The renormalon ambiguity, by definition, is an artifact that arises from the summation prescription of the perturbative series in the QCD coupling $\alpha_s$.
Therefore, we use the Wilson coefficients after leading renormalon resummation (LRR) given in \refcite{Zhang:2023bxs} under the large-$\beta_0$ approximation by
\begin{align}\label{eq:C0LRR}
\begin{split}
    C^\delta_{{0,\rm{LRR}}}(\alpha_s(\mu),&z^2\mu^2)=\int_{0,{\rm PV}}^\infty d\omega e^{-\frac{4\pi\omega}{\beta_0\alpha_s(\mu)}}\frac{2C_F}{\beta_0}\frac{1}{\omega}\\
    &\times\left[\frac{\Gamma(1-\omega)e^{\frac{5}{3}\omega}(z^2\mu^2/4)^\omega}{(1-2\omega)\Gamma(1+\omega)} -1\right] .
\end{split}
\end{align}
To be consistent with the known fixed-order Wilson coefficients at NLO, in practice, we use
\begin{align}
\begin{split}
    &C_0^{\delta \prime}(\alpha_s(\mu),z^2\mu^2)=C^\delta_{{0,\rm{LRR}}}(\alpha_s(\mu),z^2\mu^2)\\
    &+\left[C^\delta_{{0,\rm{NLO}}}(\alpha_s(\mu),z^2\mu^2)-C^\delta_{{0,\rm{LRR,NLO}}}(\alpha_s(\mu),z^2\mu^2) \right]
\end{split}
\end{align}
where the $C^\delta_{{0,\rm{LRR,NLO}}}$ is the NLO expansion of $C^\delta_{{0,\rm{LRR}}}$ and the fixed-order NLO Wilson coefficient is given by
\begin{align}\label{eq:C0NLO}
\begin{split}
    C^\delta_{{0,\rm{NLO}}}(\alpha_s(\mu),&z^2\mu^2)= \\
    &1 +\frac{\alpha_s(\mu)}{2\pi}C_F\left[2\ln(\frac{\mu^2z^2e^{2\gamma_E}}{4})+2\right].
\end{split}
\end{align}
In addition, we can also resum the large logarithms $\ln(\mu^2z^2e^{2\gamma_E}/4)$ by the renormalization group resummation (RGR)~\cite{Gao:2021hxl}. Using these coefficients,  the $\bar{m}_0$ determined using \Eq{Effm0} are shown in \fig{m0Fit} as a function of $z$.  The bands of NLO+LRR come from the scale variation of $\mu$ in the Wilson coefficients by a factor of $\sqrt{2}$. When using RGR, the running coupling is evolved from the physical scale $\mu_0=2ke^{-\gamma_E}/z$ to the factorization scale $\mu$~\cite{Gao:2021hxl,Holligan:2023rex}. And we vary $k \in [1/\sqrt{2}, 1, \sqrt{2}]$ to estimate the scale uncertainty. It can be observed that the scale uncertainties in the RGR case are smaller at small $z$, benefiting from the resummation, while they become larger at large $z$ as they become close to the Landau pole. In addition, plateaus can be observed after $z\geq 3a \sim 0.228$ fm when the discretization effects become negligible, though the uncertainty bands for the NLO+LRR+RGR case are larger with the running coupling when $z>0.25$ fm.
To avoid discretization effects at small $z$ and the Landau pole at large $z$, we choose values at $z=3a$ which give $\bar{m}_0 = 28(2)$ MeV and $129(2)$ MeV for NLO+LRR and NLO+LRR+RGR cases, respectively. In \Cref{fig:m0Ratio}, we show the data points defined on the left-hand side of \Cref{eq:Effm0} using the computed matrix elements and $\delta m(a)$, along with the ratios defined on the right-hand side of \Cref{eq:Effm0} using $\bar{m}_0$ chosen above and Wilson coefficients at NLO+LRR (orange bands) and NLO+LRR+RGR (red bands).

\begin{figure*}
    \centering
    \includegraphics[width=\columnwidth]{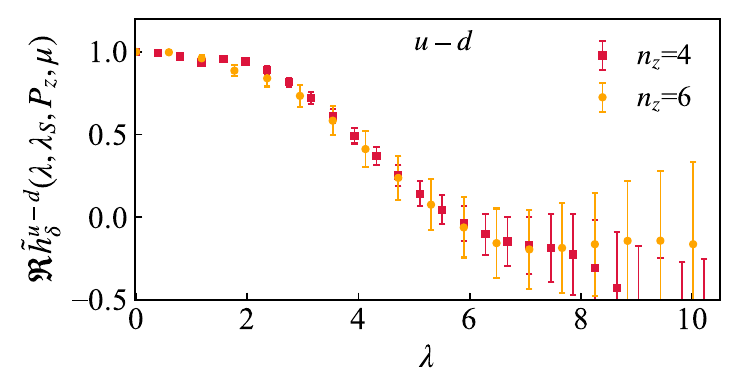}
    \includegraphics[width=\columnwidth]{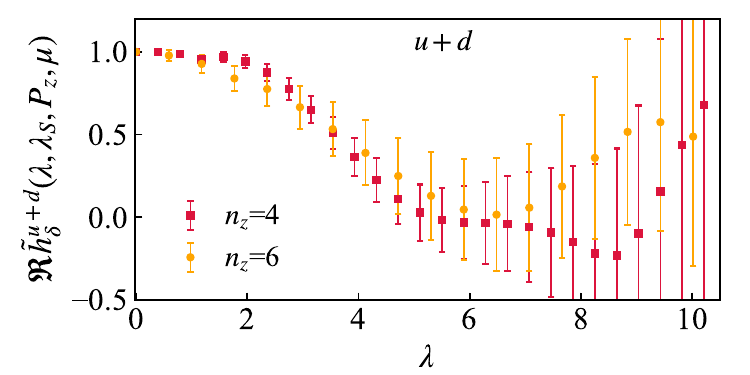}
    \includegraphics[width=\columnwidth]{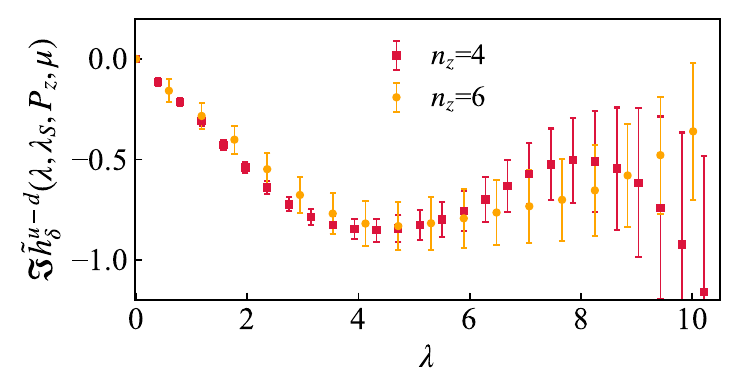}
    \includegraphics[width=\columnwidth]{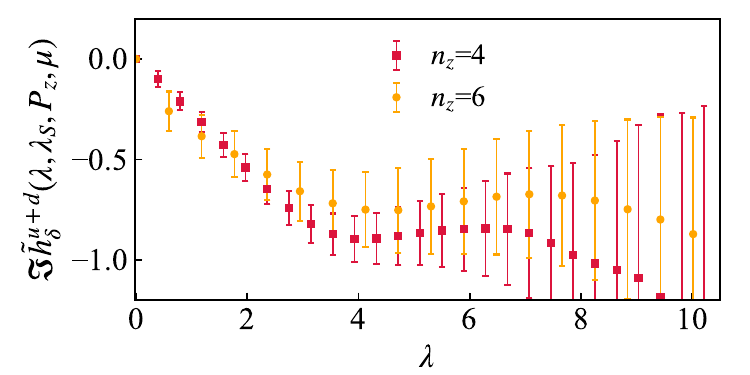}
    \caption{The (upper) real and (lower) imaginary parts of the renormalized matrix elements in the hybrid scheme for the (left) isovector and (right) isoscalar cobminations.} 
    \label{fig:Rmx_isoV_S}
\end{figure*}

The hybrid scheme renormalized matrix elements are given by
\begin{align}\label{eq:hybrid_renorm}
\begin{split}
    \tilde{h}^f_{\delta}(\lambda,\lambda_s&,P_z,\mu) = \theta(z_s-z)\frac{h^f_{\delta}(z,P_z,a)}{h^f_{\delta}(z,0,a)} \\
    &+\theta(z-z_s)\frac{h^f_{\delta}(z,P_z,a)}{h^f_{\delta}(z_s,0,a)} e^{ (\delta m(a)+ \bar{m}_0) (z-z_s)},
\end{split}
\end{align}
with $z_s=3a$.
In \fig{Rmx_isoV_S} we show the hybrid renormalized matrix elements for the isovector case (left panels) and isoscalar case (right panels) for momenta $n_z=4,6$. It can be seen that the large momentum matrix elements show a slow $P_z$ evolution and a good scaling in $\lambda$ within the statistical errors, suggesting we have good convergence in momentum.

\subsection{Extrapolation to large \texorpdfstring{$\lambda$}{lambda}}

\begin{figure}
    \centering
    \includegraphics[width=.49\columnwidth]{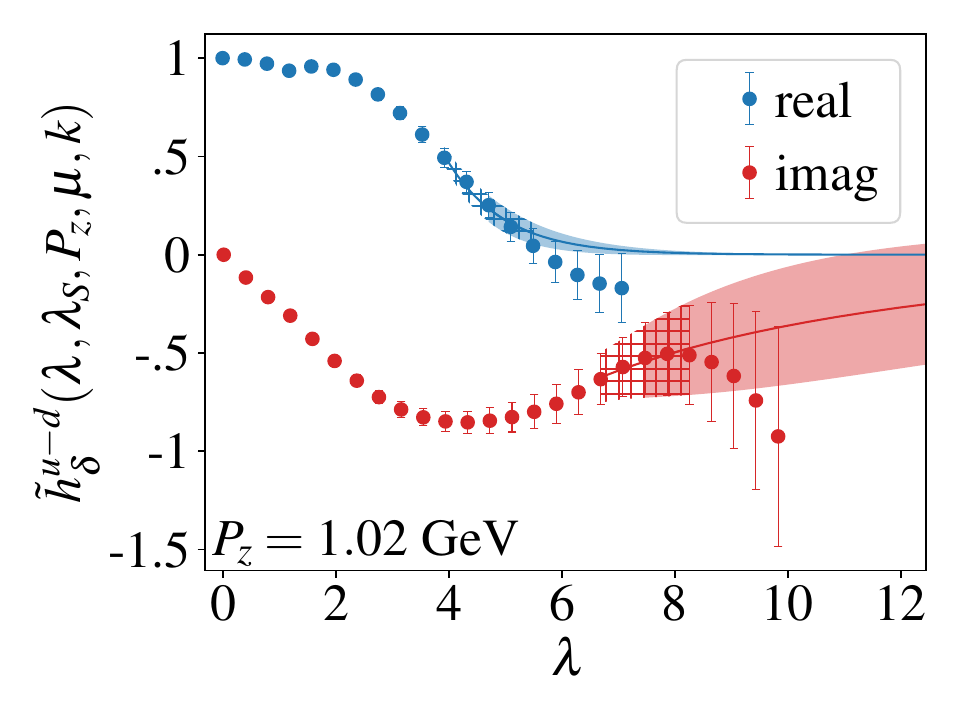}
    \includegraphics[width=.49\columnwidth]{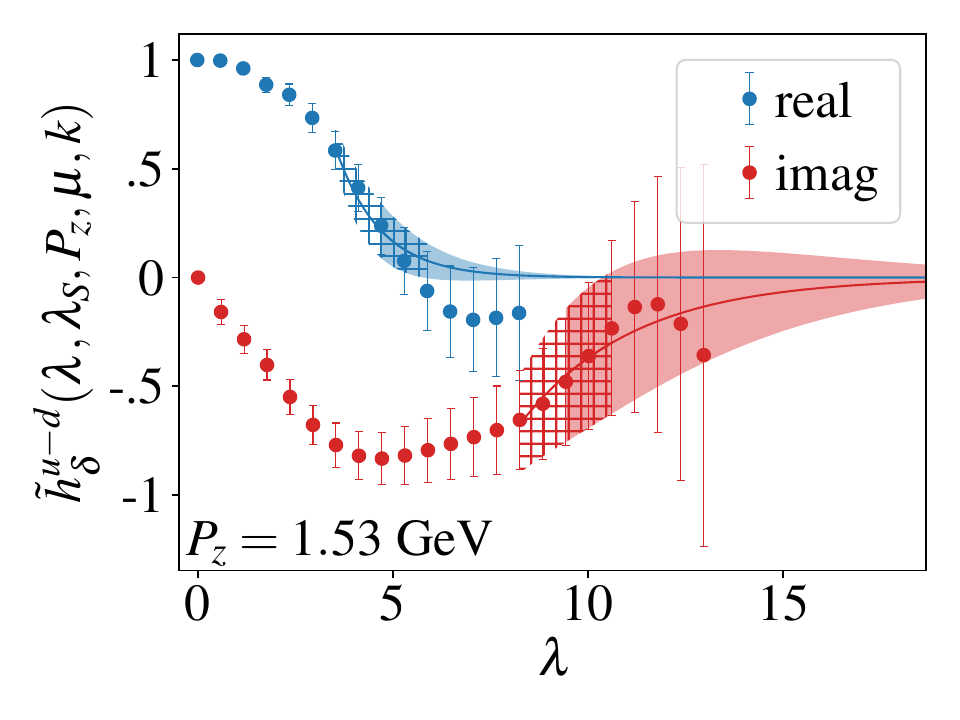}
    \includegraphics[width=.49\columnwidth]{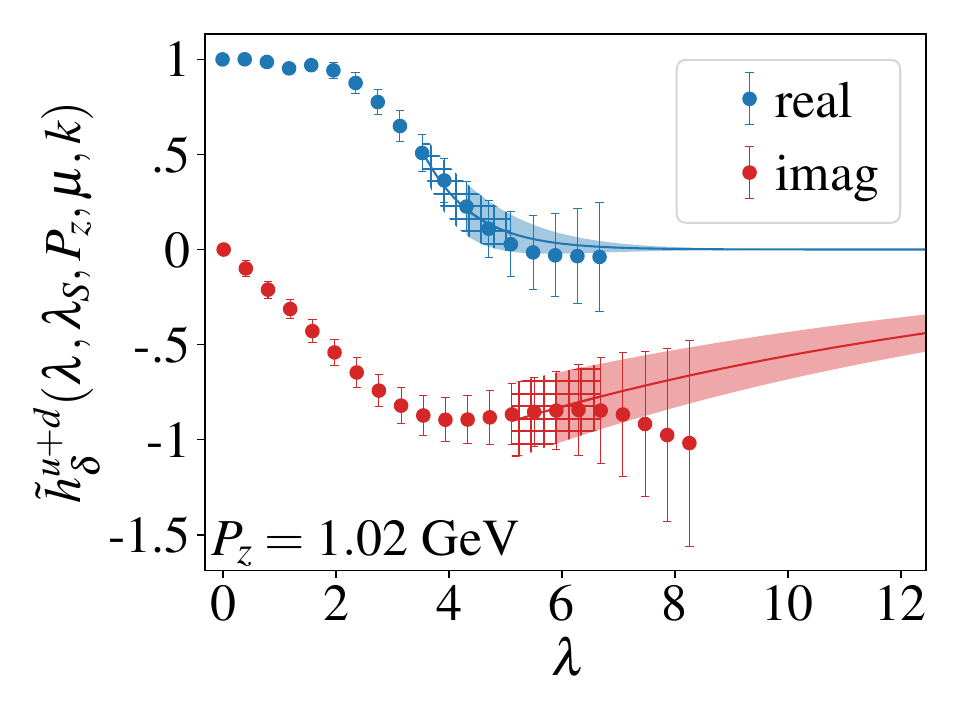}
    \includegraphics[width=.49\columnwidth]{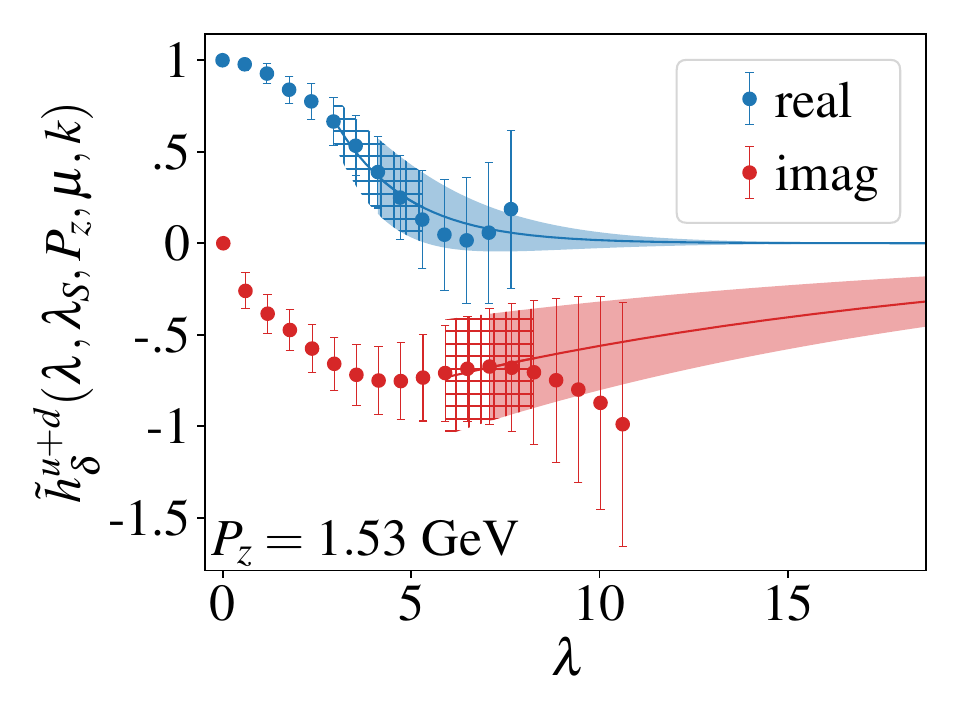}
    \caption{The (upper) isovector and (lower) isoscalar hybrid 
    renormalized matrix elements with (left) $P_z = 4 \frac{2 \pi}{L}$
    and (right) $P_z = 6 \frac{2 \pi}{L}$.
    The hatches show the range of data used for the fit to the extrapolation model
    and the bands are the result of that fit starting from $\lambda_L$.
    The hybrid renormalized data makes use of the NLO+LRR+RG Wilson coefficients computed
    at $\mu_0 = 2e^{-\gamma_E}/z$ (i.e. $k=1$) and subsequently evolved to $\mu = 2$ GeV.}
    \label{fig:hybrid_renormalized_extrap}
\end{figure}

Due to the finite extent of the lattice, one can only calculate the matrix elements up to some maximum $\lambda_{\rm max} \equiv z_{\rm max} P_z^{\rm max}$.
Further, the signal deteriorates as $\lambda$ is increased.
This poses a problem, as the matrix elements need to be Fourier transformed to obtain the quasi-PDF,
and truncating the integral will lead to unphysical oscillations in the resulting quasi-PDF (see Ref.~\cite{Karpie:2019eiq} for a discussion on this issue).
Therefore, we choose to perform an extrapolation of the data to infinity before performing the Fourier transform.
In practice, we estimate the Fourier transform with a discrete sum up to some value $\lambda_L = z_L P_z$ at which point an integral of the extrapolated function takes over.
There are a few considerations when deciding upon an appropriate value for $\lambda_L$.
In this work, we choose a value in the region where either the signal is no longer reliable or the values of the matrix elements are nearly consistent with zero.
As in our previous work in \refcite{Gao:2022uhg}, the extrapolation itself is done by performing a fit in this region using the exponential decay model
\begin{equation}
    \frac{A e^{-m_{\rm eff} \lambda / P_z}}{|\lambda|^d} ,
\end{equation}
where the fit parameters are constrained by $m_{\rm eff} > 0.1$ GeV, $A > 0$, and $d > 0$.
Using this constraint on $m_{\rm eff}$ helps to ensure the extrapolation falls off at a reasonable rate and does not significantly change the results in the regions of $x$ for which we trust the LaMET procedure.
A detailed derivation which motivates the use of this model can be found in App. B of \refcite{Gao:2021dbh}.
Results of the extrapolation fits for the largest two momenta are shown in \Cref{fig:hybrid_renormalized_extrap}.

\subsection{The quasi-PDF from a Fourier transform}

The quasi-PDF is defined as the Fourier transform of the renormalized matrix elements
\begin{equation}
    \label{eq:qPDF}
    \delta \tilde{q}^f(y,z_s,P_z,\mu) = \int \frac{dz P_z}{2 \pi} e^{iy P_z z} \tilde{h}^f_{\delta} (z, z_s, P_z, \mu) ,
\end{equation}
and is the LO approximation to the light-cone PDF within the LaMET framework.
To perform this integral, we first exploit the symmetry of the renormalized matrix elements about $z=0$, i.e. $\tilde{h}^f_\delta (z, z_s, P_z, \mu) = \tilde{h}^f_\delta (-z, z_s, P_z, \mu)^\ast$, to rewrite the integral only over positive $z$
\begin{equation}
\begin{split}
    \delta \tilde{q}^f(y,z_s,P_z,\mu) =& \int_0^\infty \frac{dz P_z}{\pi} \Re \tilde{h}^f_\delta (z, z_s, P_z, \mu) \cos (z P_z y) \\
    - \int_0^\infty &\frac{dz P_z}{\pi} \Im \tilde{h}^f_\delta (z, z_s, P_z, \mu) \sin (z P_z y) .
\end{split}
\end{equation}
Finally, we split the integrals up into two regions: i) $0 \leq z \leq z_L$ where the integral is performed via a sum over the lattice data for $\tilde{h}^f_\delta (z, z_s, P_z, \mu)$ and ii) $z_L < z < \infty$ where the integral is performed using the resulting extrapolation for $\tilde{h}^f_\delta (z, z_s, P_z, \mu)$
\begin{equation}
\label{eq:qPDF_split}
\begin{split}
    \delta \tilde{q}^f&(y,z_s,P_z,\mu) = \\
     &\Bigg[\sum_{z=0}^{z_L^{\rm re}/a} \frac{z_L^{\rm re} P_z}{\pi N_{z_L}^{\rm re}} + \int_{z_L^{\rm re}}^{\infty} \frac{dz P_z}{\pi} \Bigg] \Re \tilde{h}^f_{\delta}(z,z_s,P_z,\mu) \cos (z P_z y) \\
    - &\Bigg[\sum_{z=0}^{z_L^{\rm im}/a} \frac{z_L^{\rm im} P_z}{\pi N_{z_L}^{\rm im}} + \int_{z_L^{\rm im}}^{\infty} \frac{dz P_z}{\pi} \Bigg] \Im \tilde{h}^f_{\delta} (z,z_s,P_z,\mu) \sin (z P_z y),   
\end{split}
\end{equation}
where $z_L^{\rm re}$ and $z_L^{\rm im}$ are the values of $z$ in which the extrapolation integral takes over for the real and imaginary parts of $\tilde{h}^f_\delta (z, z_s, P_z, \mu)$, respectively, and $N^{\rm re/im} \equiv z_L^{\rm re/im}/a + 1$.

\begin{figure*}
    \centering
    \includegraphics[width=\columnwidth]{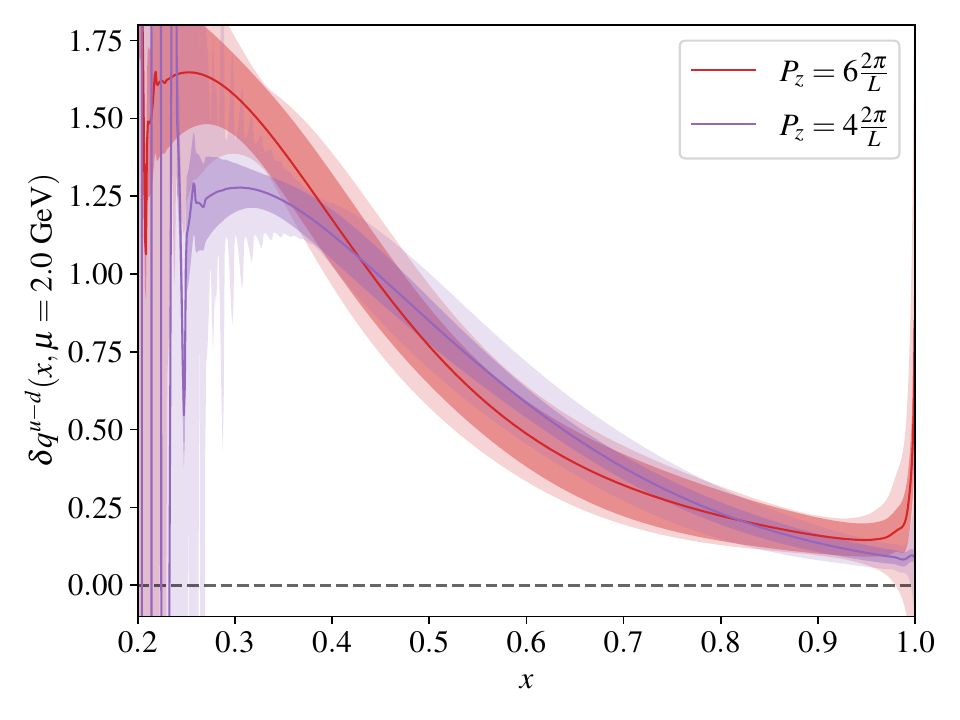}
    \includegraphics[width=\columnwidth]{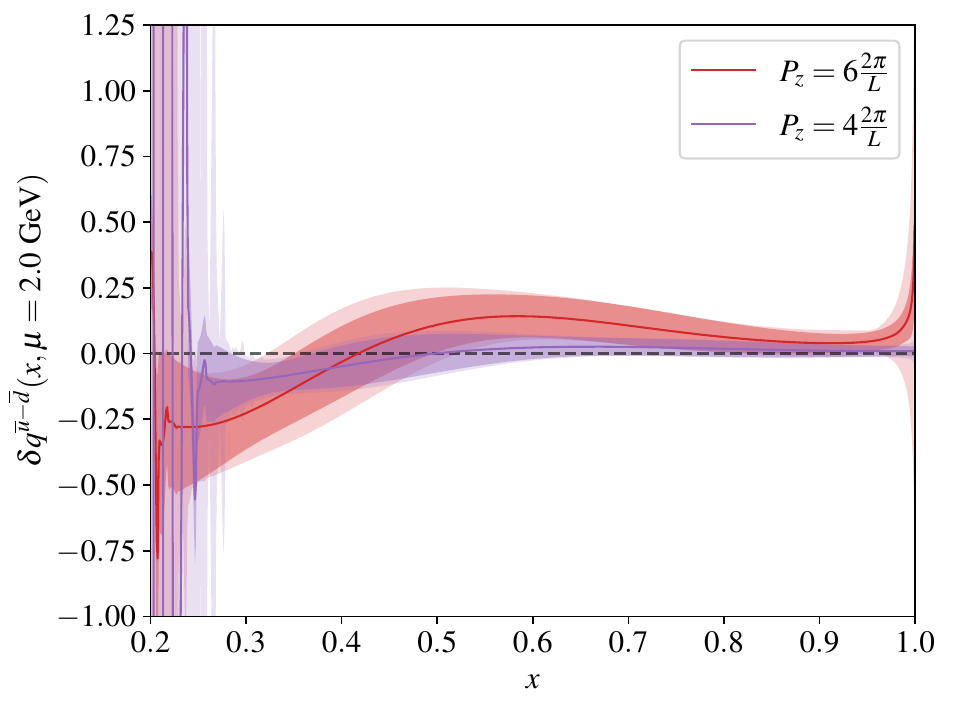}
    \includegraphics[width=\columnwidth]{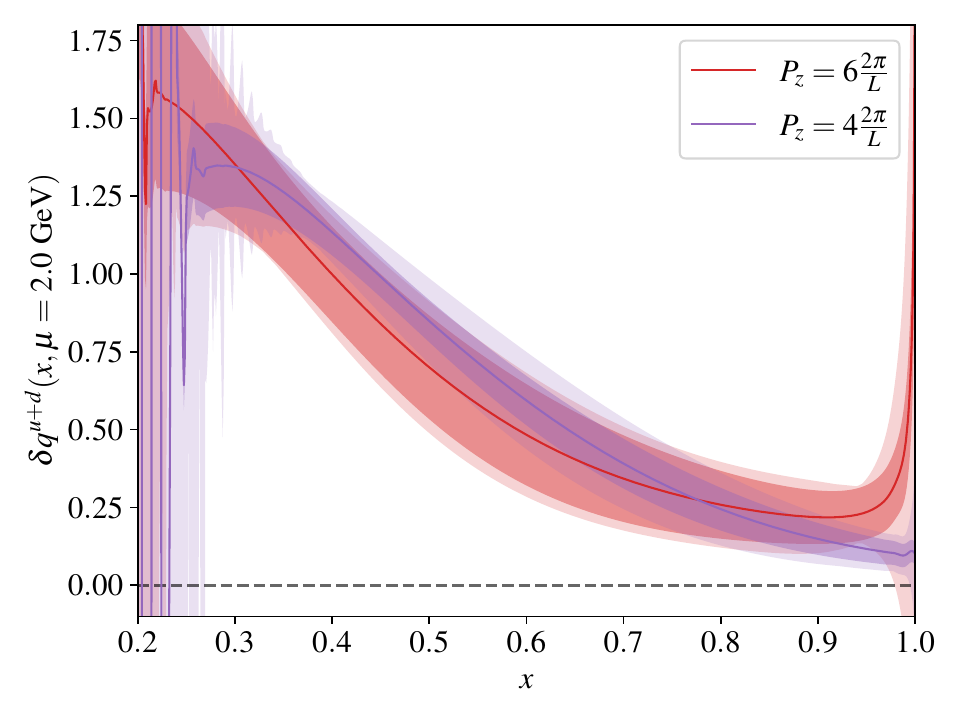}
    \includegraphics[width=\columnwidth]{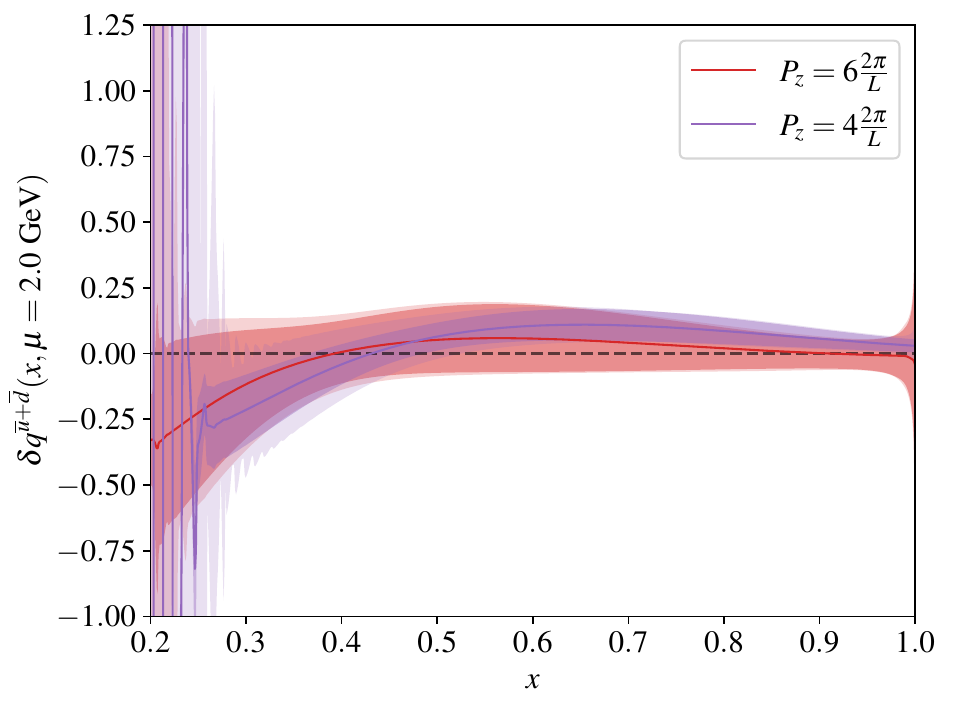}
    \caption{The $P_z$ dependence for the (upper) isovector and
    (lower) isoscalar (left) quark and (right) antiquark NLO+LRR+RGR light-cone transversity PDF $\delta q^f(x, \mu)$.
    The darker bands are the statistical errors when setting $k=1$
    and the lighter bands are the additional systematic errors associated with scale variations by additionally using $k=1/\sqrt{2}$ and $k=\sqrt{2}$.}
    \label{fig:pdf_pz_dep}
\end{figure*}

\begin{figure*}
    \centering
    \includegraphics[width=\columnwidth]{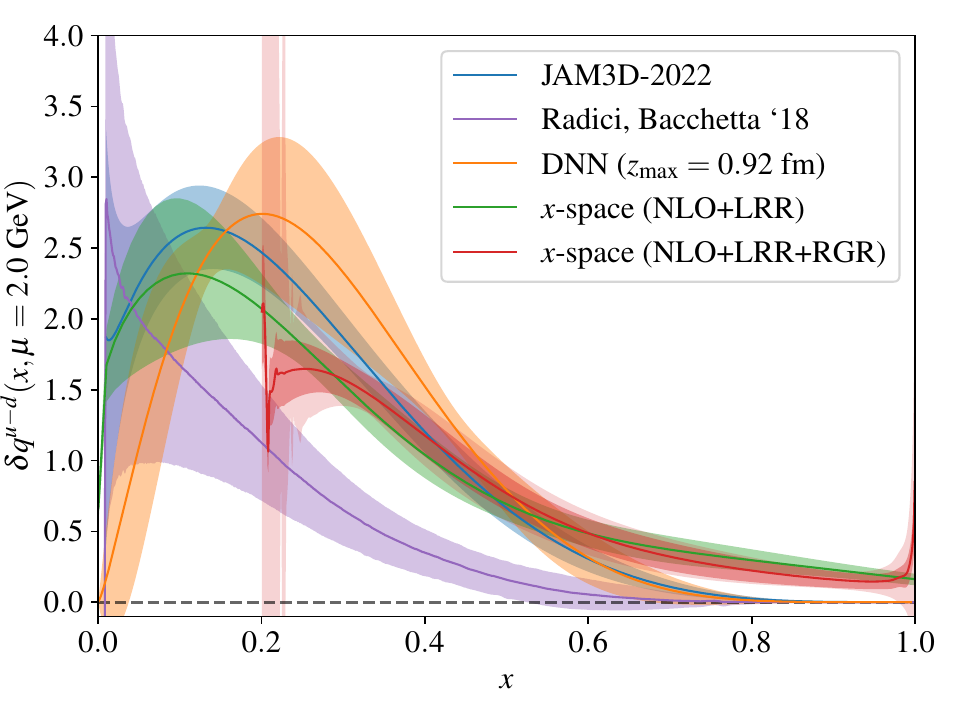}
    \includegraphics[width=\columnwidth]{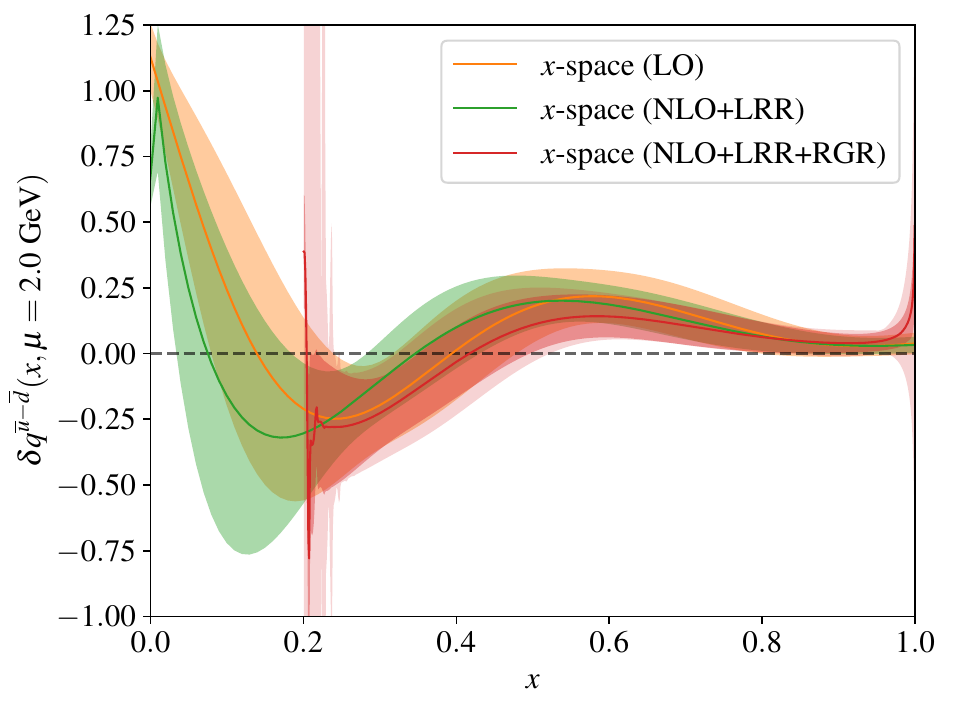}
    \includegraphics[width=\columnwidth]{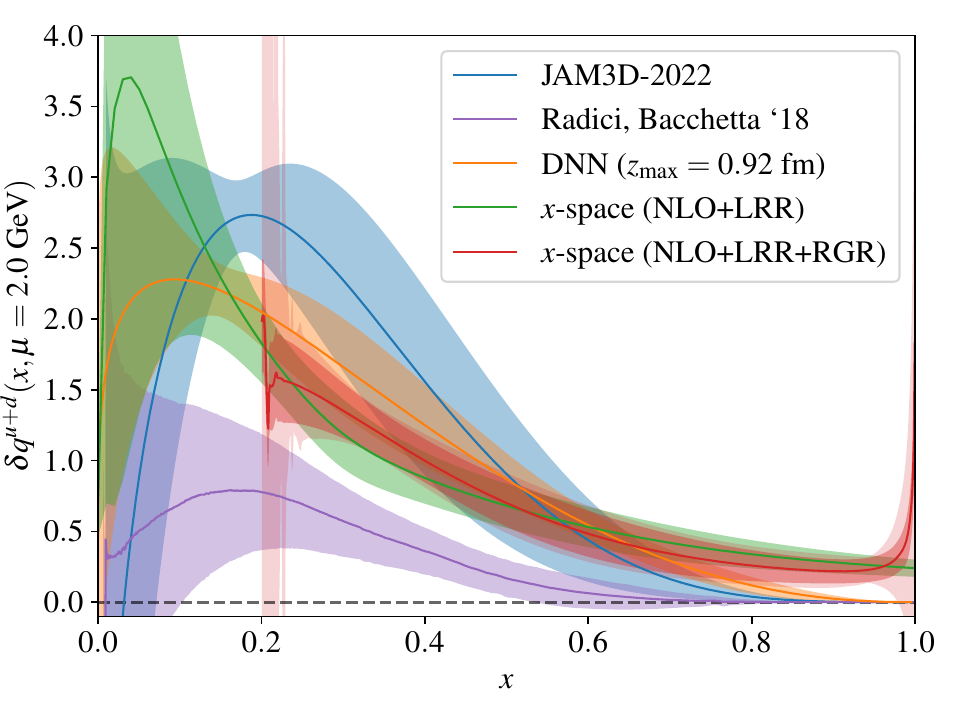}
    \includegraphics[width=\columnwidth]{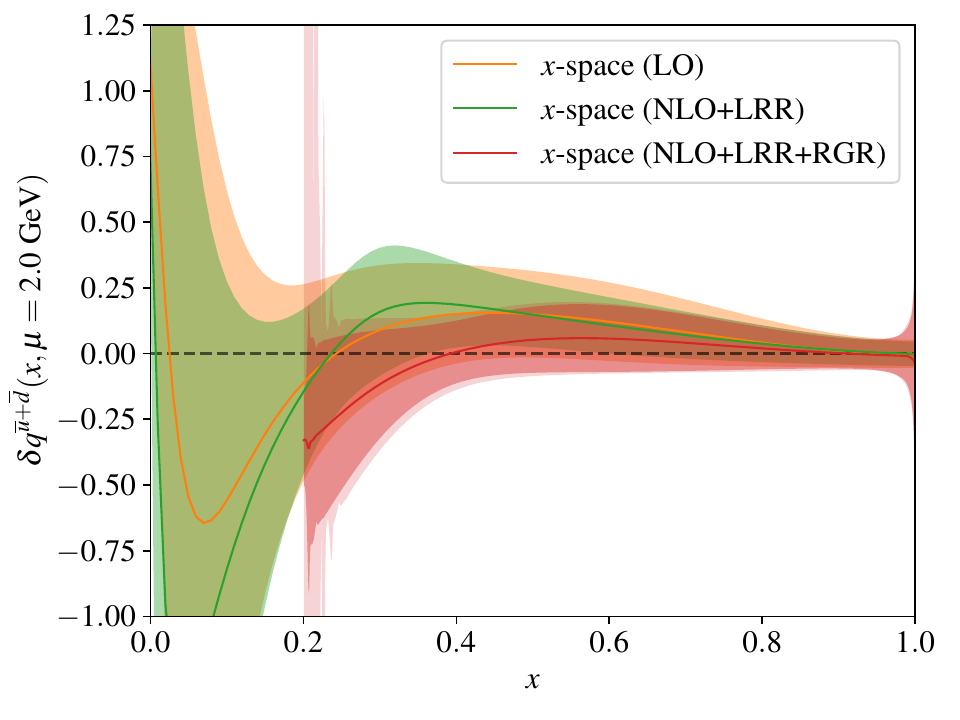}
    \caption{The (upper) isovector and (lower) isoscalar
    (left) quark and (right) antiquark transversity PDFs at NLO+LRR and NLO+LRR+RGR using the LaMET framework with the largest momentum $P_z = 6 \frac{2 \pi}{L}$.
    We also include comparisons with the DNN method of the previous section and with the
    global analysis from JAM3D-22~\cite{Gamberg:2022kdb} and Radici, Bacchetta~\cite{Radici:2018iag} which are both performed at LO.
    The darker bands for the NLO+LRR+RGR results are the statistical errors when setting $k=1$
    and the lighter bands are the additional systematic errors associated with scale variations by additionally using $k=1/\sqrt{2}$ and $k=\sqrt{2}$.}
    \label{fig:pdf_xspace_comparison}
\end{figure*}

\subsection{Matching to the light-cone PDF}

The final step in obtaining the light-cone PDF from the quasi-PDF is to match them perturbatively in $\alpha_s (\mu)$ as
\begin{equation}
\label{eq:matching}
\begin{split}
    \delta q^f(x,\mu)&=\int^{\infty}_{-\infty}\frac{d{y}}{|y|}\mathcal{C}^{-1}_\delta\left(\frac{x}{y},\frac{\mu}{yP_z},|y|\lambda_s\right)\delta\tilde{q}^f(y,z_s,P_z,\mu)\\
    &\phantom{==} +\mathcal{O}\left(\frac{\qcdcutoff^2 }{x^2P_z^2},\frac{\qcdcutoff^2}{(1-x)^2P_z^2}\right)\\
    &\equiv\mathcal{C}^{-1}_\delta \left(\frac{x}{y},\frac{\mu}{yP_z},|y|\lambda_s\right)\otimes\delta\tilde{q}^f(y,z_s,P_z,\mu)\\
    &\phantom{==} +\mathcal{O}\left(\frac{\qcdcutoff^2 }{x^2P_z^2},\frac{\qcdcutoff^2}{(1-x)^2P_z^2}\right),
\end{split}
\end{equation}
where $\mathcal{C}^{-1}_\delta (\frac{x}{y}, \frac{\mu}{y P_z}, |y|\lambda_s)$ is the inverse of the perturbative matching kernel for the transversity distribution,
and the notation $\otimes$ is used as short-hand for the integral.
One caveat of this method is that the leading power corrections to the matching can be seen to be enhanced when $x$ is near 0 or 1.
Therefore, we must be careful to estimate the range in $x$ in which these power corrections become significant and hence spoil the matching procedure.

To see how we obtain the the inverse matching kernel, we start with the perturbative expansion of the matching kernel itself
\begin{equation}\label{eq:full_matching}
    \begin{split}
        \mathcal{C}_\delta \left(\frac{x}{y},\frac{\mu}{yP_z},|y|\lambda_s\right)=\delta & \left(\frac{x}{y}-1\right)\\
        &+\sum_{n=1}^{\infty}\alpha_s^n\mathcal{C}^{(n)}_\delta \left(\frac{x}{y},\frac{\mu}{yP_z},|y|\lambda_s\right) ,
    \end{split}
\end{equation}
where only the NLO Wilson coefficients are known, i.e. we have only $\mathcal{C}^{(1)}_\delta (\frac{x}{y},\frac{\mu}{yP_z},|y|\lambda_s)$~\cite{Xiong:2013bka,Liu:2018hxv,Alexandrou:2018eet,Ji:2022thb}.
Next, by imposing the definition of the inverse matching kernel
\begin{equation}\label{eq:inverse_matching}
\mathcal{C}^{-1}_\delta \left(\frac{x}{z},\frac{\mu}{zP_z},|z|\lambda_s\right)\otimes\mathcal{C}_\delta \left(\frac{z}{y},\frac{\mu}{yP_z},|y|\lambda_s\right)=\delta\left(\frac{x}{y}-1\right) ,
\end{equation}
we find
\begin{equation}
    \begin{split}
        \mathcal{C}^{-1}_\delta & \left(\frac{x}{y},\frac{\mu}{yP_z},|y|\lambda_s\right)=\delta\left(\frac{x}{y}-1\right)\\
        &-\alpha_s\mathcal{C}^{(1)}\left(\frac{x}{y},\frac{\mu}{yP_z},|y|\lambda_s\right)+\mathcal{O}(\alpha_s^2).
    \end{split}
\end{equation}
As done in our previous work \refcite{Gao:2022uhg}, we approximate the integration by defining the integral on a finite-length grid which can be represented via matrix multiplication with a matching matrix $C^\delta_{xy}$ to obtain the light-cone PDF at NLO as
\begin{equation}
    \delta q^f(x, \mu) = \delta \tilde{q}^f (x, \mu) - \delta y \sum_{y}C^{\delta, {\rm NLO}}_{xy} \delta \tilde{q}^f (y, \mu) ,
\end{equation}
where $\delta y = 0.001$ is the grid size used for the integration.

For the matching coefficients themselves, we also implement LRR and RGR, where the RGR involves running the coupling from the physical scale $\mu_0$ with three choices of $k \in [1/\sqrt{2}, 1, \sqrt{2}]$ to $\mu = 2$ GeV which allows for assessing the systematics due to scale variation.

\subsection{Results}
As a first check regarding the significance of the power corrections, we show the momentum dependence of the NLO light-cone PDFs for both the isovector and isoscalar flavor combinations using our largest two momenta in \Cref{fig:pdf_pz_dep}.
There we see a relatively mild momentum dependence, which is expected given the observed momentum convergence in the renormalized matrix elements shown in \Cref{fig:Rmx_isoV_S}.

Finally, using the largest available momentum of $P_z = 6 \frac{2 \pi}{L}$, in \Cref{fig:pdf_xspace_comparison} we show the quark and antiquark transversity distributions from the LaMET approach for both the isovector and isoscalar flavor combinations compared to the DNN results from the previous section and with the global analysis from JAM3D-22~\cite{Gamberg:2022kdb} and Radici, Bacchetta~\cite{Radici:2018iag} which are both performed at LO.

There are a few things to note about these results.
First, recall that the global analysis and our DNN results both assume the anti-quark distribution to be zero.
Our $x$-space matching results in this section favor this assumption, at least when using an NLO matching kernel.

Further, recall that power corrections in the light-cone matching lead to a breakdown of the formalism when $x \sim 0,1$.
However, the RGR results also breakdown at small $x$, as seen by the onset of oscillations near $x \sim 0.2$, already giving a natural boundary for where we no longer trust the results.
\section{Conclusions}
\label{sec:conc}

In this paper we have presented various extractions of the transversity isovector and isoscalar quark PDFs, and their lowest few moments, of the proton from lattice QCD using a physical pion mass.
This work is a continuation towards the ultimate goal of uncovering the full structure of the proton from first principles.
Additionally, the matrix elements needed in this work also allow an estimate of the tensor charge $g_T$ to be extracted, and our results show reasonable agreement with other lattice extractions, as shown in \Cref{tab:tensor_charge}.
However, our calculations are performed at a single 
value of the lattice spacing, and for the isoscalar case we neglected the disconnected diagrams.

Regarding the transversity isovector and isoinglet PDFs, in our first method, we utilized the leading-twist OPE expansion of the reduced pseudo-ITD to extract the first few Mellin moments.
We found excellent agreement with the global analysis from JAM3D-22 for the lowest two moments and minor tensions for the next two moments.
Higher moments could not be reliably extracted.
Next, we used the pseudo-PDF approach, based on short-distance factorization, to extract an $x$-dependent PDF and utilized a deep neural network to overcome the inverse problem while remaining as unbiased as possible.
We saw some mild tension with the results from JAM3D-22 for a few small ranges of $x$ but otherwise mostly saw agreement.
Finally, we used the quasi-PDF approach, based on LaMET, to calculate the $x$-dependence PDF from hybrid-scheme renormalized matrix elements.
For this we found reasonably good agreement with JAM3D-22 in the moderate region of $x$, but there is significant tension with the results from Radici, Bacchetta.

A number of systematics are being ignored here and are left for future work.
These include the use of a single lattice spacing, NNLO corrections in $\alpha_s$, power corrections from the use of finite momentum, and isoscalar disconnected diagrams.
We can address the expected significance of these systematics, and we have good reason to expect their effects to be rather small.
Regarding the NNLO corrections, we saw in \Cref{fig:pdf_xspace_comparison} the NLO corrections were relatively mild in the middle $x$ regions, suggesting that NNLO corrections will be quite small as seen for the unpolarized proton distribution in our previous work~\cite{Gao:2022uhg}.
And, for the disconnected diagrams, we discussed earlier the expectation that the effects of these diagrams for local operator matrix elements would be smaller than the statistical error based on the study done in Refs.~\cite{Bhattacharya:2015esa,Bhattacharya:2015wna}.
Further, in our study of the unpolarized proton distribution~\cite{Gao:2022uhg}, we saw no evidence of convergence in the momentum used.
However, as seen in \Cref{fig:pdf_pz_dep}, the convergence in momentum is much more convincing for the transversity distribution.

\section*{Acknowledgments}

ADH acknowledges useful discussions with Fernando Romero-L\'opez. We would also like to thank Rui Zhang for discussions on our results.
Additionally, we would like to thank J.A. Gracey for help in implementing the four-loop conversion from RI-MOM to $\overline{\rm MS}$.

This material is based upon work supported by The U.S. Department of Energy, Office of Science, Office of Nuclear Physics through \textit{Contract No.~DE-SC0012704}, \textit{Contract No.~DE-AC02-06CH11357}, and within the frameworks of Scientific Discovery through Advanced Computing (SciDAC) award \textit{Fundamental Nuclear Physics at the Exascale and Beyond} and the Topical Collaboration in Nuclear Theory \textit{3D quark-gluon structure of hadrons: mass, spin, and tomography}.
SS is supported by the National Science Foundation under CAREER Award PHY-1847893 and by the RHIC Physics Fellow Program of the RIKEN BNL Research Center.
YZ is partially supported by the 2023 Physical Sciences and Engineering (PSE) Early Investigator Named Award program at Argonne National Laboratory.

This research used awards of computer time provided by: The INCITE program at Argonne Leadership Computing Facility, a DOE Office of Science User Facility operated under Contract No.~DE-AC02-06CH11357; the ALCC program at the Oak Ridge Leadership Computing Facility, which is a DOE Office of Science User Facility supported under Contract DE-AC05-00OR22725; the Delta system at the National Center for Supercomputing Applications through allocation PHY210071 from the ACCESS program, which is supported by National Science Foundation
grants \#2138259, \#2138286, \#2138307, \#2137603, and \#2138296.
Computations for this work were carried out in part on facilities of the USQCD
Collaboration, which are funded by the Office of Science of the
U.S. Department of Energy. Part of the data analysis are carried out on Swing, a high-performance computing cluster operated by the Laboratory Computing Resource Center at Argonne National Laboratory.

The computation of the correlators was carried out with the \texttt{Qlua} software suite~\cite{qlua}, which utilized the multigrid solver in \texttt{QUDA}~\cite{Clark:2009wm,Babich:2011np}. The analysis of the correlation functions was done with the use of \texttt{lsqfit}~\cite{lsqfit:11.5.1} and \texttt{gvar}~\cite{gvar:11.2}. Several of the plots were created with \texttt{Matplotlib}~\cite{Hunter:2007}.

\appendix

\section{Renormalization constant \texorpdfstring{$Z_T$}{ZT} in RI-MOM scheme}
\label{app:ZT}

\begin{figure}
    \centering
    \includegraphics[width=\columnwidth]{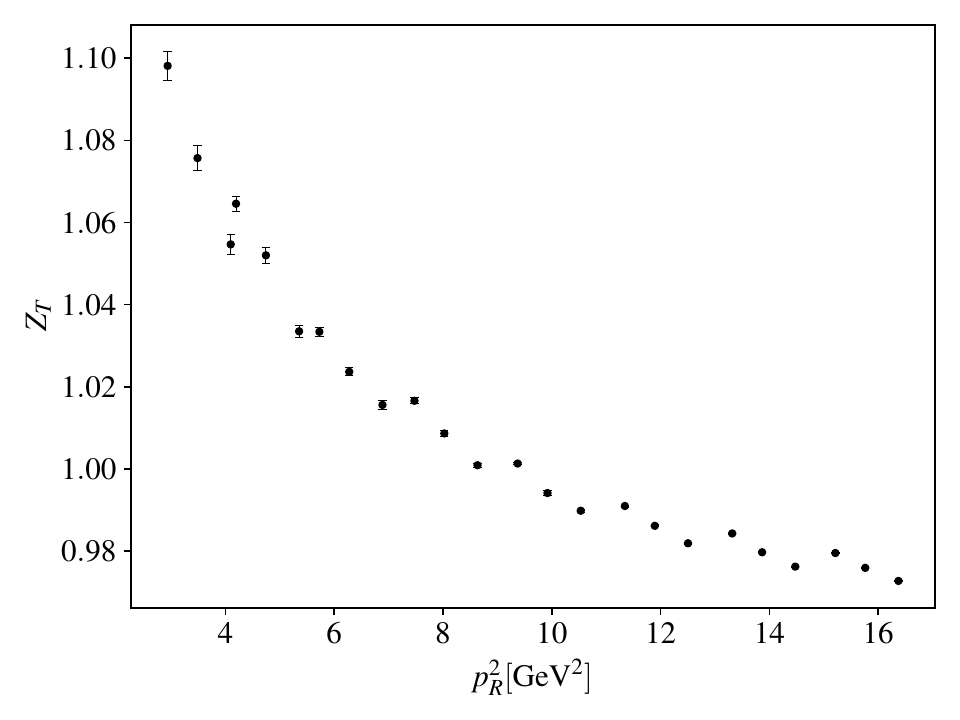}
    \caption{The tensor current renormalization factor $Z_T$ as a function of the RI-MOM
             momentum $p_R$.}
    \label{fig:ZT}
\end{figure}

Here we discuss the extraction of the renormalization constant $Z_T$ in the RI-MOM scheme and its subsequent conversion to the $\overline{\text{MS}}$ scheme at the scale $\mu = 2$ GeV.
The method starts by calculating matrix elements between off-shell quark states with lattice momenta
\begin{equation}
    a p_\mu = \frac{2 \pi}{L_\mu} (n_\mu + \frac{1}{2} \delta_{\mu 0}) ,
\end{equation}
where $L_\mu$ is the size of the lattice in the $\mu$th direction, $n_\mu \in \mathbb{Z}$, and $\mu=0$ is the temporal direction.
These matrix elements are computed in the Landau gauge.
The renormalization point is given by $(a p_R)^2 \equiv \sum_{\mu=0}^3 \sin^2 (a p_\mu)$, which is inspired by the lattice dispersion relation and helps to reduce discretization errors.
Our results for $Z_T$ are shown in \Cref{fig:ZT}.

\begin{figure}
    \centering
    \includegraphics[width=\columnwidth]{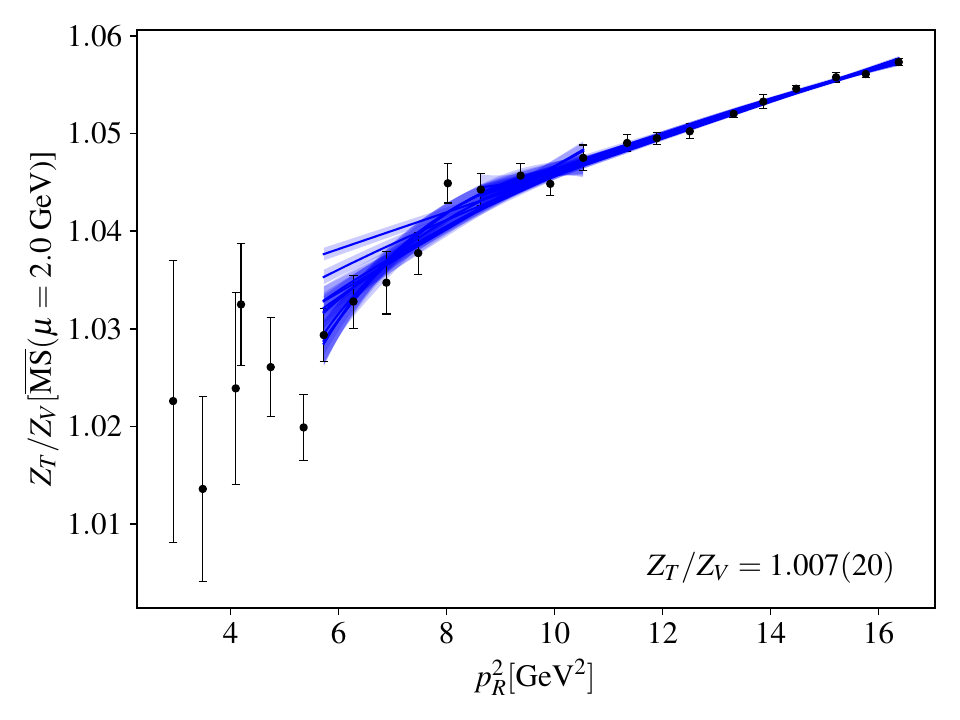}
    \caption{Ratio of the tensor to vector current renormalization factors $Z_T/Z_V$ as a function of the RI-MOM momentum $p_R$. The conversion from RI-MOM to $\overline{\rm MS}$ is done at four loops. The bands show the 18 different fits considered, all overlaid on top of one another.
    The AIC-averaged result final result for the ratio is given in the bottom right corner.}
    \label{fig:ZT-ZV_fit}
\end{figure}

There is a significant dependence on $p_R^2$ caused by non-perturbative associated with condensates and discretization errors that can be clearly seen from the ``fishbone'' structure at large $p_R^2$.
We had difficulty appropriately modeling these effects and instead chose to form ratios of $Z_T / Z_V$ in an attempt to cancel them as much as possible, similar to what was done in Ref.~\cite{Bhattacharya:2013ehc}.
We then use the conversion factor from RI-MOM to $\overline{\text{MS}}$ for the tensor current at both three loop~\cite{Gracey:2003yr} and four loop~\cite{Gracey:2022vqr} accuracy.
The resulting renormalization factors are then in the $\overline{\text{MS}}$ scheme at the scale $\mu^2 = p_R^2$, and thus we evolve them to the same scale using the evolution function computed at two loops in Ref.~\cite{Chetyrkin:1997dh}.
We evolve $Z_T$ to the scale $\mu = 2$ GeV, as this is a commonly used scale for reporting results of nucleon charges.
The resulting ratio $Z_T / Z_V$ after conversion to $\overline{\text{MS}}$ at $\mu = 2$ GeV is then fit to
\begin{equation}
        Z_T/Z_V + B / p_R^2 + D_1 p_R + D_2 p_R^2 ,
\end{equation}
where last two terms incorporate discretization effects.
In order to remove bias from our choice of fit,
we consider six different variations of this fit form, corresponding to setting various terms to zero.
Specifically, we consider a linear form (i.e. $D_2 = 0$), a quadratic form (i.e. $D_1 = 0$), and a linear+quadratic form (i.e. $D_1 \neq 0$ and $D_2 \neq 0$).
Then for each of these three, we also consider fits in which $B$ is zero and non-zero.
To further give variation to our fits, we use three ranges of the data.
The first includes all but the smallest values of $p_R^2$, which is always left out.
Then we consider removing more of the small $p_R^2$ data, and finally removing the largest $p_R^2$ data.
This gives a total of 18 fits we consider.
To give a final estimate, we simply take an AIC average over all fits, giving
\begin{equation}
\begin{split}
    Z_T / Z_V = 1.018(18) , \;\; \overline{\text{MS}} (\mu = 2 \; \text{GeV}), \;\; 3\text{-loop}, \\
    Z_T / Z_V = 1.007(20) , \;\; \overline{\text{MS}} (\mu = 2 \; \text{GeV}), \;\; 4\text{-loop} ,
\end{split}
\end{equation}
using the three-loop and four-loop conversion to $\overline{\rm MS}$, respectively.
The results of all these fits, using the four-loop conversion to $\overline{\rm MS}$, are shown in \Cref{fig:ZT-ZV_fit}.
This, rather conservative, method for estimating the systematic error is justified for this observable which is likely affected by large systematics.

\section{Three-point function fits}
\label{app:three_point_fits}

\begin{figure*}
    \centering
    \includegraphics[width=\columnwidth]{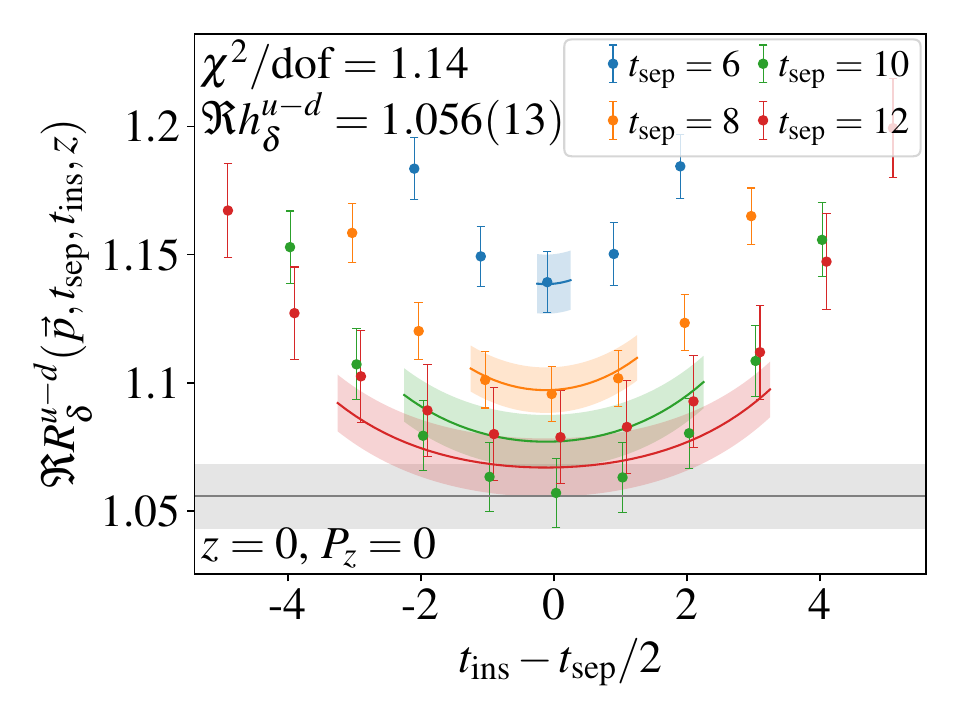}
    \includegraphics[width=\columnwidth]{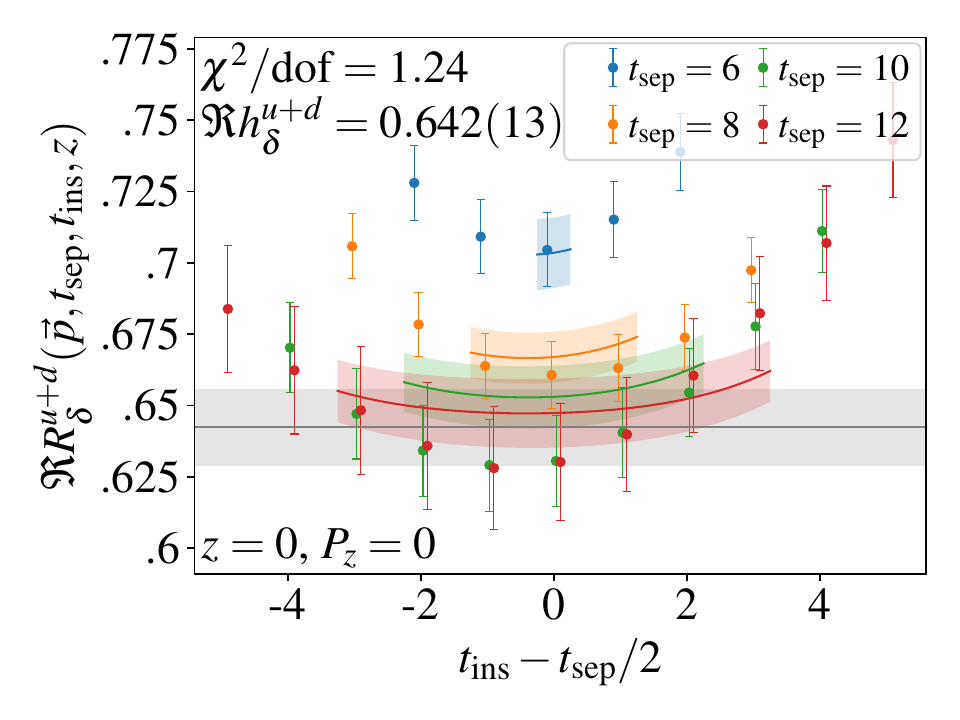}
    \caption{The real parts of the ratio of local (left) isovector and (right) isoscalar three-point to two-point functions for $P_z = 0$. The $\chi^2/\textrm{dof}$ reported, estimate
             for the ground-state bare matrix element (also represented by a gray band),
             and $\ts$ fit bands come from the preferred fit strategy,
             i.e. the two-state fit to the ratio $R_{\delta}$ with $n_{\rm exc} = 2$,
             where $n_{\rm exc}$ is the number of data points nearest both the source and sink that are not included in the fit.
             The range in $\ti$ of the $\ts$ bands
             covers the included data points in the fit.}
    \label{fig:c3pt_loc_u-d_and_u+d_repr_fits_re}
\end{figure*}

Here we show a handful of fits to the ratios of three-point to two-point functions used in the main text.
All of these fits utilize our preferred fit strategy, i.e. the two-state fit to the ratio $R_{\delta}$ in \eqref{eq:ratio} with $n_{\rm exc} = 2$, where $n_{\rm exc}$ is the number of data points nearest both the source and sink that are not included in the fit.

\begin{figure*}
    \centering
    \includegraphics[width=0.32\textwidth]{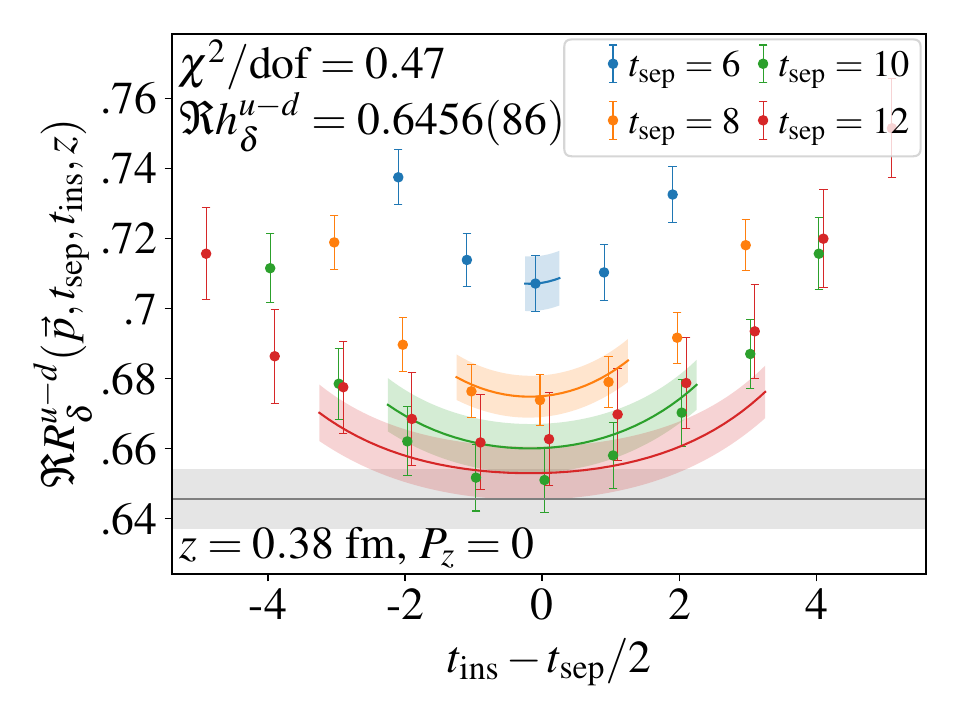}
    \includegraphics[width=0.32\textwidth]{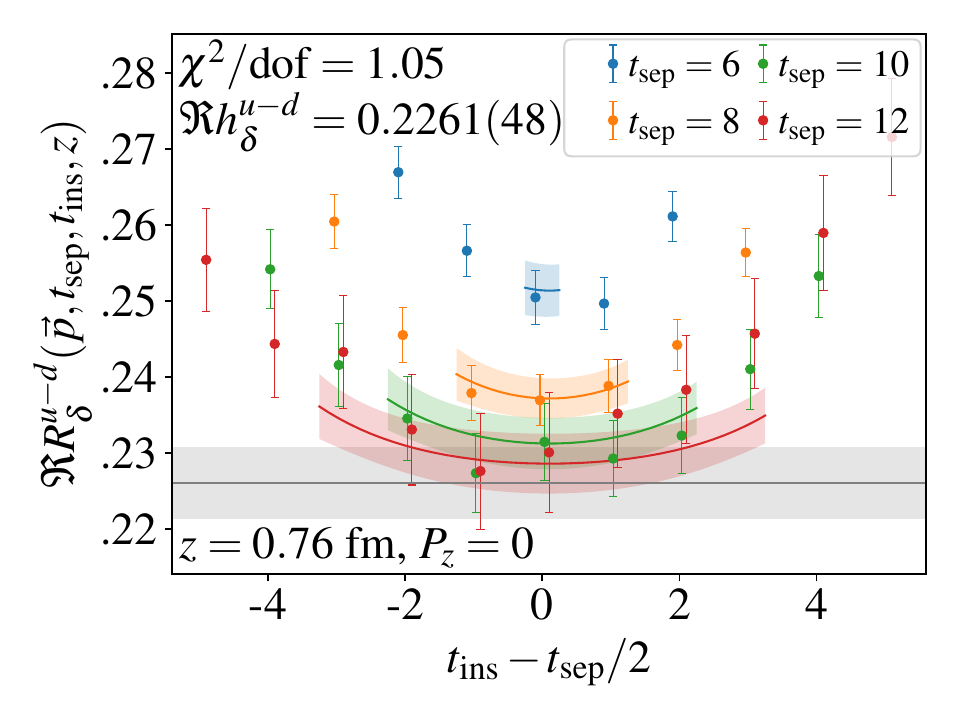}
    \includegraphics[width=0.32\textwidth]{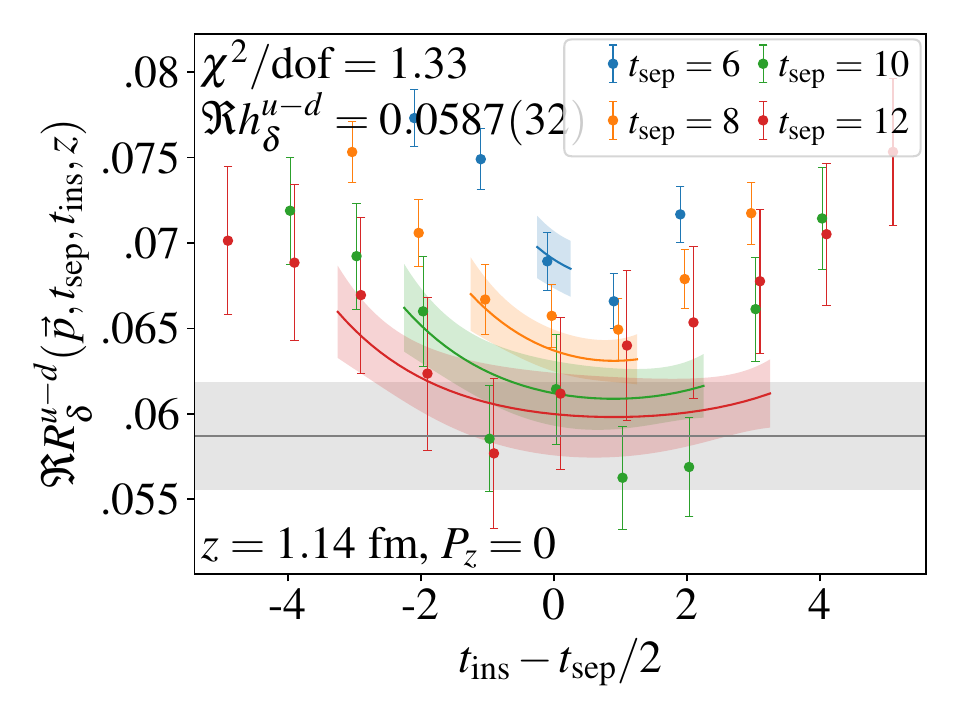}
    \includegraphics[width=0.32\textwidth]{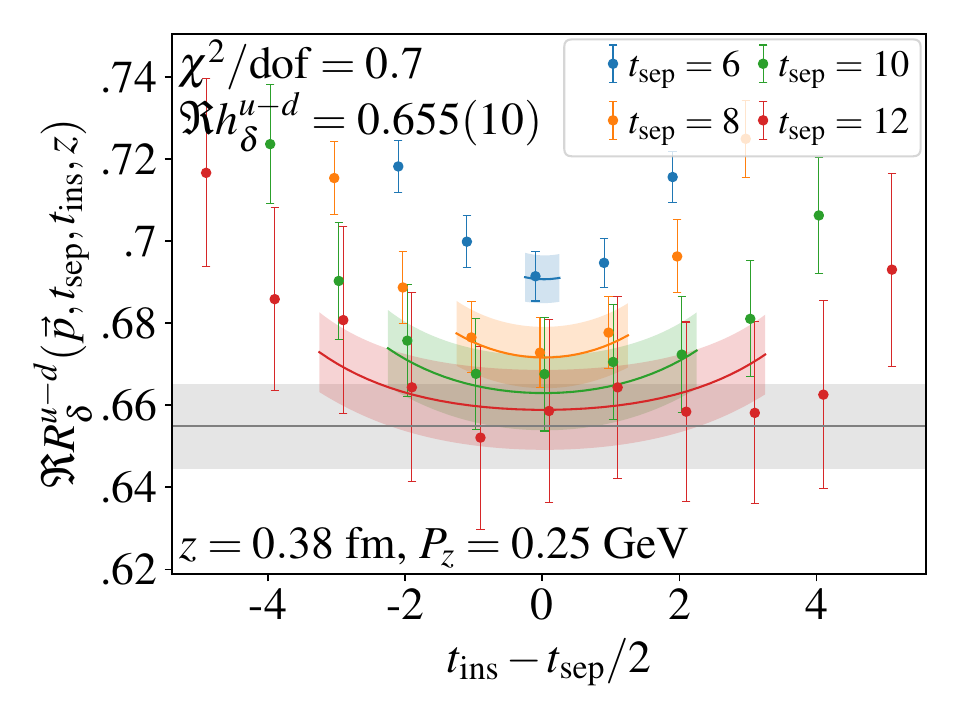}
    \includegraphics[width=0.32\textwidth]{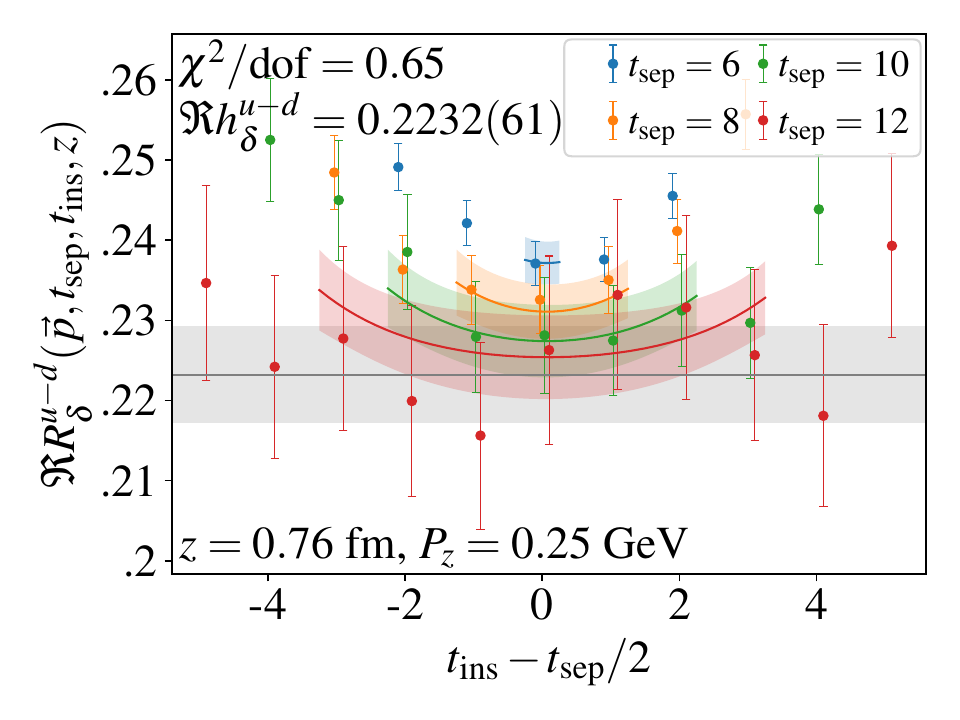}
    \includegraphics[width=0.32\textwidth]{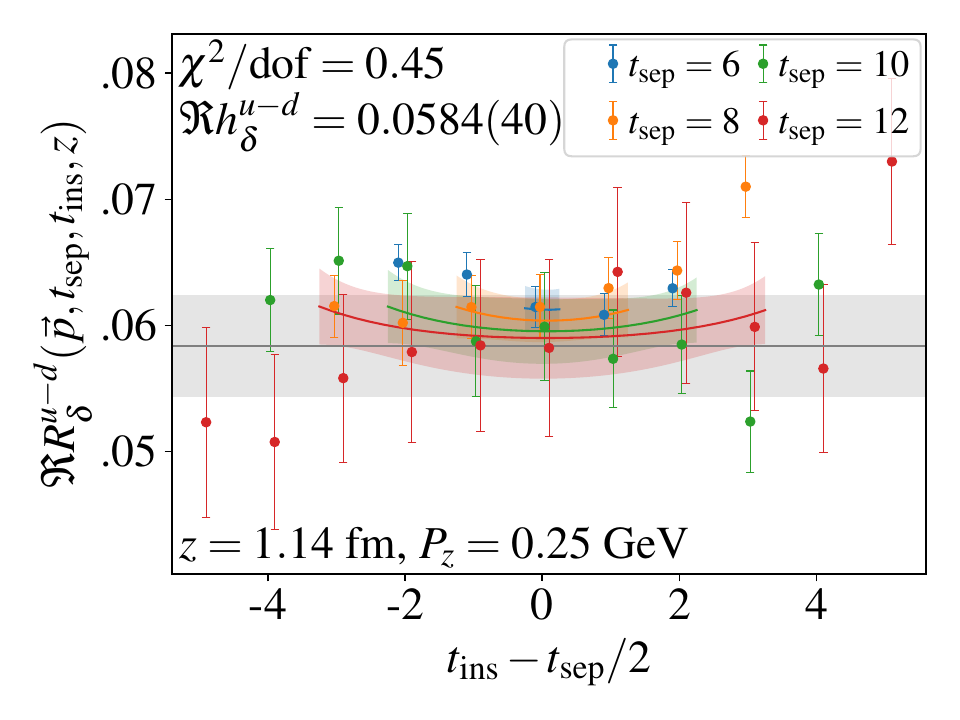}
    \includegraphics[width=0.32\textwidth]{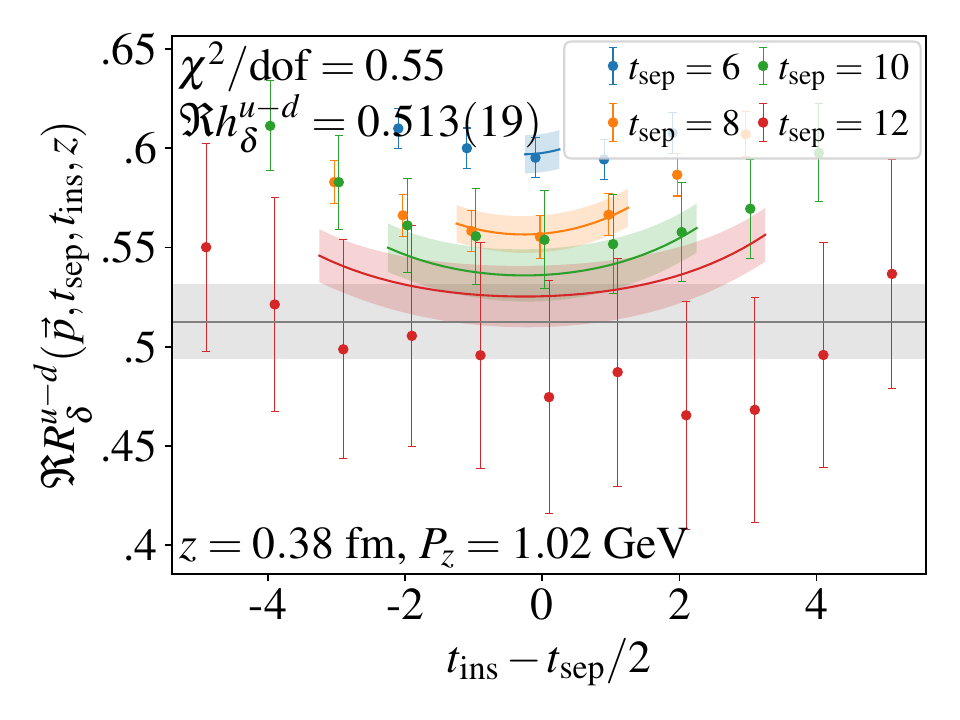}
    \includegraphics[width=0.32\textwidth]{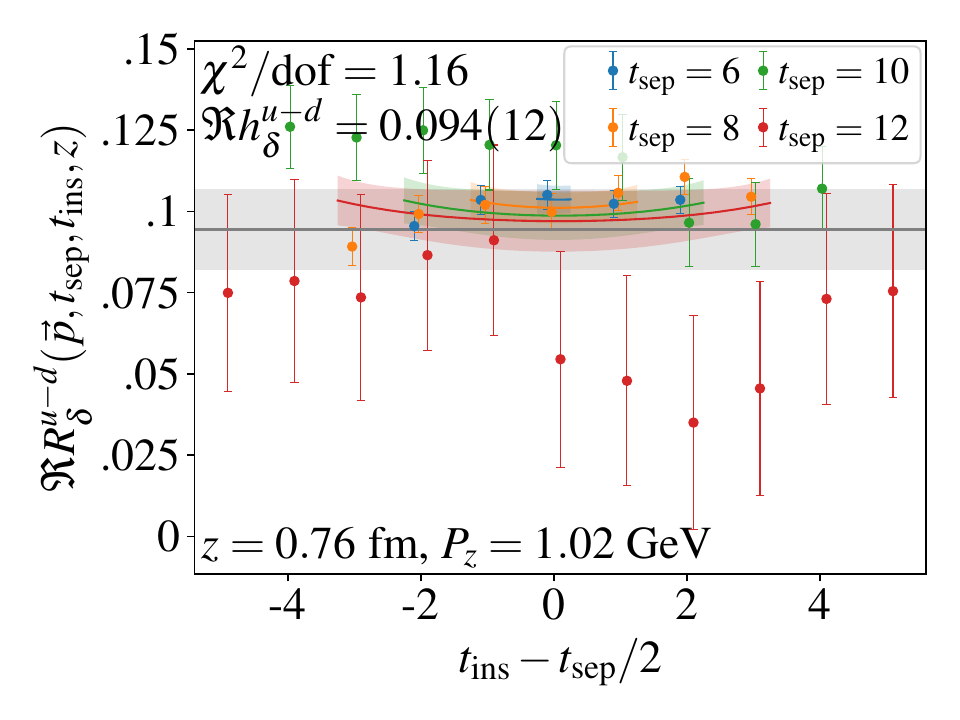}
    \includegraphics[width=0.32\textwidth]{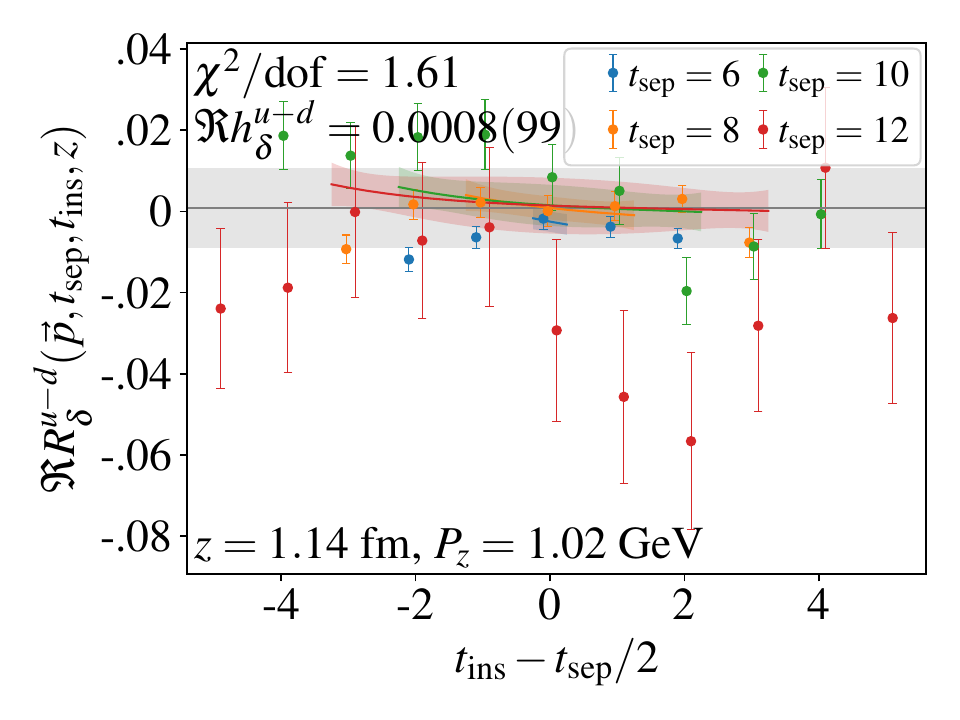}
    \includegraphics[width=0.32\textwidth]{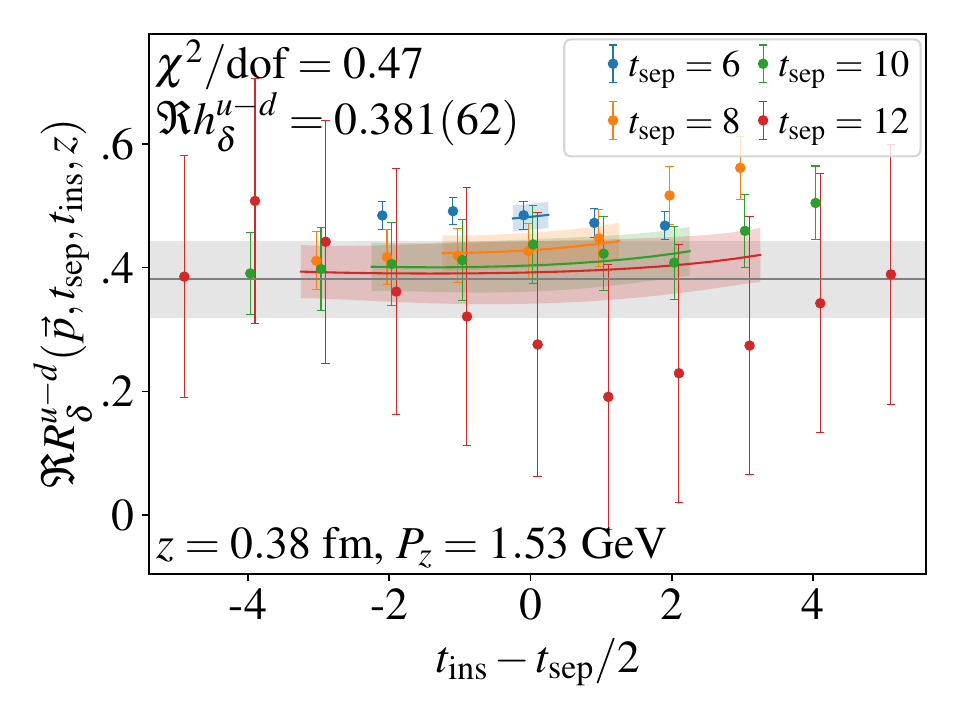}
    \includegraphics[width=0.32\textwidth]{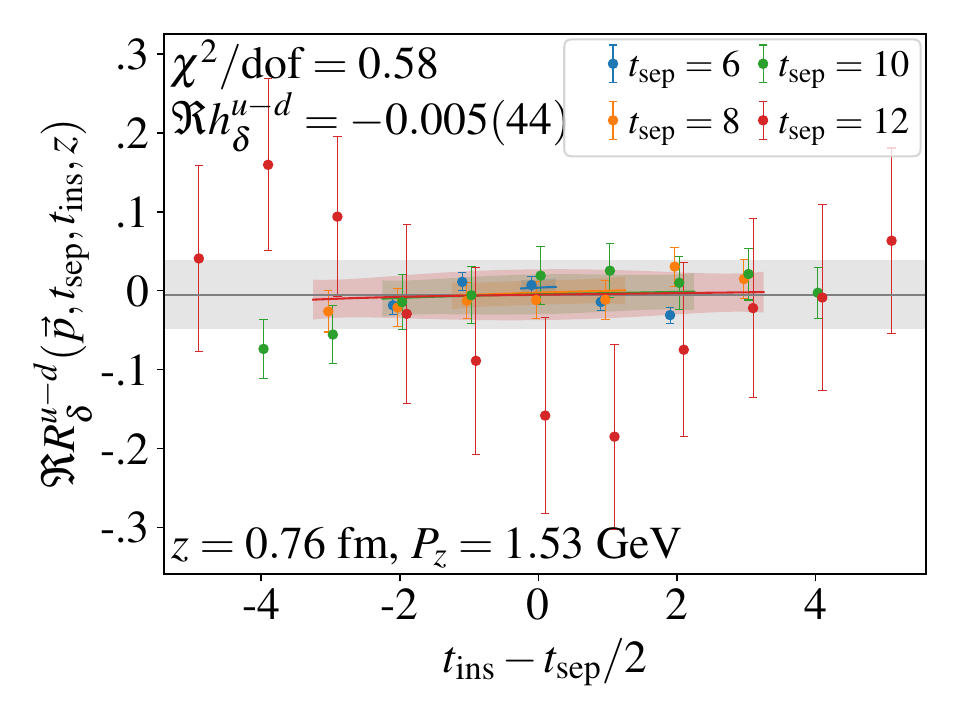}
    \includegraphics[width=0.32\textwidth]{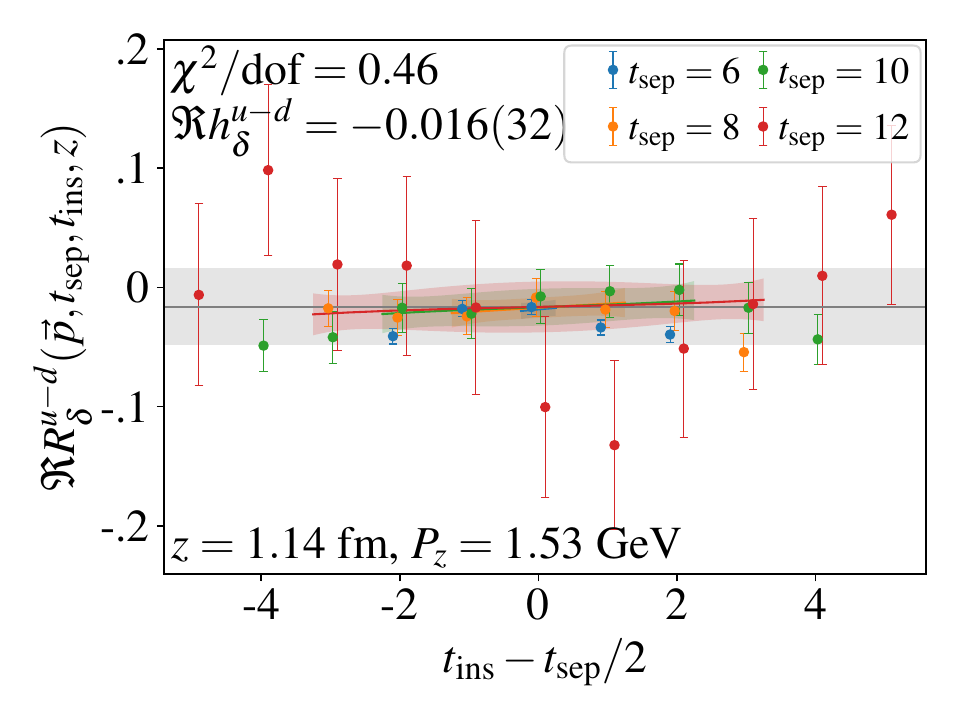}
    \caption{The real parts of the ratio of isovector three-point to two-point functions for all values
             of momentum (one for each row) and a few representative values of Wilson-line
             length $z$ (one for each column). The $\chi^2/\textrm{dof}$ reported, estimate
             for the ground-state bare matrix element (also represented by a gray band),
             and $\ts$ fit bands come from the preferred fit strategy,
             i.e. the two-state fit to the ratio $R_{\delta}$ with $n_{\rm exc} = 2$,
             where $n_{\rm exc}$ is the number of data points nearest both the source and sink that are not included in the fit.
             The range in $\ti$ of the $\ts$ bands
             covers the included data points in the fit.}
    \label{fig:c3pt_u-d_repr_fits_re}
\end{figure*}

\begin{figure*}
    \centering
    \includegraphics[width=0.32\textwidth]{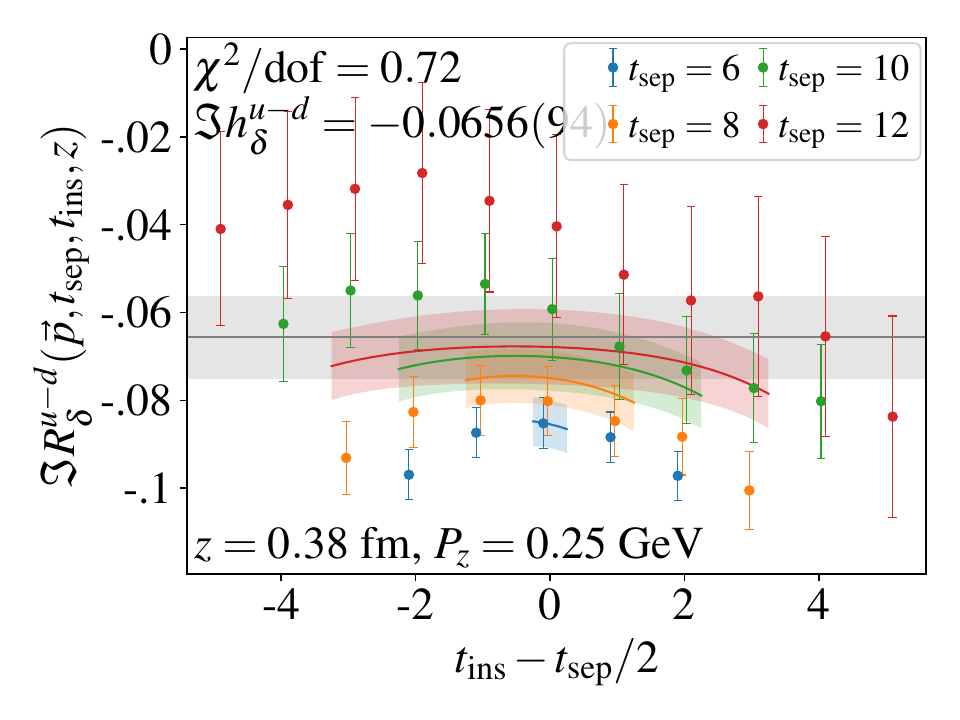}
    \includegraphics[width=0.32\textwidth]{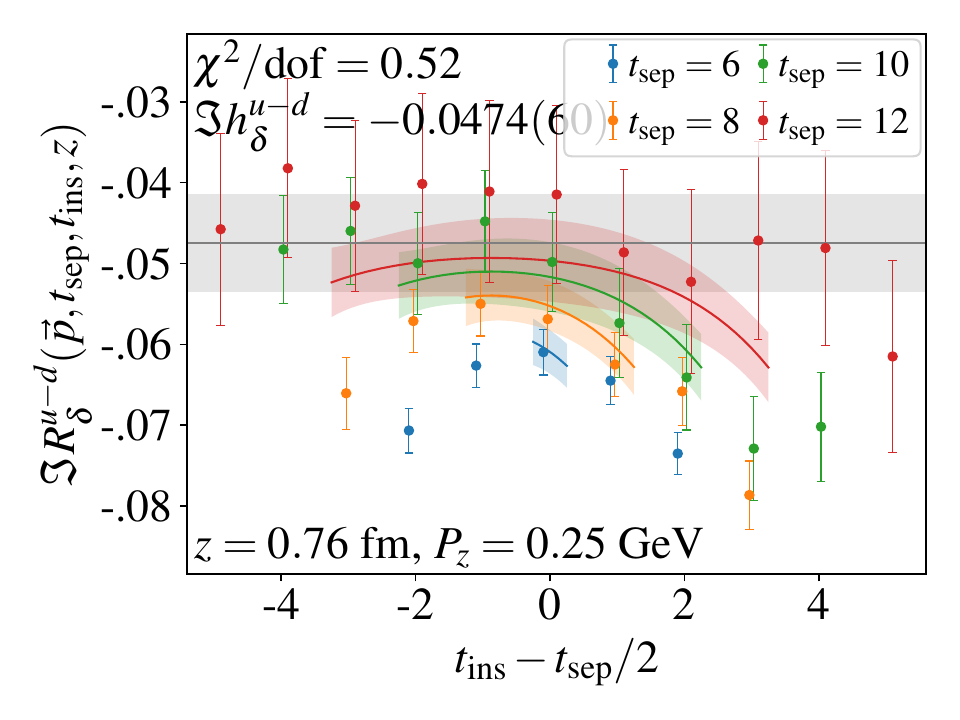}
    \includegraphics[width=0.32\textwidth]{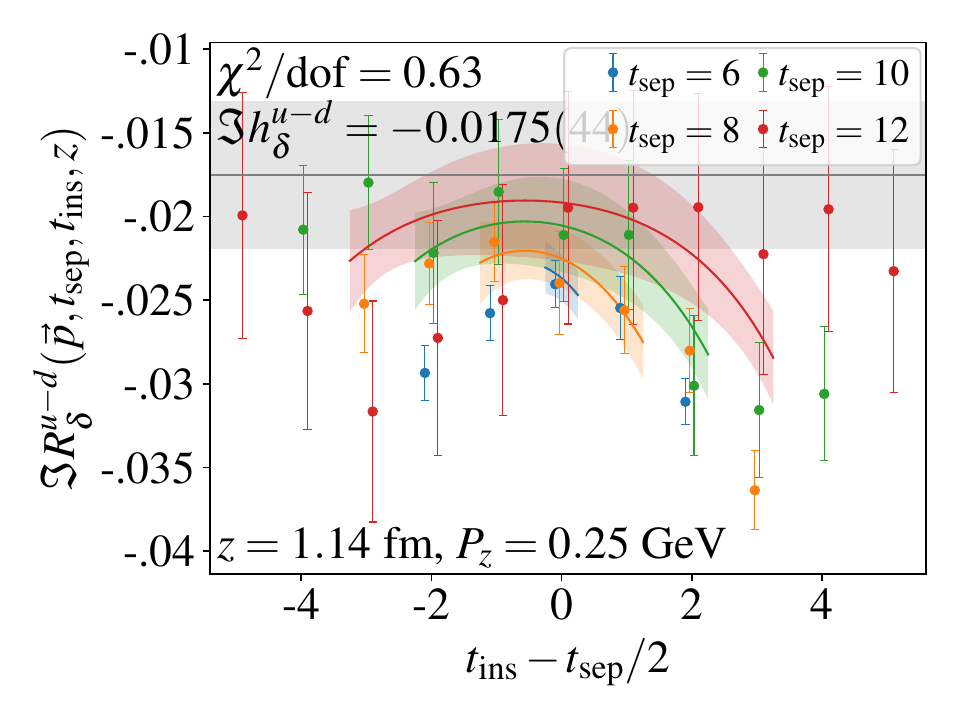}
    \includegraphics[width=0.32\textwidth]{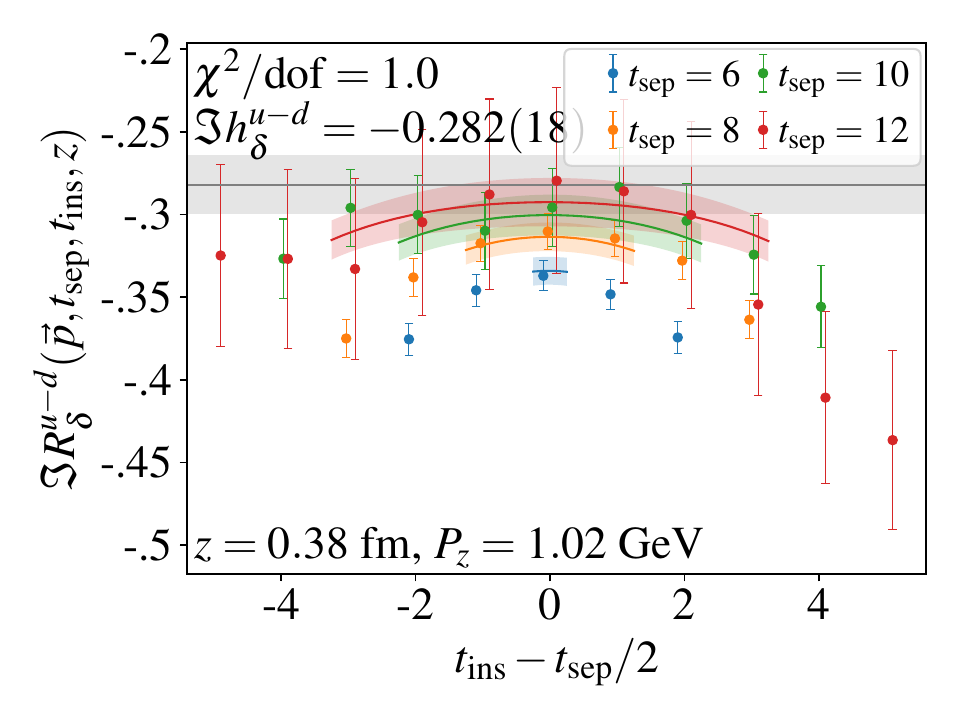}
    \includegraphics[width=0.32\textwidth]{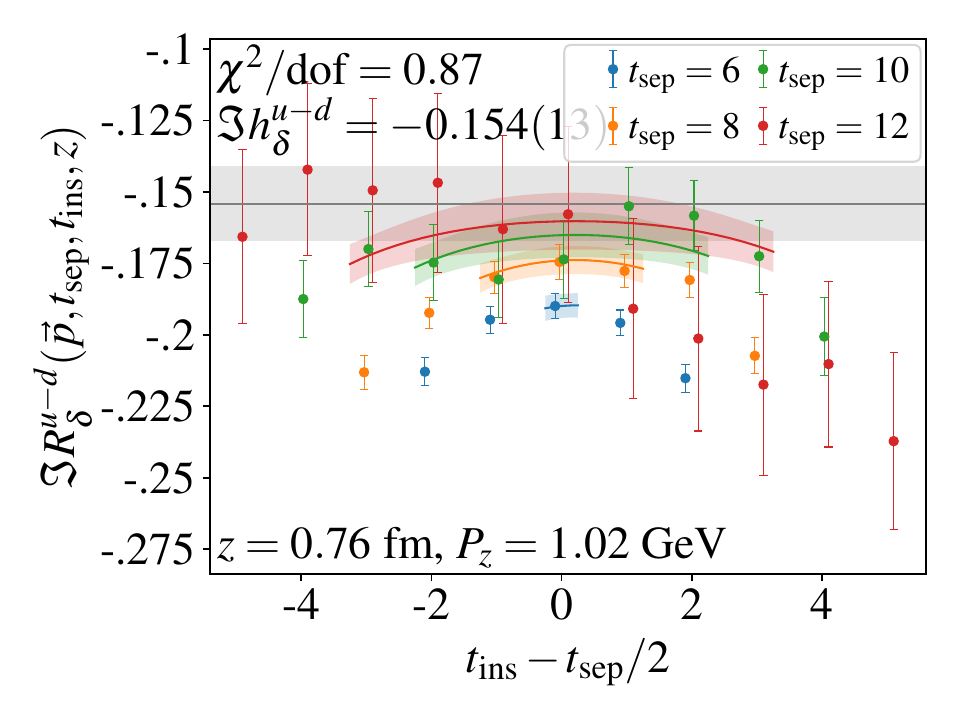}
    \includegraphics[width=0.32\textwidth]{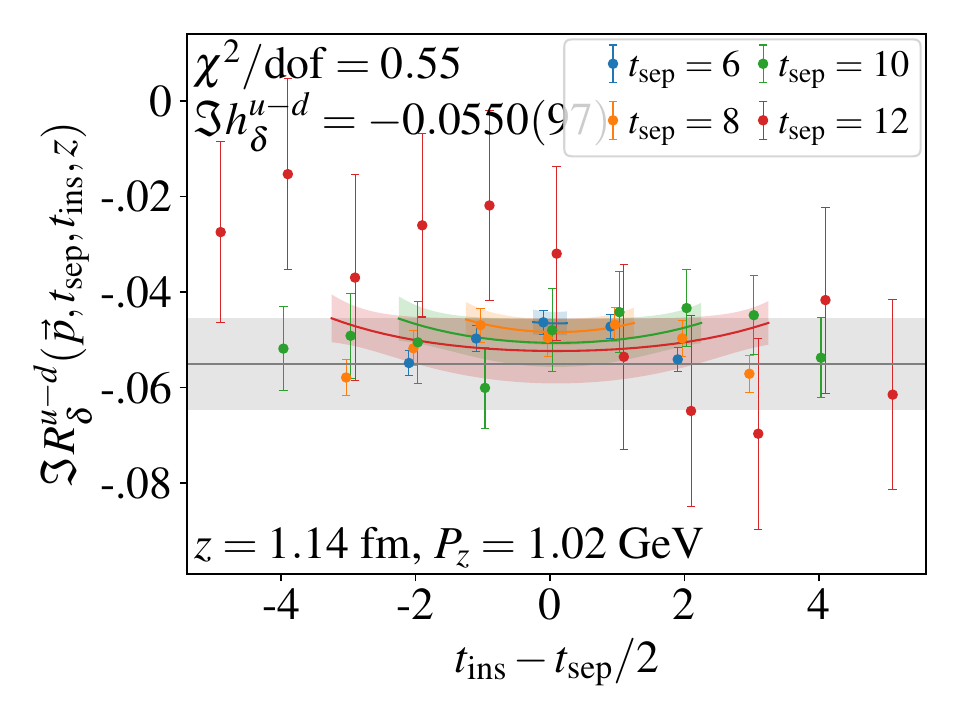}
    \includegraphics[width=0.32\textwidth]{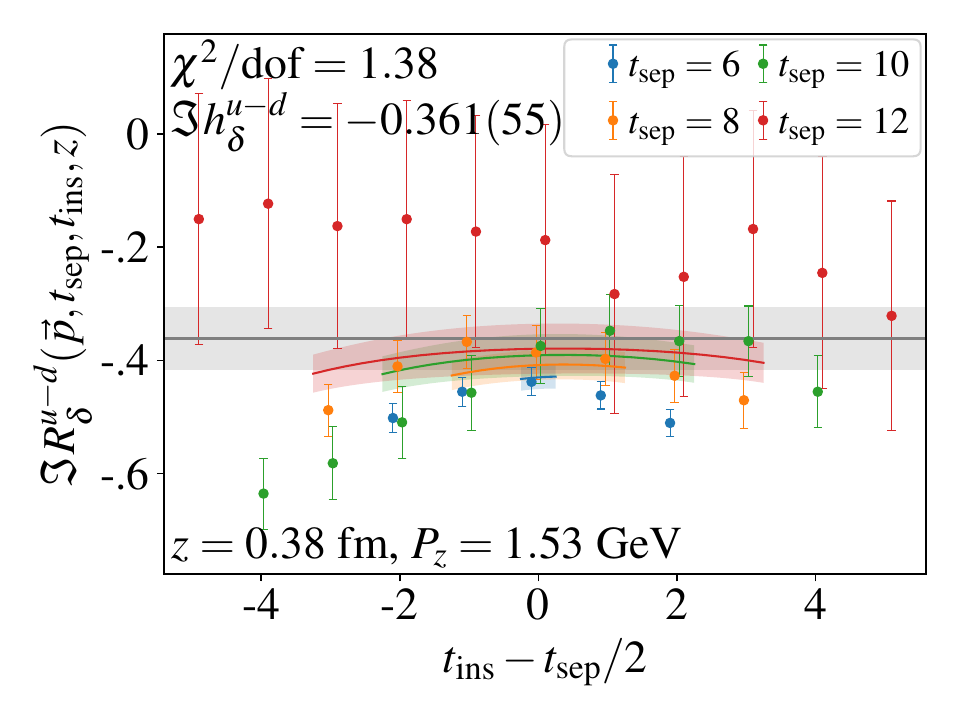}
    \includegraphics[width=0.32\textwidth]{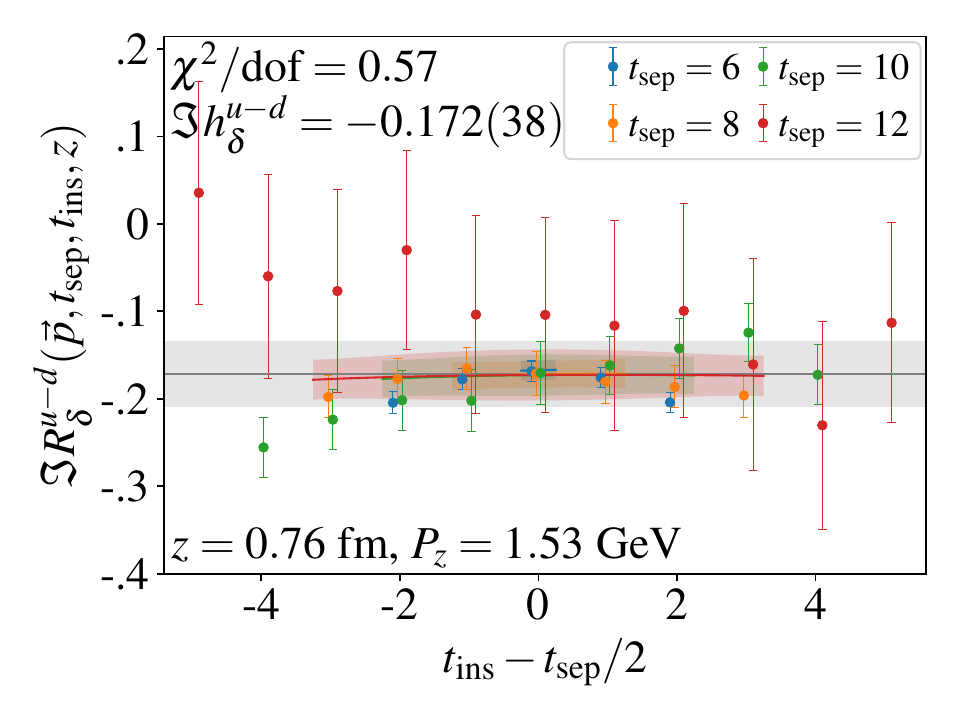}
    \includegraphics[width=0.32\textwidth]{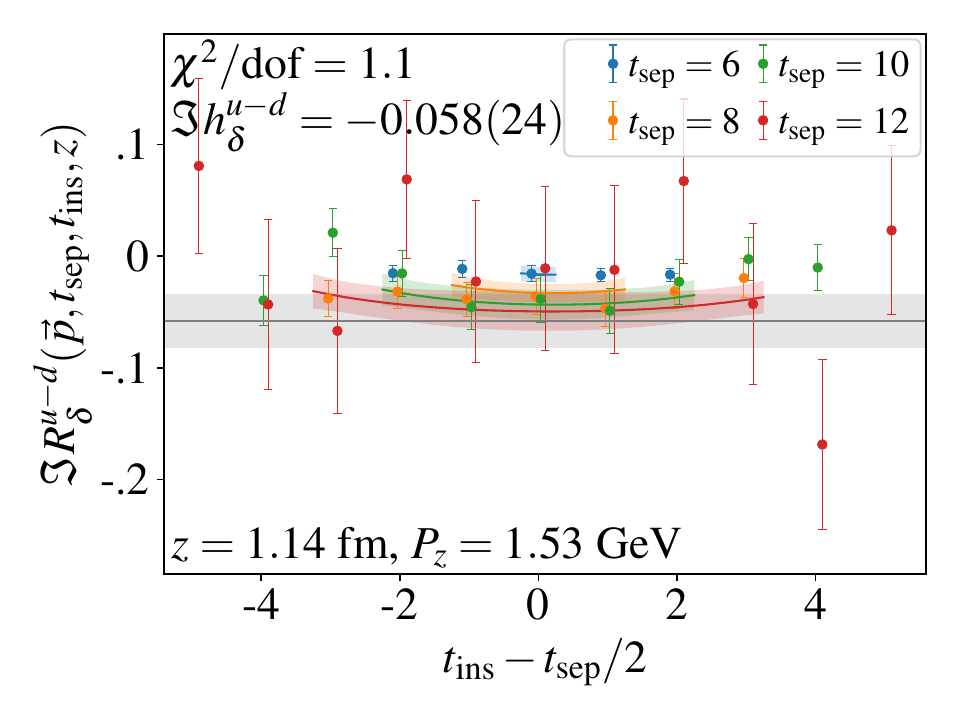}
    \caption{The same as \Cref{fig:c3pt_u-d_repr_fits_re}, but for the imaginary parts.
             Additionally, the zero-momentum matrix elements are not shown, as these are
             all consistent with zero (as they are expected to be).}
    \label{fig:c3pt_u-d_repr_fits_im}
\end{figure*}

\begin{figure*}
    \centering
    \includegraphics[width=0.32\textwidth]{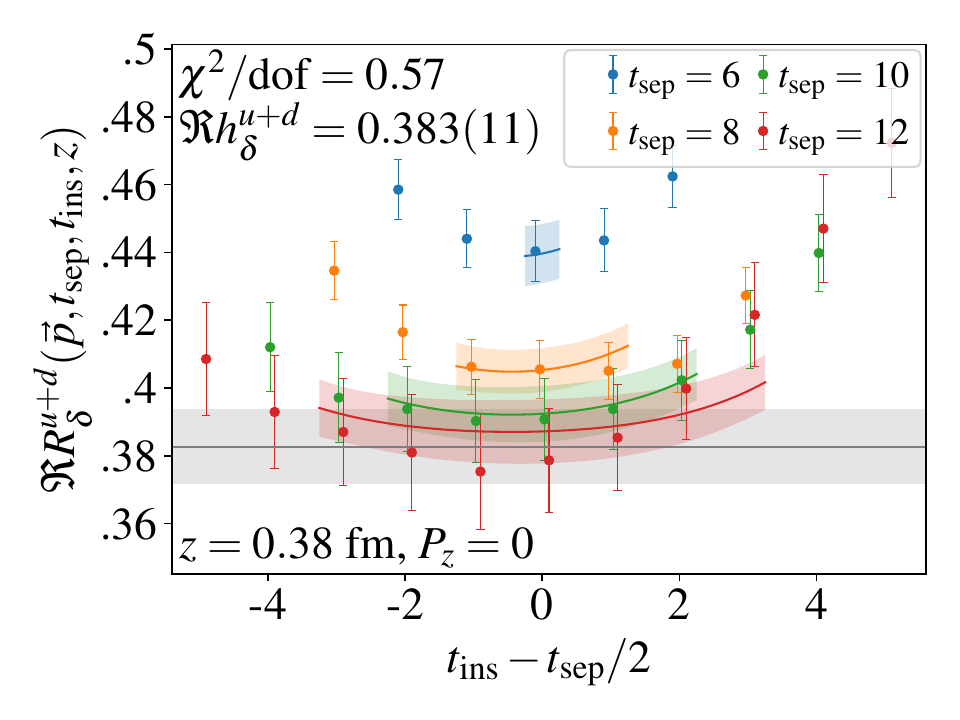}
    \includegraphics[width=0.32\textwidth]{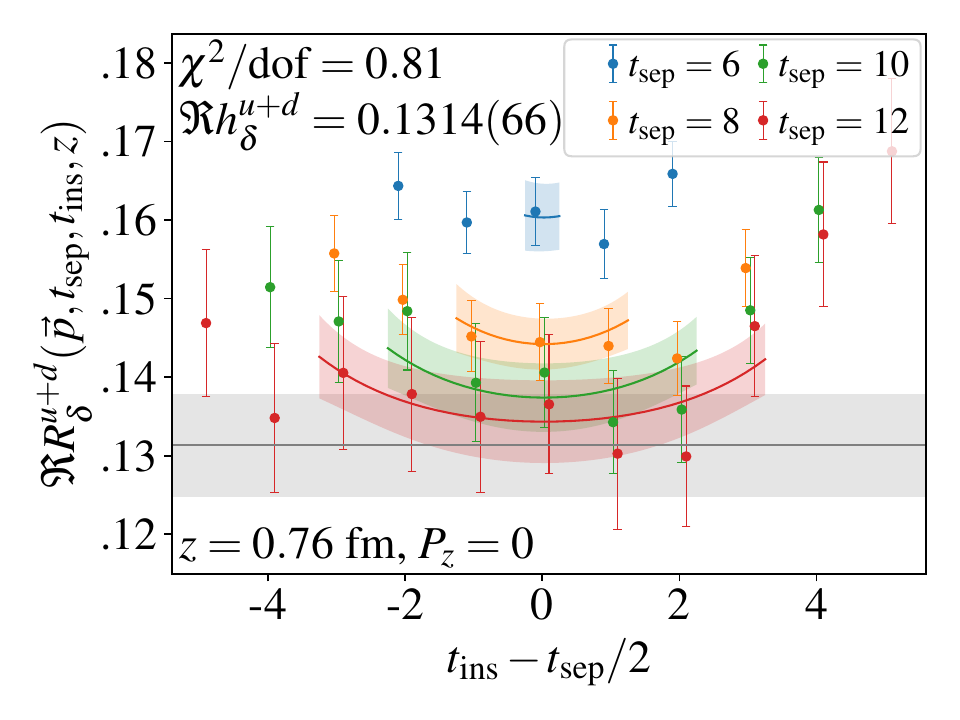}
    \includegraphics[width=0.32\textwidth]{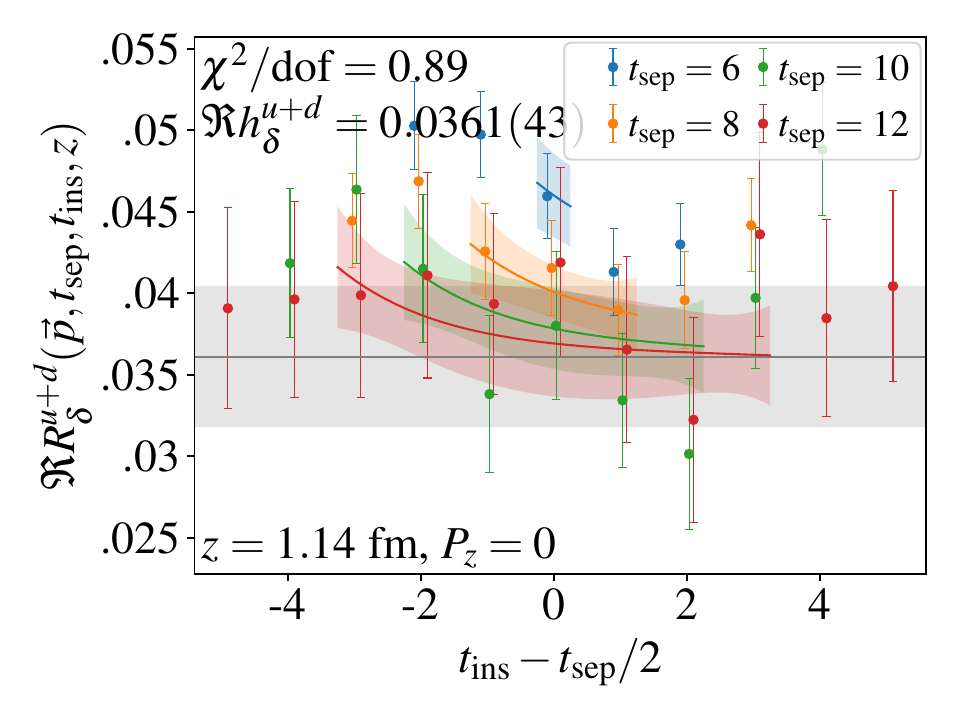}
    \includegraphics[width=0.32\textwidth]{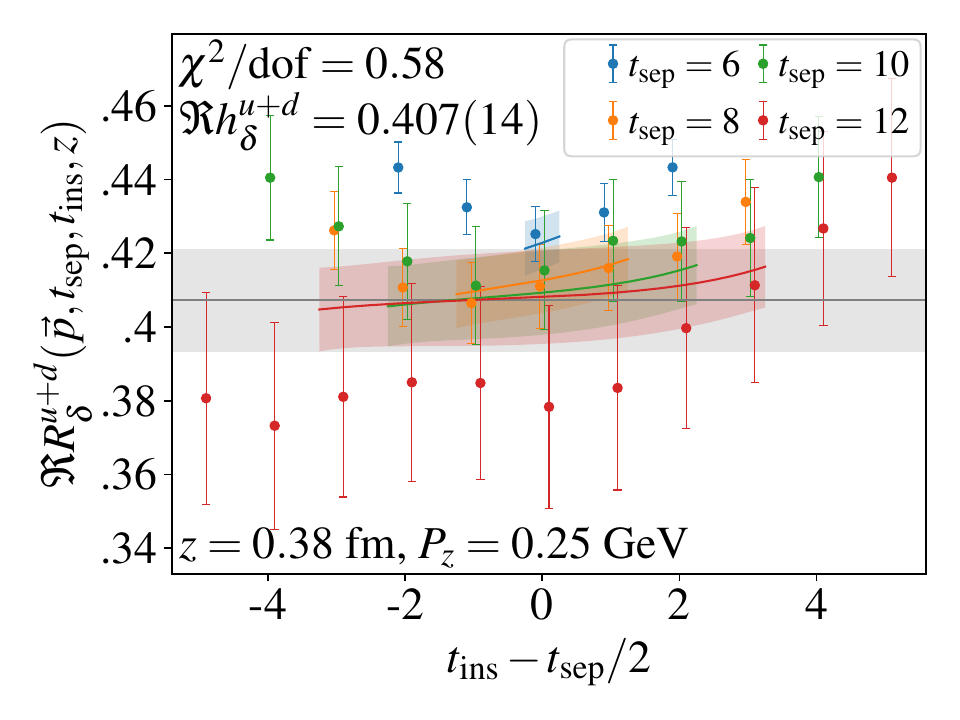}
    \includegraphics[width=0.32\textwidth]{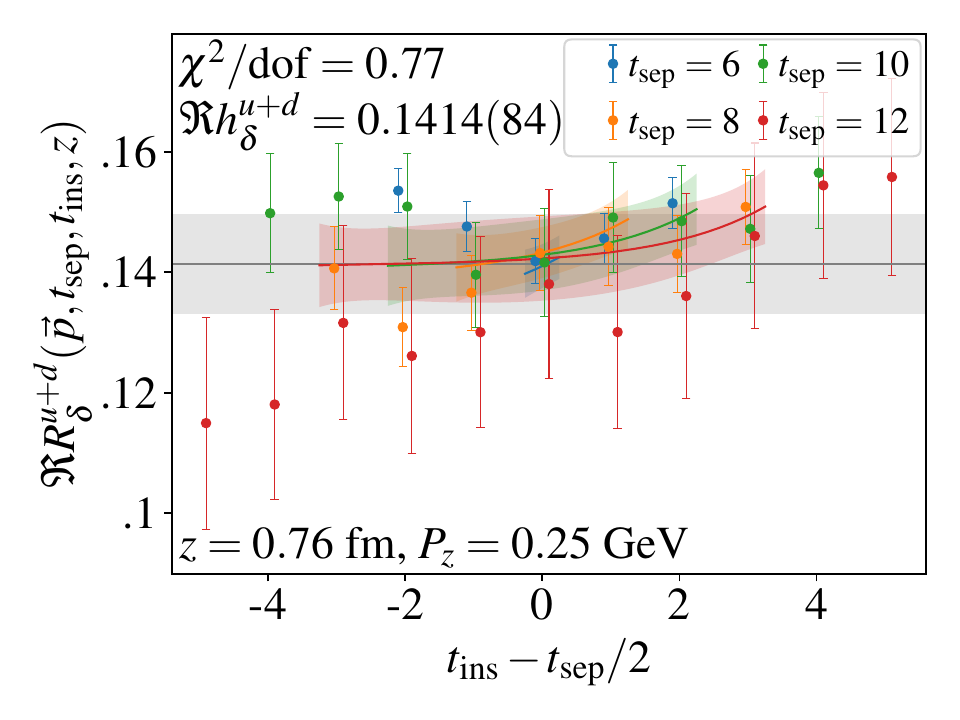}
    \includegraphics[width=0.32\textwidth]{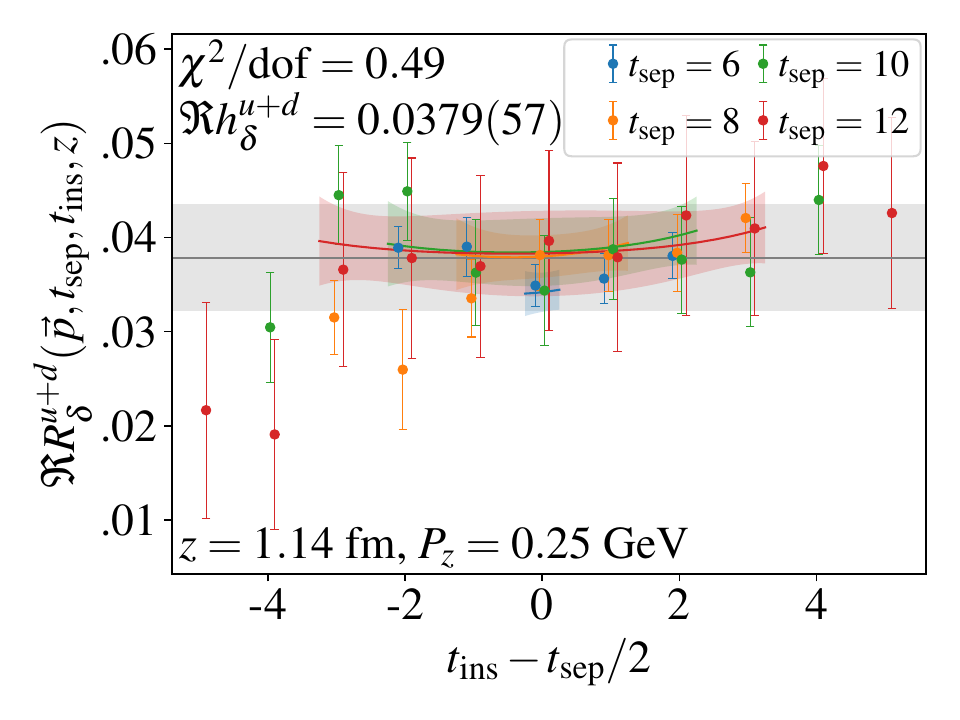}
    \includegraphics[width=0.32\textwidth]{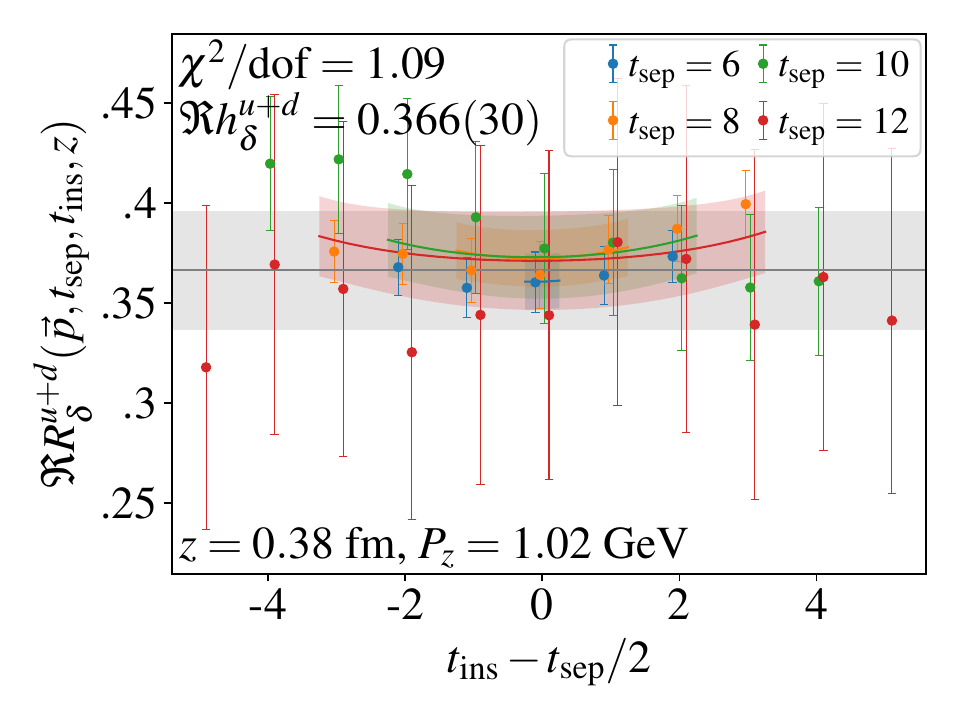}
    \includegraphics[width=0.32\textwidth]{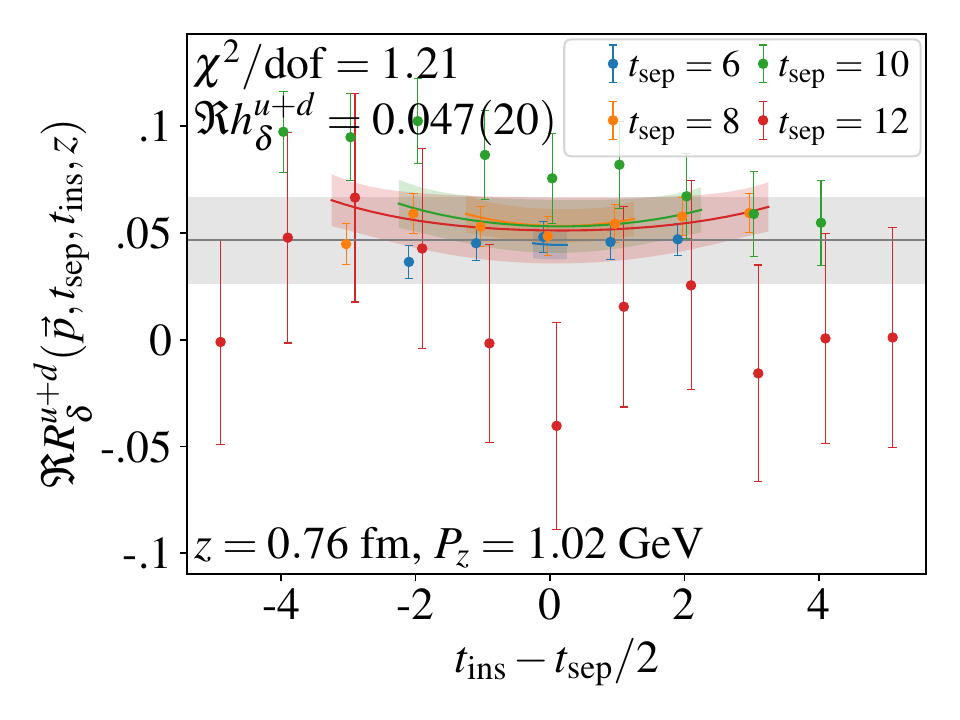}
    \includegraphics[width=0.32\textwidth]{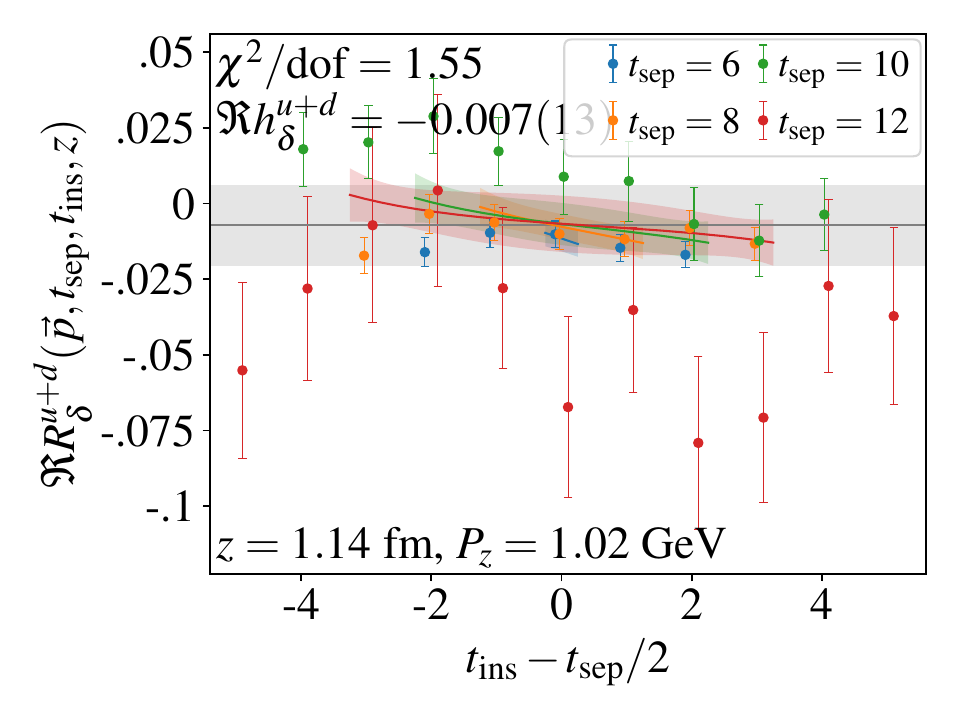}
    \includegraphics[width=0.32\textwidth]{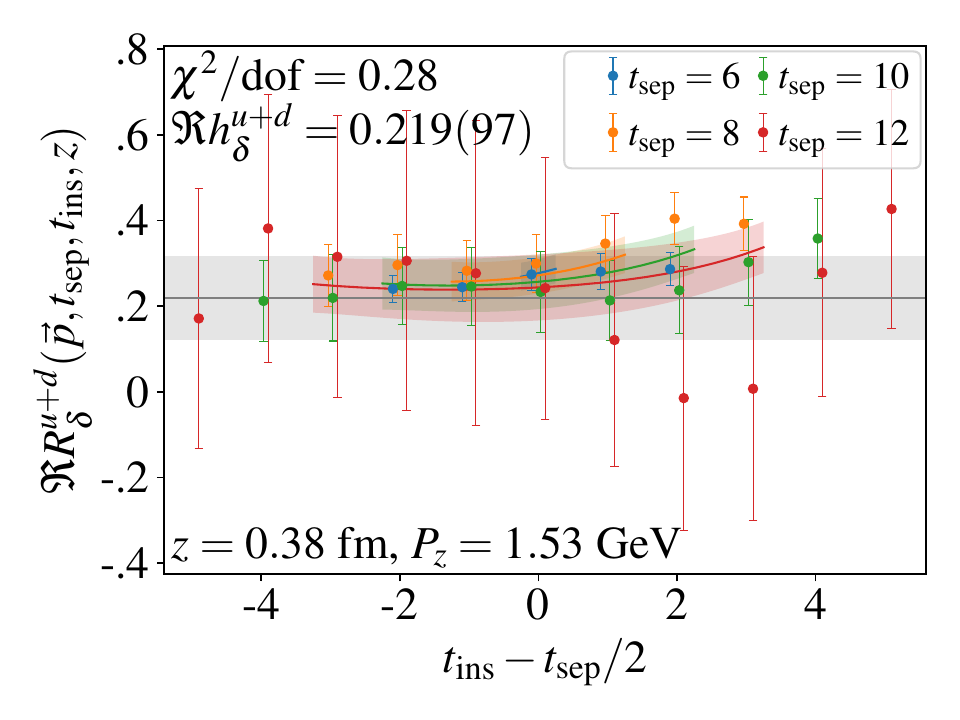}
    \includegraphics[width=0.32\textwidth]{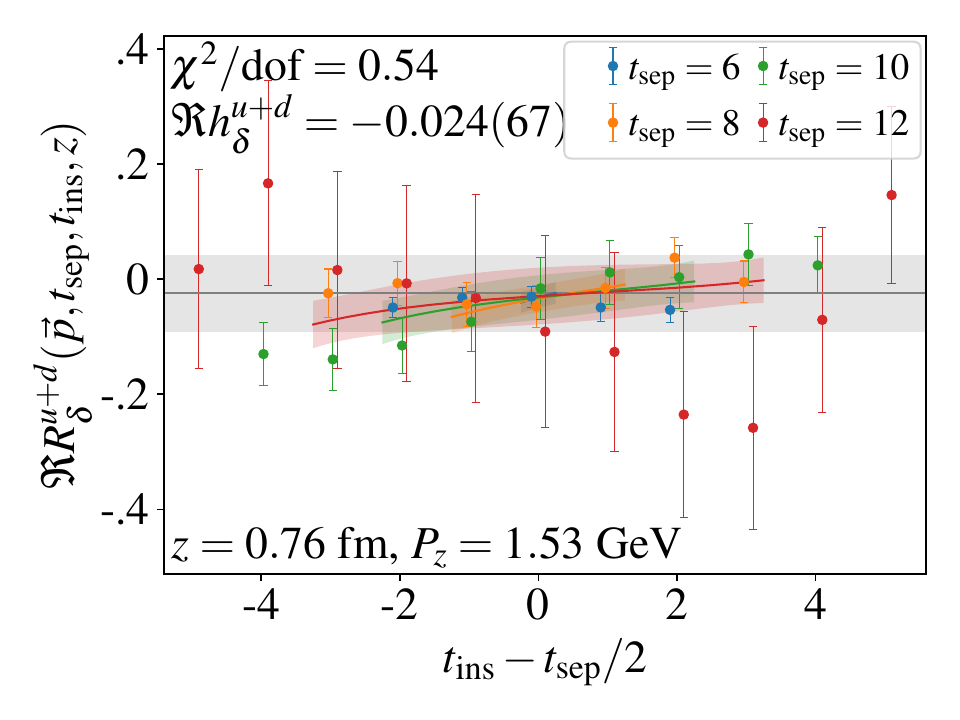}
    \includegraphics[width=0.32\textwidth]{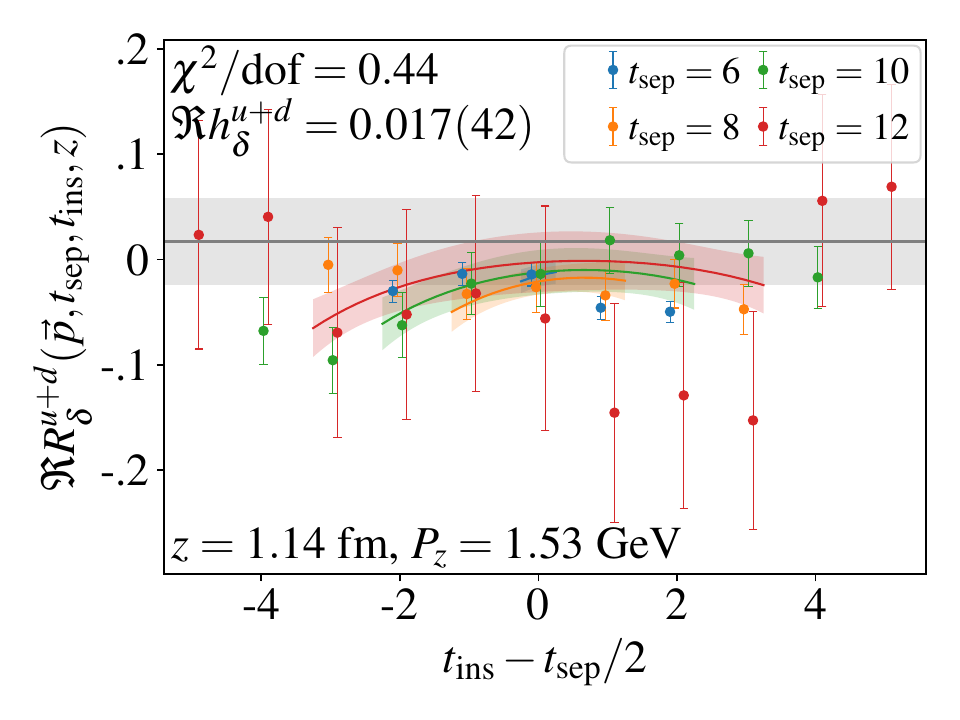}
    \caption{The same as \Cref{fig:c3pt_u-d_repr_fits_re}, but for the isoscalar
             matrix elements.}
    \label{fig:c3pt_u+d_repr_fits_re}
\end{figure*}

\begin{figure*}
    \centering
    \includegraphics[width=0.32\textwidth]{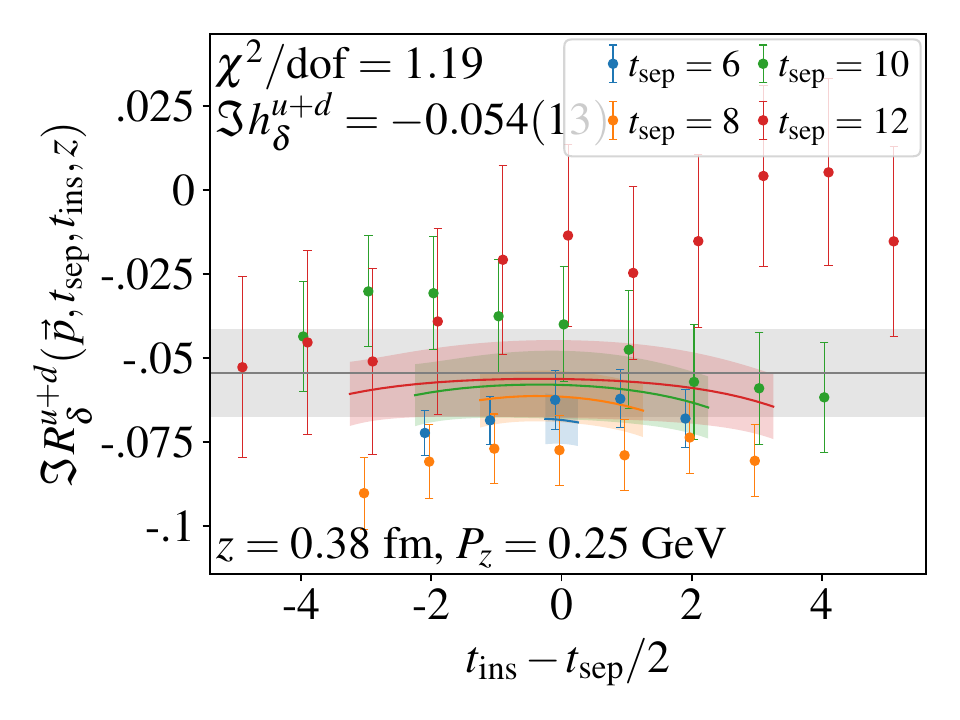}
    \includegraphics[width=0.32\textwidth]{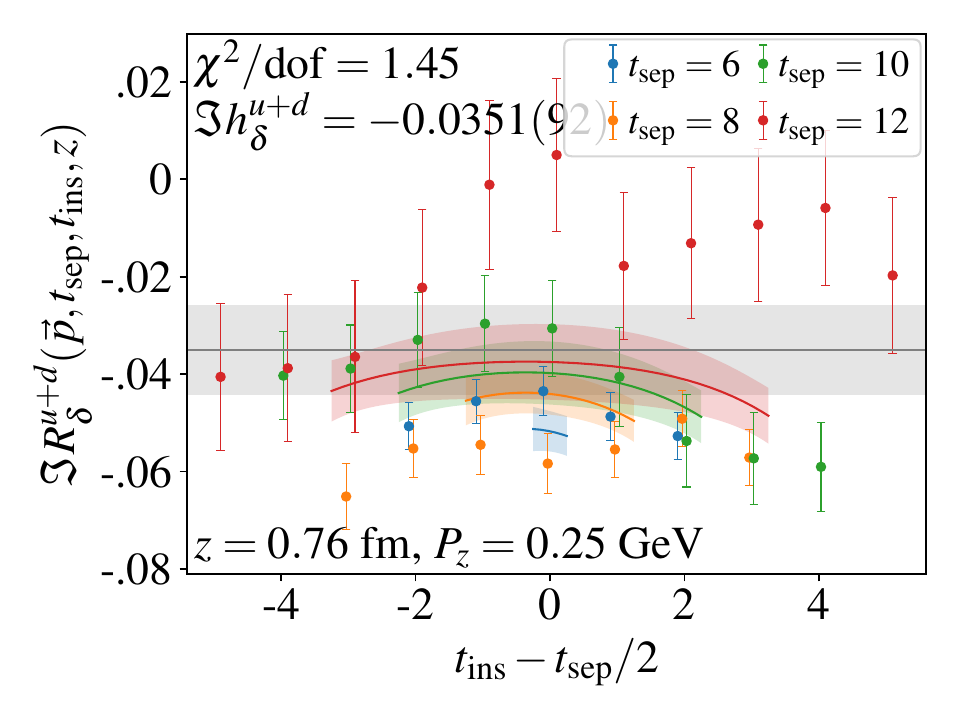}
    \includegraphics[width=0.32\textwidth]{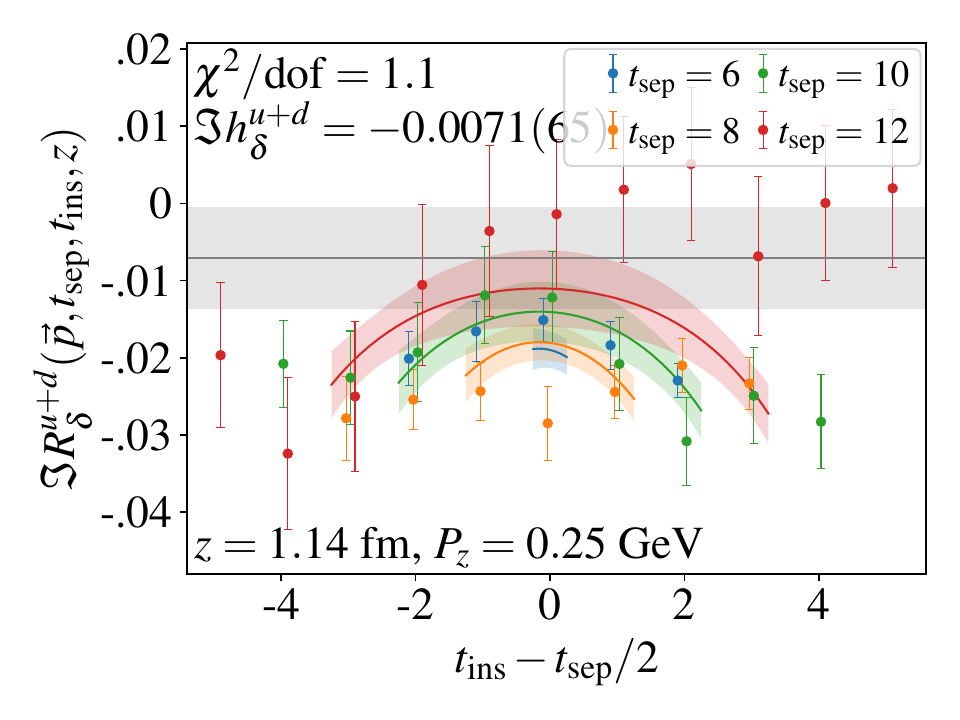}
    \includegraphics[width=0.32\textwidth]{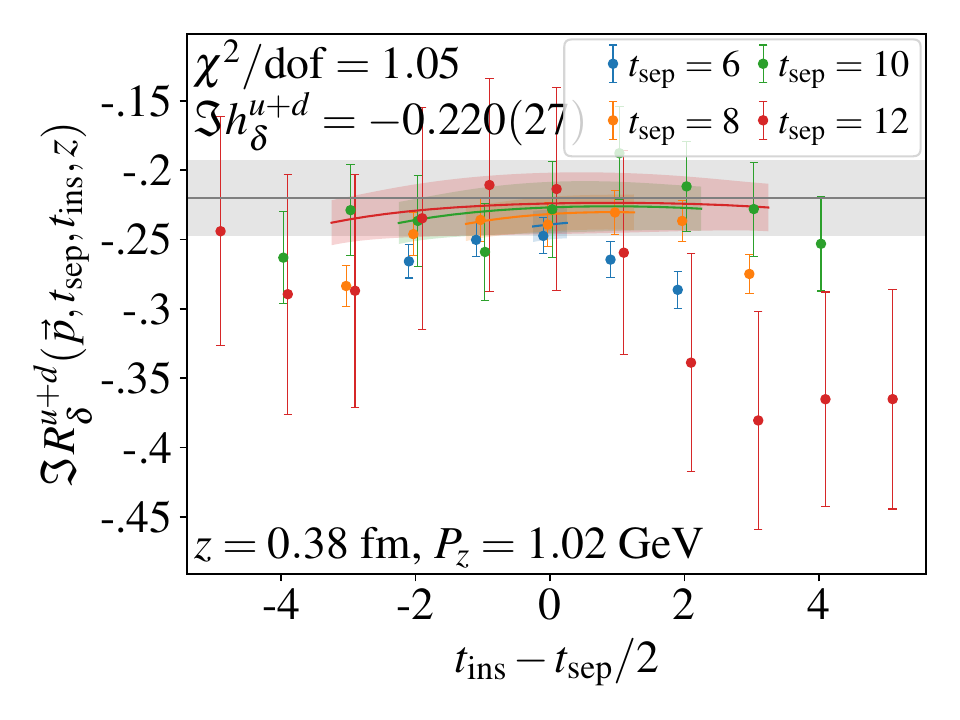}
    \includegraphics[width=0.32\textwidth]{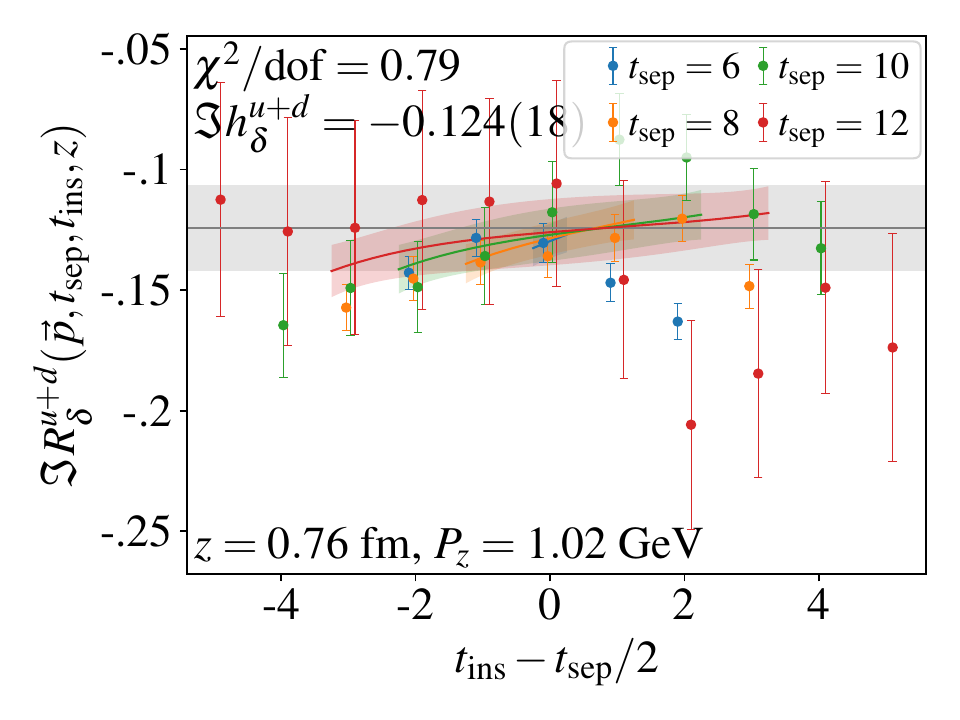}
    \includegraphics[width=0.32\textwidth]{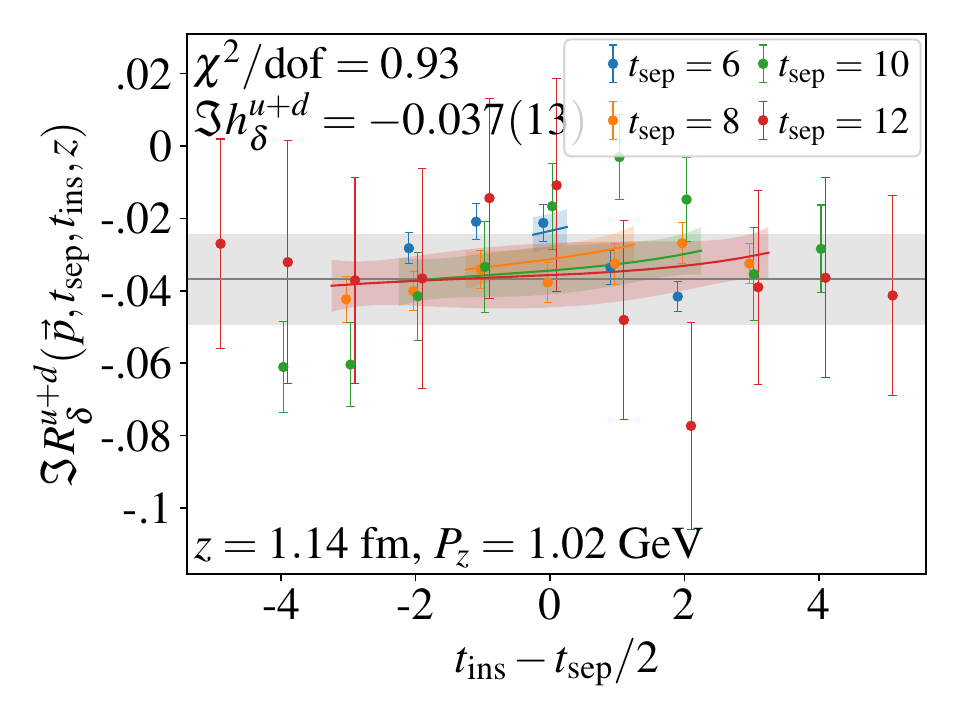}
    \includegraphics[width=0.32\textwidth]{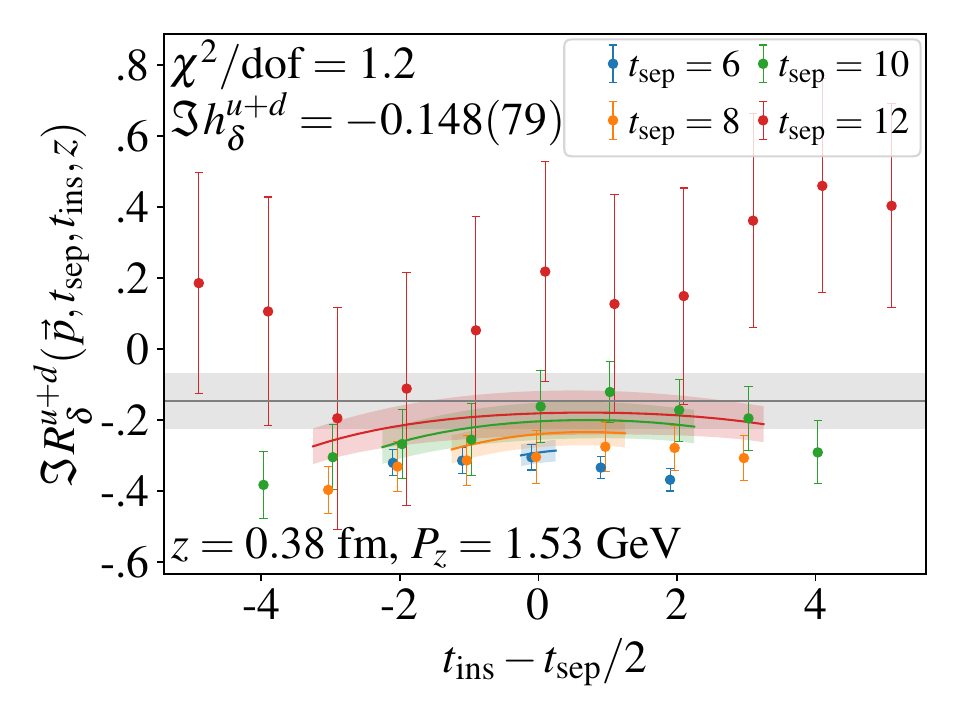}
    \includegraphics[width=0.32\textwidth]{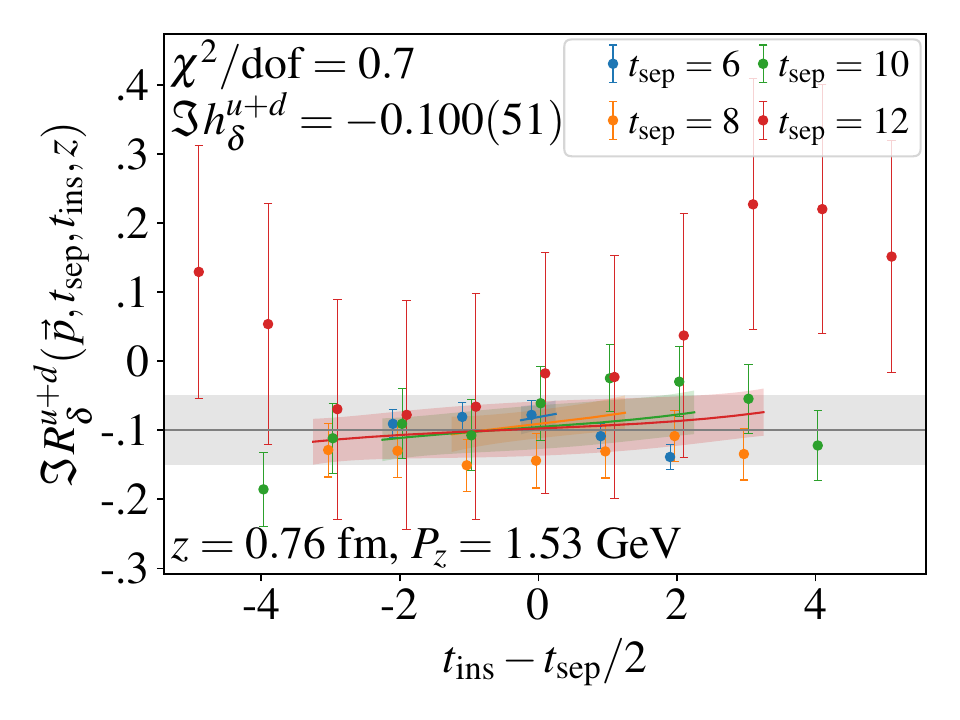}
    \includegraphics[width=0.32\textwidth]{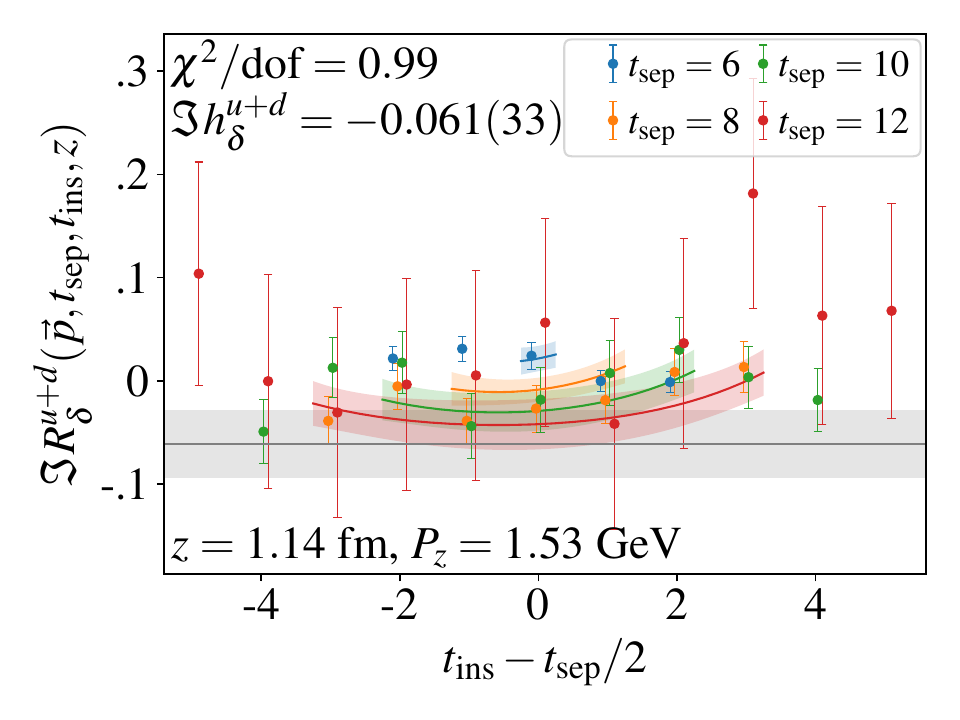}
    \caption{The same as \Cref{fig:c3pt_u+d_repr_fits_re}, but for the imaginary parts.
             Additionally, the zero-momentum matrix elements are not shown, as these are
             all consistent with zero (as they are expected to be).}
    \label{fig:c3pt_u+d_repr_fits_im}
\end{figure*}

The fits included here are to the local zero-momentum three to two-point function ratios, shown in \Cref{fig:c3pt_loc_u-d_and_u+d_repr_fits_re}, and several non-local three to two-point function ratios, shown in \Cref{fig:c3pt_u-d_repr_fits_re,fig:c3pt_u-d_repr_fits_im,fig:c3pt_u+d_repr_fits_re,fig:c3pt_u+d_repr_fits_im}.
These include both isovector and isoscalar combinations.

\bibliography{refs}

\begin{thebibliography}{114}
\expandafter\ifx\csname natexlab\endcsname\relax\def\natexlab#1{#1}\fi
\expandafter\ifx\csname bibnamefont\endcsname\relax
  \def\bibnamefont#1{#1}\fi
\expandafter\ifx\csname bibfnamefont\endcsname\relax
  \def\bibfnamefont#1{#1}\fi
\expandafter\ifx\csname citenamefont\endcsname\relax
  \def\citenamefont#1{#1}\fi
\expandafter\ifx\csname url\endcsname\relax
  \def\url#1{\texttt{#1}}\fi
\expandafter\ifx\csname urlprefix\endcsname\relax\def\urlprefix{URL }\fi
\providecommand{\bibinfo}[2]{#2}
\providecommand{\eprint}[2][]{\url{#2}}

\bibitem[{\citenamefont{Alekhin et~al.}(2017)\citenamefont{Alekhin, Bl\"umlein,
  Moch, and Placakyte}}]{Alekhin:2017kpj}
\bibinfo{author}{\bibfnamefont{S.}~\bibnamefont{Alekhin}},
  \bibinfo{author}{\bibfnamefont{J.}~\bibnamefont{Bl\"umlein}},
  \bibinfo{author}{\bibfnamefont{S.}~\bibnamefont{Moch}}, \bibnamefont{and}
  \bibinfo{author}{\bibfnamefont{R.}~\bibnamefont{Placakyte}},
  \bibinfo{journal}{Phys. Rev. D} \textbf{\bibinfo{volume}{96}},
  \bibinfo{pages}{014011} (\bibinfo{year}{2017}), \eprint{1701.05838}.

\bibitem[{\citenamefont{Hou et~al.}(2021)}]{Hou:2019efy}
\bibinfo{author}{\bibfnamefont{T.-J.} \bibnamefont{Hou}} \bibnamefont{et~al.},
  \bibinfo{journal}{Phys. Rev. D} \textbf{\bibinfo{volume}{103}},
  \bibinfo{pages}{014013} (\bibinfo{year}{2021}), \eprint{1912.10053}.

\bibitem[{\citenamefont{Bailey et~al.}(2021)\citenamefont{Bailey, Cridge,
  Harland-Lang, Martin, and Thorne}}]{Bailey:2020ooq}
\bibinfo{author}{\bibfnamefont{S.}~\bibnamefont{Bailey}},
  \bibinfo{author}{\bibfnamefont{T.}~\bibnamefont{Cridge}},
  \bibinfo{author}{\bibfnamefont{L.~A.} \bibnamefont{Harland-Lang}},
  \bibinfo{author}{\bibfnamefont{A.~D.} \bibnamefont{Martin}},
  \bibnamefont{and} \bibinfo{author}{\bibfnamefont{R.~S.}
  \bibnamefont{Thorne}}, \bibinfo{journal}{Eur. Phys. J. C}
  \textbf{\bibinfo{volume}{81}}, \bibinfo{pages}{341} (\bibinfo{year}{2021}),
  \eprint{2012.04684}.

\bibitem[{\citenamefont{Ball et~al.}(2022)}]{NNPDF:2021njg}
\bibinfo{author}{\bibfnamefont{R.~D.} \bibnamefont{Ball}} \bibnamefont{et~al.}
  (\bibinfo{collaboration}{NNPDF}), \bibinfo{journal}{Eur. Phys. J. C}
  \textbf{\bibinfo{volume}{82}}, \bibinfo{pages}{428} (\bibinfo{year}{2022}),
  \eprint{2109.02653}.

\bibitem[{\citenamefont{Adolph et~al.}(2015)}]{COMPASS:2014bze}
\bibinfo{author}{\bibfnamefont{C.}~\bibnamefont{Adolph}} \bibnamefont{et~al.}
  (\bibinfo{collaboration}{COMPASS}), \bibinfo{journal}{Phys. Lett. B}
  \textbf{\bibinfo{volume}{744}}, \bibinfo{pages}{250} (\bibinfo{year}{2015}),
  \eprint{1408.4405}.

\bibitem[{\citenamefont{Airapetian et~al.}(2020)}]{HERMES:2020ifk}
\bibinfo{author}{\bibfnamefont{A.}~\bibnamefont{Airapetian}}
  \bibnamefont{et~al.} (\bibinfo{collaboration}{HERMES}),
  \bibinfo{journal}{JHEP} \textbf{\bibinfo{volume}{12}}, \bibinfo{pages}{010}
  (\bibinfo{year}{2020}), \eprint{2007.07755}.

\bibitem[{\citenamefont{Adolph et~al.}(2012)}]{COMPASS:2012bfl}
\bibinfo{author}{\bibfnamefont{C.}~\bibnamefont{Adolph}} \bibnamefont{et~al.}
  (\bibinfo{collaboration}{COMPASS}), \bibinfo{journal}{Phys. Lett. B}
  \textbf{\bibinfo{volume}{713}}, \bibinfo{pages}{10} (\bibinfo{year}{2012}),
  \eprint{1202.6150}.

\bibitem[{\citenamefont{Adolph et~al.}(2014)}]{COMPASS:2014ysd}
\bibinfo{author}{\bibfnamefont{C.}~\bibnamefont{Adolph}} \bibnamefont{et~al.}
  (\bibinfo{collaboration}{COMPASS}), \bibinfo{journal}{Phys. Lett. B}
  \textbf{\bibinfo{volume}{736}}, \bibinfo{pages}{124} (\bibinfo{year}{2014}),
  \eprint{1401.7873}.

\bibitem[{\citenamefont{Airapetian et~al.}(2008)}]{HERMES:2008mcr}
\bibinfo{author}{\bibfnamefont{A.}~\bibnamefont{Airapetian}}
  \bibnamefont{et~al.} (\bibinfo{collaboration}{HERMES}),
  \bibinfo{journal}{JHEP} \textbf{\bibinfo{volume}{06}}, \bibinfo{pages}{017}
  (\bibinfo{year}{2008}), \eprint{0803.2367}.

\bibitem[{\citenamefont{Yang}(2009)}]{Yang:2009zzr}
\bibinfo{author}{\bibfnamefont{R.}~\bibnamefont{Yang}}
  (\bibinfo{collaboration}{PHENIX}), \bibinfo{journal}{AIP Conf. Proc.}
  \textbf{\bibinfo{volume}{1182}}, \bibinfo{pages}{569} (\bibinfo{year}{2009}).

\bibitem[{\citenamefont{Fatemi}(2012)}]{Fatemi:2012ry}
\bibinfo{author}{\bibfnamefont{R.}~\bibnamefont{Fatemi}}
  (\bibinfo{collaboration}{STAR}), \bibinfo{journal}{AIP Conf. Proc.}
  \textbf{\bibinfo{volume}{1441}}, \bibinfo{pages}{233} (\bibinfo{year}{2012}),
  \eprint{1206.3861}.

\bibitem[{\citenamefont{Adamczyk et~al.}(2015)}]{STAR:2015jkc}
\bibinfo{author}{\bibfnamefont{L.}~\bibnamefont{Adamczyk}} \bibnamefont{et~al.}
  (\bibinfo{collaboration}{STAR}), \bibinfo{journal}{Phys. Rev. Lett.}
  \textbf{\bibinfo{volume}{115}}, \bibinfo{pages}{242501}
  (\bibinfo{year}{2015}), \eprint{1504.00415}.

\bibitem[{\citenamefont{Bacchetta et~al.}(2011)\citenamefont{Bacchetta,
  Courtoy, and Radici}}]{Bacchetta:2011ip}
\bibinfo{author}{\bibfnamefont{A.}~\bibnamefont{Bacchetta}},
  \bibinfo{author}{\bibfnamefont{A.}~\bibnamefont{Courtoy}}, \bibnamefont{and}
  \bibinfo{author}{\bibfnamefont{M.}~\bibnamefont{Radici}},
  \bibinfo{journal}{Phys. Rev. Lett.} \textbf{\bibinfo{volume}{107}},
  \bibinfo{pages}{012001} (\bibinfo{year}{2011}), \eprint{1104.3855}.

\bibitem[{\citenamefont{Bacchetta et~al.}(2013)\citenamefont{Bacchetta,
  Courtoy, and Radici}}]{Bacchetta:2012ty}
\bibinfo{author}{\bibfnamefont{A.}~\bibnamefont{Bacchetta}},
  \bibinfo{author}{\bibfnamefont{A.}~\bibnamefont{Courtoy}}, \bibnamefont{and}
  \bibinfo{author}{\bibfnamefont{M.}~\bibnamefont{Radici}},
  \bibinfo{journal}{JHEP} \textbf{\bibinfo{volume}{03}}, \bibinfo{pages}{119}
  (\bibinfo{year}{2013}), \eprint{1212.3568}.

\bibitem[{\citenamefont{Anselmino et~al.}(2013)\citenamefont{Anselmino,
  Boglione, D'Alesio, Melis, Murgia, and Prokudin}}]{Anselmino:2013vqa}
\bibinfo{author}{\bibfnamefont{M.}~\bibnamefont{Anselmino}},
  \bibinfo{author}{\bibfnamefont{M.}~\bibnamefont{Boglione}},
  \bibinfo{author}{\bibfnamefont{U.}~\bibnamefont{D'Alesio}},
  \bibinfo{author}{\bibfnamefont{S.}~\bibnamefont{Melis}},
  \bibinfo{author}{\bibfnamefont{F.}~\bibnamefont{Murgia}}, \bibnamefont{and}
  \bibinfo{author}{\bibfnamefont{A.}~\bibnamefont{Prokudin}},
  \bibinfo{journal}{Phys. Rev. D} \textbf{\bibinfo{volume}{87}},
  \bibinfo{pages}{094019} (\bibinfo{year}{2013}), \eprint{1303.3822}.

\bibitem[{\citenamefont{Kang et~al.}(2015)\citenamefont{Kang, Prokudin, Sun,
  and Yuan}}]{Kang:2014zza}
\bibinfo{author}{\bibfnamefont{Z.-B.} \bibnamefont{Kang}},
  \bibinfo{author}{\bibfnamefont{A.}~\bibnamefont{Prokudin}},
  \bibinfo{author}{\bibfnamefont{P.}~\bibnamefont{Sun}}, \bibnamefont{and}
  \bibinfo{author}{\bibfnamefont{F.}~\bibnamefont{Yuan}},
  \bibinfo{journal}{Phys. Rev. D} \textbf{\bibinfo{volume}{91}},
  \bibinfo{pages}{071501} (\bibinfo{year}{2015}), \eprint{1410.4877}.

\bibitem[{\citenamefont{Martin et~al.}(2015)\citenamefont{Martin, Bradamante,
  and Barone}}]{Martin:2014wua}
\bibinfo{author}{\bibfnamefont{A.}~\bibnamefont{Martin}},
  \bibinfo{author}{\bibfnamefont{F.}~\bibnamefont{Bradamante}},
  \bibnamefont{and} \bibinfo{author}{\bibfnamefont{V.}~\bibnamefont{Barone}},
  \bibinfo{journal}{Phys. Rev. D} \textbf{\bibinfo{volume}{91}},
  \bibinfo{pages}{014034} (\bibinfo{year}{2015}), \eprint{1412.5946}.

\bibitem[{\citenamefont{Radici et~al.}(2015)\citenamefont{Radici, Courtoy,
  Bacchetta, and Guagnelli}}]{Radici:2015mwa}
\bibinfo{author}{\bibfnamefont{M.}~\bibnamefont{Radici}},
  \bibinfo{author}{\bibfnamefont{A.}~\bibnamefont{Courtoy}},
  \bibinfo{author}{\bibfnamefont{A.}~\bibnamefont{Bacchetta}},
  \bibnamefont{and}
  \bibinfo{author}{\bibfnamefont{M.}~\bibnamefont{Guagnelli}},
  \bibinfo{journal}{JHEP} \textbf{\bibinfo{volume}{05}}, \bibinfo{pages}{123}
  (\bibinfo{year}{2015}), \eprint{1503.03495}.

\bibitem[{\citenamefont{Anselmino et~al.}(2015)\citenamefont{Anselmino,
  Boglione, D'Alesio, Gonzalez~Hernandez, Melis, Murgia, and
  Prokudin}}]{Anselmino:2015sxa}
\bibinfo{author}{\bibfnamefont{M.}~\bibnamefont{Anselmino}},
  \bibinfo{author}{\bibfnamefont{M.}~\bibnamefont{Boglione}},
  \bibinfo{author}{\bibfnamefont{U.}~\bibnamefont{D'Alesio}},
  \bibinfo{author}{\bibfnamefont{J.~O.} \bibnamefont{Gonzalez~Hernandez}},
  \bibinfo{author}{\bibfnamefont{S.}~\bibnamefont{Melis}},
  \bibinfo{author}{\bibfnamefont{F.}~\bibnamefont{Murgia}}, \bibnamefont{and}
  \bibinfo{author}{\bibfnamefont{A.}~\bibnamefont{Prokudin}},
  \bibinfo{journal}{Phys. Rev. D} \textbf{\bibinfo{volume}{92}},
  \bibinfo{pages}{114023} (\bibinfo{year}{2015}), \eprint{1510.05389}.

\bibitem[{\citenamefont{Kang et~al.}(2016)\citenamefont{Kang, Prokudin, Sun,
  and Yuan}}]{Kang:2015msa}
\bibinfo{author}{\bibfnamefont{Z.-B.} \bibnamefont{Kang}},
  \bibinfo{author}{\bibfnamefont{A.}~\bibnamefont{Prokudin}},
  \bibinfo{author}{\bibfnamefont{P.}~\bibnamefont{Sun}}, \bibnamefont{and}
  \bibinfo{author}{\bibfnamefont{F.}~\bibnamefont{Yuan}},
  \bibinfo{journal}{Phys. Rev. D} \textbf{\bibinfo{volume}{93}},
  \bibinfo{pages}{014009} (\bibinfo{year}{2016}), \eprint{1505.05589}.

\bibitem[{\citenamefont{Radici et~al.}(2016)\citenamefont{Radici, Ricci,
  Bacchetta, and Mukherjee}}]{Radici:2016lam}
\bibinfo{author}{\bibfnamefont{M.}~\bibnamefont{Radici}},
  \bibinfo{author}{\bibfnamefont{A.~M.} \bibnamefont{Ricci}},
  \bibinfo{author}{\bibfnamefont{A.}~\bibnamefont{Bacchetta}},
  \bibnamefont{and}
  \bibinfo{author}{\bibfnamefont{A.}~\bibnamefont{Mukherjee}},
  \bibinfo{journal}{Phys. Rev. D} \textbf{\bibinfo{volume}{94}},
  \bibinfo{pages}{034012} (\bibinfo{year}{2016}), \eprint{1604.06585}.

\bibitem[{\citenamefont{Lin et~al.}(2018)\citenamefont{Lin, Melnitchouk,
  Prokudin, Sato, and Shows}}]{Lin:2017stx}
\bibinfo{author}{\bibfnamefont{H.-W.} \bibnamefont{Lin}},
  \bibinfo{author}{\bibfnamefont{W.}~\bibnamefont{Melnitchouk}},
  \bibinfo{author}{\bibfnamefont{A.}~\bibnamefont{Prokudin}},
  \bibinfo{author}{\bibfnamefont{N.}~\bibnamefont{Sato}}, \bibnamefont{and}
  \bibinfo{author}{\bibfnamefont{H.}~\bibnamefont{Shows}},
  \bibinfo{journal}{Phys. Rev. Lett.} \textbf{\bibinfo{volume}{120}},
  \bibinfo{pages}{152502} (\bibinfo{year}{2018}), \eprint{1710.09858}.

\bibitem[{\citenamefont{Radici and Bacchetta}(2018)}]{Radici:2018iag}
\bibinfo{author}{\bibfnamefont{M.}~\bibnamefont{Radici}} \bibnamefont{and}
  \bibinfo{author}{\bibfnamefont{A.}~\bibnamefont{Bacchetta}},
  \bibinfo{journal}{Phys. Rev. Lett.} \textbf{\bibinfo{volume}{120}},
  \bibinfo{pages}{192001} (\bibinfo{year}{2018}), \eprint{1802.05212}.

\bibitem[{\citenamefont{Benel et~al.}(2020)\citenamefont{Benel, Courtoy, and
  Ferro-Hernandez}}]{Benel:2019mcq}
\bibinfo{author}{\bibfnamefont{J.}~\bibnamefont{Benel}},
  \bibinfo{author}{\bibfnamefont{A.}~\bibnamefont{Courtoy}}, \bibnamefont{and}
  \bibinfo{author}{\bibfnamefont{R.}~\bibnamefont{Ferro-Hernandez}},
  \bibinfo{journal}{Eur. Phys. J. C} \textbf{\bibinfo{volume}{80}},
  \bibinfo{pages}{465} (\bibinfo{year}{2020}), \eprint{1912.03289}.

\bibitem[{\citenamefont{Anselmino et~al.}(2020)\citenamefont{Anselmino,
  Kishore, and Mukherjee}}]{Anselmino:2020nrk}
\bibinfo{author}{\bibfnamefont{M.}~\bibnamefont{Anselmino}},
  \bibinfo{author}{\bibfnamefont{R.}~\bibnamefont{Kishore}}, \bibnamefont{and}
  \bibinfo{author}{\bibfnamefont{A.}~\bibnamefont{Mukherjee}},
  \bibinfo{journal}{Phys. Rev. D} \textbf{\bibinfo{volume}{102}},
  \bibinfo{pages}{096012} (\bibinfo{year}{2020}), \eprint{2009.03148}.

\bibitem[{\citenamefont{D'Alesio et~al.}(2020)\citenamefont{D'Alesio, Flore,
  and Prokudin}}]{DAlesio:2020vtw}
\bibinfo{author}{\bibfnamefont{U.}~\bibnamefont{D'Alesio}},
  \bibinfo{author}{\bibfnamefont{C.}~\bibnamefont{Flore}}, \bibnamefont{and}
  \bibinfo{author}{\bibfnamefont{A.}~\bibnamefont{Prokudin}},
  \bibinfo{journal}{Phys. Lett. B} \textbf{\bibinfo{volume}{803}},
  \bibinfo{pages}{135347} (\bibinfo{year}{2020}), \eprint{2001.01573}.

\bibitem[{\citenamefont{Cammarota et~al.}(2020)\citenamefont{Cammarota,
  Gamberg, Kang, Miller, Pitonyak, Prokudin, Rogers, and
  Sato}}]{Cammarota:2020qcw}
\bibinfo{author}{\bibfnamefont{J.}~\bibnamefont{Cammarota}},
  \bibinfo{author}{\bibfnamefont{L.}~\bibnamefont{Gamberg}},
  \bibinfo{author}{\bibfnamefont{Z.-B.} \bibnamefont{Kang}},
  \bibinfo{author}{\bibfnamefont{J.~A.} \bibnamefont{Miller}},
  \bibinfo{author}{\bibfnamefont{D.}~\bibnamefont{Pitonyak}},
  \bibinfo{author}{\bibfnamefont{A.}~\bibnamefont{Prokudin}},
  \bibinfo{author}{\bibfnamefont{T.~C.} \bibnamefont{Rogers}},
  \bibnamefont{and} \bibinfo{author}{\bibfnamefont{N.}~\bibnamefont{Sato}}
  (\bibinfo{collaboration}{Jefferson Lab Angular Momentum}),
  \bibinfo{journal}{Phys. Rev. D} \textbf{\bibinfo{volume}{102}},
  \bibinfo{pages}{054002} (\bibinfo{year}{2020}), \eprint{2002.08384}.

\bibitem[{\citenamefont{Gamberg et~al.}(2022)\citenamefont{Gamberg, Malda,
  Miller, Pitonyak, Prokudin, and Sato}}]{Gamberg:2022kdb}
\bibinfo{author}{\bibfnamefont{L.}~\bibnamefont{Gamberg}},
  \bibinfo{author}{\bibfnamefont{M.}~\bibnamefont{Malda}},
  \bibinfo{author}{\bibfnamefont{J.~A.} \bibnamefont{Miller}},
  \bibinfo{author}{\bibfnamefont{D.}~\bibnamefont{Pitonyak}},
  \bibinfo{author}{\bibfnamefont{A.}~\bibnamefont{Prokudin}}, \bibnamefont{and}
  \bibinfo{author}{\bibfnamefont{N.}~\bibnamefont{Sato}}
  (\bibinfo{collaboration}{Jefferson Lab Angular Momentum (JAM), Jefferson Lab
  Angular Momentum}), \bibinfo{journal}{Phys. Rev. D}
  \textbf{\bibinfo{volume}{106}}, \bibinfo{pages}{034014}
  (\bibinfo{year}{2022}), \eprint{2205.00999}.

\bibitem[{\citenamefont{Cocuzza
  et~al.}(2023{\natexlab{a}})\citenamefont{Cocuzza, Metz, Pitonyak, Prokudin,
  Sato, and Seidl}}]{Cocuzza:2023oam}
\bibinfo{author}{\bibfnamefont{C.}~\bibnamefont{Cocuzza}},
  \bibinfo{author}{\bibfnamefont{A.}~\bibnamefont{Metz}},
  \bibinfo{author}{\bibfnamefont{D.}~\bibnamefont{Pitonyak}},
  \bibinfo{author}{\bibfnamefont{A.}~\bibnamefont{Prokudin}},
  \bibinfo{author}{\bibfnamefont{N.}~\bibnamefont{Sato}}, \bibnamefont{and}
  \bibinfo{author}{\bibfnamefont{R.}~\bibnamefont{Seidl}}
  (\bibinfo{collaboration}{JAM}) (\bibinfo{year}{2023}{\natexlab{a}}),
  \eprint{2306.12998}.

\bibitem[{\citenamefont{Cocuzza
  et~al.}(2023{\natexlab{b}})\citenamefont{Cocuzza, Metz, Pitonyak, Prokudin,
  Sato, and Seidl}}]{Cocuzza:2023vqs}
\bibinfo{author}{\bibfnamefont{C.}~\bibnamefont{Cocuzza}},
  \bibinfo{author}{\bibfnamefont{A.}~\bibnamefont{Metz}},
  \bibinfo{author}{\bibfnamefont{D.}~\bibnamefont{Pitonyak}},
  \bibinfo{author}{\bibfnamefont{A.}~\bibnamefont{Prokudin}},
  \bibinfo{author}{\bibfnamefont{N.}~\bibnamefont{Sato}}, \bibnamefont{and}
  \bibinfo{author}{\bibfnamefont{R.}~\bibnamefont{Seidl}}
  (\bibinfo{year}{2023}{\natexlab{b}}), \eprint{2308.14857}.

\bibitem[{\citenamefont{Dudek et~al.}(2012)}]{Dudek:2012vr}
\bibinfo{author}{\bibfnamefont{J.}~\bibnamefont{Dudek}} \bibnamefont{et~al.},
  \bibinfo{journal}{Eur. Phys. J. A} \textbf{\bibinfo{volume}{48}},
  \bibinfo{pages}{187} (\bibinfo{year}{2012}), \eprint{1208.1244}.

\bibitem[{\citenamefont{Burkert et~al.}(2023)}]{Burkert:2022hjz}
\bibinfo{author}{\bibfnamefont{V.~D.} \bibnamefont{Burkert}}
  \bibnamefont{et~al.}, \bibinfo{journal}{Prog. Part. Nucl. Phys.}
  \textbf{\bibinfo{volume}{131}}, \bibinfo{pages}{104032}
  (\bibinfo{year}{2023}), \eprint{2211.15746}.

\bibitem[{\citenamefont{Cichy and Constantinou}(2019)}]{Cichy:2018mum}
\bibinfo{author}{\bibfnamefont{K.}~\bibnamefont{Cichy}} \bibnamefont{and}
  \bibinfo{author}{\bibfnamefont{M.}~\bibnamefont{Constantinou}},
  \bibinfo{journal}{Adv. High Energy Phys.} \textbf{\bibinfo{volume}{2019}},
  \bibinfo{pages}{3036904} (\bibinfo{year}{2019}), \eprint{1811.07248}.

\bibitem[{\citenamefont{Zhao}(2019)}]{Zhao:2018fyu}
\bibinfo{author}{\bibfnamefont{Y.}~\bibnamefont{Zhao}}, \bibinfo{journal}{Int.
  J. Mod. Phys. A} \textbf{\bibinfo{volume}{33}}, \bibinfo{pages}{1830033}
  (\bibinfo{year}{2019}), \eprint{1812.07192}.

\bibitem[{\citenamefont{Radyushkin}(2020)}]{Radyushkin:2019mye}
\bibinfo{author}{\bibfnamefont{A.~V.} \bibnamefont{Radyushkin}},
  \bibinfo{journal}{Int. J. Mod. Phys. A} \textbf{\bibinfo{volume}{35}},
  \bibinfo{pages}{2030002} (\bibinfo{year}{2020}), \eprint{1912.04244}.

\bibitem[{\citenamefont{Ji et~al.}(2021{\natexlab{a}})\citenamefont{Ji, Liu,
  Liu, Zhang, and Zhao}}]{Ji:2020ect}
\bibinfo{author}{\bibfnamefont{X.}~\bibnamefont{Ji}},
  \bibinfo{author}{\bibfnamefont{Y.-S.} \bibnamefont{Liu}},
  \bibinfo{author}{\bibfnamefont{Y.}~\bibnamefont{Liu}},
  \bibinfo{author}{\bibfnamefont{J.-H.} \bibnamefont{Zhang}}, \bibnamefont{and}
  \bibinfo{author}{\bibfnamefont{Y.}~\bibnamefont{Zhao}},
  \bibinfo{journal}{Rev. Mod. Phys.} \textbf{\bibinfo{volume}{93}},
  \bibinfo{pages}{035005} (\bibinfo{year}{2021}{\natexlab{a}}),
  \eprint{2004.03543}.

\bibitem[{\citenamefont{Constantinou}(2021)}]{Constantinou:2020pek}
\bibinfo{author}{\bibfnamefont{M.}~\bibnamefont{Constantinou}},
  \bibinfo{journal}{Eur. Phys. J. A} \textbf{\bibinfo{volume}{57}},
  \bibinfo{pages}{77} (\bibinfo{year}{2021}), \eprint{2010.02445}.

\bibitem[{\citenamefont{Constantinou et~al.}(2021)}]{Constantinou:2020hdm}
\bibinfo{author}{\bibfnamefont{M.}~\bibnamefont{Constantinou}}
  \bibnamefont{et~al.}, \bibinfo{journal}{Prog. Part. Nucl. Phys.}
  \textbf{\bibinfo{volume}{121}}, \bibinfo{pages}{103908}
  (\bibinfo{year}{2021}), \eprint{2006.08636}.

\bibitem[{\citenamefont{Cichy}(2022)}]{Cichy:2021lih}
\bibinfo{author}{\bibfnamefont{K.}~\bibnamefont{Cichy}}, \bibinfo{journal}{PoS}
  \textbf{\bibinfo{volume}{LATTICE2021}}, \bibinfo{pages}{017}
  (\bibinfo{year}{2022}), \eprint{2110.07440}.

\bibitem[{\citenamefont{Ji}(2013)}]{Ji:2013dva}
\bibinfo{author}{\bibfnamefont{X.}~\bibnamefont{Ji}}, \bibinfo{journal}{Phys.
  Rev. Lett.} \textbf{\bibinfo{volume}{110}}, \bibinfo{pages}{262002}
  (\bibinfo{year}{2013}), \eprint{1305.1539}.

\bibitem[{\citenamefont{Ji}(2014)}]{Ji:2014gla}
\bibinfo{author}{\bibfnamefont{X.}~\bibnamefont{Ji}}, \bibinfo{journal}{Sci.
  China Phys. Mech. Astron.} \textbf{\bibinfo{volume}{57}},
  \bibinfo{pages}{1407} (\bibinfo{year}{2014}), \eprint{1404.6680}.

\bibitem[{\citenamefont{Radyushkin}(2017)}]{Radyushkin:2017cyf}
\bibinfo{author}{\bibfnamefont{A.~V.} \bibnamefont{Radyushkin}},
  \bibinfo{journal}{Phys. Rev. D} \textbf{\bibinfo{volume}{96}},
  \bibinfo{pages}{034025} (\bibinfo{year}{2017}), \eprint{1705.01488}.

\bibitem[{\citenamefont{Orginos et~al.}(2017)\citenamefont{Orginos, Radyushkin,
  Karpie, and Zafeiropoulos}}]{Orginos:2017kos}
\bibinfo{author}{\bibfnamefont{K.}~\bibnamefont{Orginos}},
  \bibinfo{author}{\bibfnamefont{A.}~\bibnamefont{Radyushkin}},
  \bibinfo{author}{\bibfnamefont{J.}~\bibnamefont{Karpie}}, \bibnamefont{and}
  \bibinfo{author}{\bibfnamefont{S.}~\bibnamefont{Zafeiropoulos}},
  \bibinfo{journal}{Phys. Rev. D} \textbf{\bibinfo{volume}{96}},
  \bibinfo{pages}{094503} (\bibinfo{year}{2017}), \eprint{1706.05373}.

\bibitem[{\citenamefont{Ji et~al.}(2017)\citenamefont{Ji, Zhang, and
  Zhao}}]{Ji:2017rah}
\bibinfo{author}{\bibfnamefont{X.}~\bibnamefont{Ji}},
  \bibinfo{author}{\bibfnamefont{J.-H.} \bibnamefont{Zhang}}, \bibnamefont{and}
  \bibinfo{author}{\bibfnamefont{Y.}~\bibnamefont{Zhao}},
  \bibinfo{journal}{Nucl. Phys. B} \textbf{\bibinfo{volume}{924}},
  \bibinfo{pages}{366} (\bibinfo{year}{2017}), \eprint{1706.07416}.

\bibitem[{\citenamefont{Radyushkin}(2018)}]{Radyushkin:2017lvu}
\bibinfo{author}{\bibfnamefont{A.~V.} \bibnamefont{Radyushkin}},
  \bibinfo{journal}{Phys. Lett. B} \textbf{\bibinfo{volume}{781}},
  \bibinfo{pages}{433} (\bibinfo{year}{2018}), \eprint{1710.08813}.

\bibitem[{\citenamefont{Izubuchi et~al.}(2018)\citenamefont{Izubuchi, Ji, Jin,
  Stewart, and Zhao}}]{Izubuchi:2018srq}
\bibinfo{author}{\bibfnamefont{T.}~\bibnamefont{Izubuchi}},
  \bibinfo{author}{\bibfnamefont{X.}~\bibnamefont{Ji}},
  \bibinfo{author}{\bibfnamefont{L.}~\bibnamefont{Jin}},
  \bibinfo{author}{\bibfnamefont{I.~W.} \bibnamefont{Stewart}},
  \bibnamefont{and} \bibinfo{author}{\bibfnamefont{Y.}~\bibnamefont{Zhao}},
  \bibinfo{journal}{Phys. Rev. D} \textbf{\bibinfo{volume}{98}},
  \bibinfo{pages}{056004} (\bibinfo{year}{2018}), \eprint{1801.03917}.

\bibitem[{\citenamefont{Ji}(2022)}]{Ji:2022ezo}
\bibinfo{author}{\bibfnamefont{X.}~\bibnamefont{Ji}} (\bibinfo{year}{2022}),
  \eprint{2209.09332}.

\bibitem[{\citenamefont{Holligan et~al.}(2023)\citenamefont{Holligan, Ji, Lin,
  Su, and Zhang}}]{Holligan:2023rex}
\bibinfo{author}{\bibfnamefont{J.}~\bibnamefont{Holligan}},
  \bibinfo{author}{\bibfnamefont{X.}~\bibnamefont{Ji}},
  \bibinfo{author}{\bibfnamefont{H.-W.} \bibnamefont{Lin}},
  \bibinfo{author}{\bibfnamefont{Y.}~\bibnamefont{Su}}, \bibnamefont{and}
  \bibinfo{author}{\bibfnamefont{R.}~\bibnamefont{Zhang}},
  \bibinfo{journal}{Nucl. Phys. B} \textbf{\bibinfo{volume}{993}},
  \bibinfo{pages}{116282} (\bibinfo{year}{2023}), \eprint{2301.10372}.

\bibitem[{\citenamefont{Chen et~al.}(2016)\citenamefont{Chen, Cohen, Ji, Lin,
  and Zhang}}]{Chen:2016utp}
\bibinfo{author}{\bibfnamefont{J.-W.} \bibnamefont{Chen}},
  \bibinfo{author}{\bibfnamefont{S.~D.} \bibnamefont{Cohen}},
  \bibinfo{author}{\bibfnamefont{X.}~\bibnamefont{Ji}},
  \bibinfo{author}{\bibfnamefont{H.-W.} \bibnamefont{Lin}}, \bibnamefont{and}
  \bibinfo{author}{\bibfnamefont{J.-H.} \bibnamefont{Zhang}},
  \bibinfo{journal}{Nucl. Phys. B} \textbf{\bibinfo{volume}{911}},
  \bibinfo{pages}{246} (\bibinfo{year}{2016}), \eprint{1603.06664}.

\bibitem[{\citenamefont{Alexandrou et~al.}(2018)\citenamefont{Alexandrou,
  Cichy, Constantinou, Jansen, Scapellato, and Steffens}}]{Alexandrou:2018eet}
\bibinfo{author}{\bibfnamefont{C.}~\bibnamefont{Alexandrou}},
  \bibinfo{author}{\bibfnamefont{K.}~\bibnamefont{Cichy}},
  \bibinfo{author}{\bibfnamefont{M.}~\bibnamefont{Constantinou}},
  \bibinfo{author}{\bibfnamefont{K.}~\bibnamefont{Jansen}},
  \bibinfo{author}{\bibfnamefont{A.}~\bibnamefont{Scapellato}},
  \bibnamefont{and} \bibinfo{author}{\bibfnamefont{F.}~\bibnamefont{Steffens}},
  \bibinfo{journal}{Phys. Rev. D} \textbf{\bibinfo{volume}{98}},
  \bibinfo{pages}{091503} (\bibinfo{year}{2018}), \eprint{1807.00232}.

\bibitem[{\citenamefont{Liu et~al.}(2018)\citenamefont{Liu, Chen, Jin, Li, Lin,
  Yang, Zhang, and Zhao}}]{Liu:2018hxv}
\bibinfo{author}{\bibfnamefont{Y.-S.} \bibnamefont{Liu}},
  \bibinfo{author}{\bibfnamefont{J.-W.} \bibnamefont{Chen}},
  \bibinfo{author}{\bibfnamefont{L.}~\bibnamefont{Jin}},
  \bibinfo{author}{\bibfnamefont{R.}~\bibnamefont{Li}},
  \bibinfo{author}{\bibfnamefont{H.-W.} \bibnamefont{Lin}},
  \bibinfo{author}{\bibfnamefont{Y.-B.} \bibnamefont{Yang}},
  \bibinfo{author}{\bibfnamefont{J.-H.} \bibnamefont{Zhang}}, \bibnamefont{and}
  \bibinfo{author}{\bibfnamefont{Y.}~\bibnamefont{Zhao}}
  (\bibinfo{year}{2018}), \eprint{1810.05043}.

\bibitem[{\citenamefont{Alexandrou et~al.}(2019)\citenamefont{Alexandrou,
  Cichy, Constantinou, Hadjiyiannakou, Jansen, Scapellato, and
  Steffens}}]{Alexandrou:2019lfo}
\bibinfo{author}{\bibfnamefont{C.}~\bibnamefont{Alexandrou}},
  \bibinfo{author}{\bibfnamefont{K.}~\bibnamefont{Cichy}},
  \bibinfo{author}{\bibfnamefont{M.}~\bibnamefont{Constantinou}},
  \bibinfo{author}{\bibfnamefont{K.}~\bibnamefont{Hadjiyiannakou}},
  \bibinfo{author}{\bibfnamefont{K.}~\bibnamefont{Jansen}},
  \bibinfo{author}{\bibfnamefont{A.}~\bibnamefont{Scapellato}},
  \bibnamefont{and} \bibinfo{author}{\bibfnamefont{F.}~\bibnamefont{Steffens}},
  \bibinfo{journal}{Phys. Rev. D} \textbf{\bibinfo{volume}{99}},
  \bibinfo{pages}{114504} (\bibinfo{year}{2019}), \eprint{1902.00587}.

\bibitem[{\citenamefont{Alexandrou et~al.}(2021)\citenamefont{Alexandrou,
  Constantinou, Hadjiyiannakou, Jansen, and Manigrasso}}]{Alexandrou:2021oih}
\bibinfo{author}{\bibfnamefont{C.}~\bibnamefont{Alexandrou}},
  \bibinfo{author}{\bibfnamefont{M.}~\bibnamefont{Constantinou}},
  \bibinfo{author}{\bibfnamefont{K.}~\bibnamefont{Hadjiyiannakou}},
  \bibinfo{author}{\bibfnamefont{K.}~\bibnamefont{Jansen}}, \bibnamefont{and}
  \bibinfo{author}{\bibfnamefont{F.}~\bibnamefont{Manigrasso}},
  \bibinfo{journal}{Phys. Rev. D} \textbf{\bibinfo{volume}{104}},
  \bibinfo{pages}{054503} (\bibinfo{year}{2021}), \eprint{2106.16065}.

\bibitem[{\citenamefont{Yao et~al.}(2022)}]{LatticeParton:2022xsd}
\bibinfo{author}{\bibfnamefont{F.}~\bibnamefont{Yao}} \bibnamefont{et~al.}
  (\bibinfo{collaboration}{Lattice Parton}) (\bibinfo{year}{2022}),
  \eprint{2208.08008}.

\bibitem[{\citenamefont{Egerer et~al.}(2022)}]{HadStruc:2021qdf}
\bibinfo{author}{\bibfnamefont{C.}~\bibnamefont{Egerer}} \bibnamefont{et~al.}
  (\bibinfo{collaboration}{HadStruc}), \bibinfo{journal}{Phys. Rev. D}
  \textbf{\bibinfo{volume}{105}}, \bibinfo{pages}{034507}
  (\bibinfo{year}{2022}), \eprint{2111.01808}.

\bibitem[{\citenamefont{Constantinou and
  Panagopoulos}(2017)}]{Constantinou:2017sej}
\bibinfo{author}{\bibfnamefont{M.}~\bibnamefont{Constantinou}}
  \bibnamefont{and}
  \bibinfo{author}{\bibfnamefont{H.}~\bibnamefont{Panagopoulos}},
  \bibinfo{journal}{Phys. Rev. D} \textbf{\bibinfo{volume}{96}},
  \bibinfo{pages}{054506} (\bibinfo{year}{2017}), \eprint{1705.11193}.

\bibitem[{\citenamefont{Alexandrou et~al.}(2017)\citenamefont{Alexandrou,
  Cichy, Constantinou, Hadjiyiannakou, Jansen, Panagopoulos, and
  Steffens}}]{Alexandrou:2017huk}
\bibinfo{author}{\bibfnamefont{C.}~\bibnamefont{Alexandrou}},
  \bibinfo{author}{\bibfnamefont{K.}~\bibnamefont{Cichy}},
  \bibinfo{author}{\bibfnamefont{M.}~\bibnamefont{Constantinou}},
  \bibinfo{author}{\bibfnamefont{K.}~\bibnamefont{Hadjiyiannakou}},
  \bibinfo{author}{\bibfnamefont{K.}~\bibnamefont{Jansen}},
  \bibinfo{author}{\bibfnamefont{H.}~\bibnamefont{Panagopoulos}},
  \bibnamefont{and} \bibinfo{author}{\bibfnamefont{F.}~\bibnamefont{Steffens}},
  \bibinfo{journal}{Nucl. Phys. B} \textbf{\bibinfo{volume}{923}},
  \bibinfo{pages}{394} (\bibinfo{year}{2017}), \eprint{1706.00265}.

\bibitem[{\citenamefont{Chen et~al.}(2018)\citenamefont{Chen, Ishikawa, Jin,
  Lin, Yang, Zhang, and Zhao}}]{Chen:2017mzz}
\bibinfo{author}{\bibfnamefont{J.-W.} \bibnamefont{Chen}},
  \bibinfo{author}{\bibfnamefont{T.}~\bibnamefont{Ishikawa}},
  \bibinfo{author}{\bibfnamefont{L.}~\bibnamefont{Jin}},
  \bibinfo{author}{\bibfnamefont{H.-W.} \bibnamefont{Lin}},
  \bibinfo{author}{\bibfnamefont{Y.-B.} \bibnamefont{Yang}},
  \bibinfo{author}{\bibfnamefont{J.-H.} \bibnamefont{Zhang}}, \bibnamefont{and}
  \bibinfo{author}{\bibfnamefont{Y.}~\bibnamefont{Zhao}},
  \bibinfo{journal}{Phys. Rev. D} \textbf{\bibinfo{volume}{97}},
  \bibinfo{pages}{014505} (\bibinfo{year}{2018}), \eprint{1706.01295}.

\bibitem[{\citenamefont{Stewart and Zhao}(2018)}]{Stewart:2017tvs}
\bibinfo{author}{\bibfnamefont{I.~W.} \bibnamefont{Stewart}} \bibnamefont{and}
  \bibinfo{author}{\bibfnamefont{Y.}~\bibnamefont{Zhao}},
  \bibinfo{journal}{Phys. Rev. D} \textbf{\bibinfo{volume}{97}},
  \bibinfo{pages}{054512} (\bibinfo{year}{2018}), \eprint{1709.04933}.

\bibitem[{\citenamefont{Ji et~al.}(2021{\natexlab{b}})\citenamefont{Ji, Liu,
  Sch\"afer, Wang, Yang, Zhang, and Zhao}}]{Ji:2020brr}
\bibinfo{author}{\bibfnamefont{X.}~\bibnamefont{Ji}},
  \bibinfo{author}{\bibfnamefont{Y.}~\bibnamefont{Liu}},
  \bibinfo{author}{\bibfnamefont{A.}~\bibnamefont{Sch\"afer}},
  \bibinfo{author}{\bibfnamefont{W.}~\bibnamefont{Wang}},
  \bibinfo{author}{\bibfnamefont{Y.-B.} \bibnamefont{Yang}},
  \bibinfo{author}{\bibfnamefont{J.-H.} \bibnamefont{Zhang}}, \bibnamefont{and}
  \bibinfo{author}{\bibfnamefont{Y.}~\bibnamefont{Zhao}},
  \bibinfo{journal}{Nucl. Phys. B} \textbf{\bibinfo{volume}{964}},
  \bibinfo{pages}{115311} (\bibinfo{year}{2021}{\natexlab{b}}),
  \eprint{2008.03886}.

\bibitem[{\citenamefont{Zhang et~al.}(2023)\citenamefont{Zhang, Holligan, Ji,
  and Su}}]{Zhang:2023bxs}
\bibinfo{author}{\bibfnamefont{R.}~\bibnamefont{Zhang}},
  \bibinfo{author}{\bibfnamefont{J.}~\bibnamefont{Holligan}},
  \bibinfo{author}{\bibfnamefont{X.}~\bibnamefont{Ji}}, \bibnamefont{and}
  \bibinfo{author}{\bibfnamefont{Y.}~\bibnamefont{Su}}, \bibinfo{journal}{Phys.
  Lett. B} \textbf{\bibinfo{volume}{844}}, \bibinfo{pages}{138081}
  (\bibinfo{year}{2023}), \eprint{2305.05212}.

\bibitem[{\citenamefont{Gao et~al.}(2023)\citenamefont{Gao, Hanlon, Holligan,
  Karthik, Mukherjee, Petreczky, Syritsyn, and Zhao}}]{Gao:2022uhg}
\bibinfo{author}{\bibfnamefont{X.}~\bibnamefont{Gao}},
  \bibinfo{author}{\bibfnamefont{A.~D.} \bibnamefont{Hanlon}},
  \bibinfo{author}{\bibfnamefont{J.}~\bibnamefont{Holligan}},
  \bibinfo{author}{\bibfnamefont{N.}~\bibnamefont{Karthik}},
  \bibinfo{author}{\bibfnamefont{S.}~\bibnamefont{Mukherjee}},
  \bibinfo{author}{\bibfnamefont{P.}~\bibnamefont{Petreczky}},
  \bibinfo{author}{\bibfnamefont{S.}~\bibnamefont{Syritsyn}}, \bibnamefont{and}
  \bibinfo{author}{\bibfnamefont{Y.}~\bibnamefont{Zhao}},
  \bibinfo{journal}{Phys. Rev. D} \textbf{\bibinfo{volume}{107}},
  \bibinfo{pages}{074509} (\bibinfo{year}{2023}), \eprint{2212.12569}.

\bibitem[{\citenamefont{Gao et~al.}(2022{\natexlab{a}})\citenamefont{Gao,
  Hanlon, Mukherjee, Petreczky, Scior, Syritsyn, and Zhao}}]{Gao:2021dbh}
\bibinfo{author}{\bibfnamefont{X.}~\bibnamefont{Gao}},
  \bibinfo{author}{\bibfnamefont{A.~D.} \bibnamefont{Hanlon}},
  \bibinfo{author}{\bibfnamefont{S.}~\bibnamefont{Mukherjee}},
  \bibinfo{author}{\bibfnamefont{P.}~\bibnamefont{Petreczky}},
  \bibinfo{author}{\bibfnamefont{P.}~\bibnamefont{Scior}},
  \bibinfo{author}{\bibfnamefont{S.}~\bibnamefont{Syritsyn}}, \bibnamefont{and}
  \bibinfo{author}{\bibfnamefont{Y.}~\bibnamefont{Zhao}},
  \bibinfo{journal}{Phys. Rev. Lett.} \textbf{\bibinfo{volume}{128}},
  \bibinfo{pages}{142003} (\bibinfo{year}{2022}{\natexlab{a}}),
  \eprint{2112.02208}.

\bibitem[{\citenamefont{Gao et~al.}(2022{\natexlab{b}})\citenamefont{Gao,
  Hanlon, Karthik, Mukherjee, Petreczky, Scior, Shi, Syritsyn, Zhao, and
  Zhou}}]{Gao:2022iex}
\bibinfo{author}{\bibfnamefont{X.}~\bibnamefont{Gao}},
  \bibinfo{author}{\bibfnamefont{A.~D.} \bibnamefont{Hanlon}},
  \bibinfo{author}{\bibfnamefont{N.}~\bibnamefont{Karthik}},
  \bibinfo{author}{\bibfnamefont{S.}~\bibnamefont{Mukherjee}},
  \bibinfo{author}{\bibfnamefont{P.}~\bibnamefont{Petreczky}},
  \bibinfo{author}{\bibfnamefont{P.}~\bibnamefont{Scior}},
  \bibinfo{author}{\bibfnamefont{S.}~\bibnamefont{Shi}},
  \bibinfo{author}{\bibfnamefont{S.}~\bibnamefont{Syritsyn}},
  \bibinfo{author}{\bibfnamefont{Y.}~\bibnamefont{Zhao}}, \bibnamefont{and}
  \bibinfo{author}{\bibfnamefont{K.}~\bibnamefont{Zhou}}
  (\bibinfo{year}{2022}{\natexlab{b}}), \eprint{2208.02297}.

\bibitem[{\citenamefont{Follana et~al.}(2007)\citenamefont{Follana, Mason,
  Davies, Hornbostel, Lepage, Shigemitsu, Trottier, and Wong}}]{Follana:2006rc}
\bibinfo{author}{\bibfnamefont{E.}~\bibnamefont{Follana}},
  \bibinfo{author}{\bibfnamefont{Q.}~\bibnamefont{Mason}},
  \bibinfo{author}{\bibfnamefont{C.}~\bibnamefont{Davies}},
  \bibinfo{author}{\bibfnamefont{K.}~\bibnamefont{Hornbostel}},
  \bibinfo{author}{\bibfnamefont{G.~P.} \bibnamefont{Lepage}},
  \bibinfo{author}{\bibfnamefont{J.}~\bibnamefont{Shigemitsu}},
  \bibinfo{author}{\bibfnamefont{H.}~\bibnamefont{Trottier}}, \bibnamefont{and}
  \bibinfo{author}{\bibfnamefont{K.}~\bibnamefont{Wong}}
  (\bibinfo{collaboration}{HPQCD, UKQCD}), \bibinfo{journal}{Phys. Rev. D}
  \textbf{\bibinfo{volume}{75}}, \bibinfo{pages}{054502}
  (\bibinfo{year}{2007}), \eprint{hep-lat/0610092}.

\bibitem[{\citenamefont{Bazavov et~al.}(2019)}]{Bazavov:2019www}
\bibinfo{author}{\bibfnamefont{A.}~\bibnamefont{Bazavov}} \bibnamefont{et~al.},
  \bibinfo{journal}{Phys. Rev. D} \textbf{\bibinfo{volume}{100}},
  \bibinfo{pages}{094510} (\bibinfo{year}{2019}), \eprint{1908.09552}.

\bibitem[{\citenamefont{Hasenfratz and Knechtli}(2001)}]{Hasenfratz:2001hp}
\bibinfo{author}{\bibfnamefont{A.}~\bibnamefont{Hasenfratz}} \bibnamefont{and}
  \bibinfo{author}{\bibfnamefont{F.}~\bibnamefont{Knechtli}},
  \bibinfo{journal}{Phys. Rev. D} \textbf{\bibinfo{volume}{64}},
  \bibinfo{pages}{034504} (\bibinfo{year}{2001}), \eprint{hep-lat/0103029}.

\bibitem[{\citenamefont{Bhattacharya
  et~al.}(2015{\natexlab{a}})\citenamefont{Bhattacharya, Cirigliano, Cohen,
  Gupta, Joseph, Lin, and Yoon}}]{Bhattacharya:2015wna}
\bibinfo{author}{\bibfnamefont{T.}~\bibnamefont{Bhattacharya}},
  \bibinfo{author}{\bibfnamefont{V.}~\bibnamefont{Cirigliano}},
  \bibinfo{author}{\bibfnamefont{S.}~\bibnamefont{Cohen}},
  \bibinfo{author}{\bibfnamefont{R.}~\bibnamefont{Gupta}},
  \bibinfo{author}{\bibfnamefont{A.}~\bibnamefont{Joseph}},
  \bibinfo{author}{\bibfnamefont{H.-W.} \bibnamefont{Lin}}, \bibnamefont{and}
  \bibinfo{author}{\bibfnamefont{B.}~\bibnamefont{Yoon}}
  (\bibinfo{collaboration}{PNDME}), \bibinfo{journal}{Phys. Rev. D}
  \textbf{\bibinfo{volume}{92}}, \bibinfo{pages}{094511}
  (\bibinfo{year}{2015}{\natexlab{a}}), \eprint{1506.06411}.

\bibitem[{\citenamefont{Mondal et~al.}(2020)\citenamefont{Mondal, Gupta, Park,
  Yoon, Bhattacharya, and Lin}}]{Mondal:2020cmt}
\bibinfo{author}{\bibfnamefont{S.}~\bibnamefont{Mondal}},
  \bibinfo{author}{\bibfnamefont{R.}~\bibnamefont{Gupta}},
  \bibinfo{author}{\bibfnamefont{S.}~\bibnamefont{Park}},
  \bibinfo{author}{\bibfnamefont{B.}~\bibnamefont{Yoon}},
  \bibinfo{author}{\bibfnamefont{T.}~\bibnamefont{Bhattacharya}},
  \bibnamefont{and} \bibinfo{author}{\bibfnamefont{H.-W.} \bibnamefont{Lin}},
  \bibinfo{journal}{Phys. Rev. D} \textbf{\bibinfo{volume}{102}},
  \bibinfo{pages}{054512} (\bibinfo{year}{2020}), \eprint{2005.13779}.

\bibitem[{\citenamefont{Bali et~al.}(2016)\citenamefont{Bali, Lang, Musch, and
  Sch\"afer}}]{Bali:2016lva}
\bibinfo{author}{\bibfnamefont{G.~S.} \bibnamefont{Bali}},
  \bibinfo{author}{\bibfnamefont{B.}~\bibnamefont{Lang}},
  \bibinfo{author}{\bibfnamefont{B.~U.} \bibnamefont{Musch}}, \bibnamefont{and}
  \bibinfo{author}{\bibfnamefont{A.}~\bibnamefont{Sch\"afer}},
  \bibinfo{journal}{Phys. Rev. D} \textbf{\bibinfo{volume}{93}},
  \bibinfo{pages}{094515} (\bibinfo{year}{2016}), \eprint{1602.05525}.

\bibitem[{\citenamefont{Izubuchi et~al.}(2019)\citenamefont{Izubuchi, Jin,
  Kallidonis, Karthik, Mukherjee, Petreczky, Shugert, and
  Syritsyn}}]{Izubuchi:2019lyk}
\bibinfo{author}{\bibfnamefont{T.}~\bibnamefont{Izubuchi}},
  \bibinfo{author}{\bibfnamefont{L.}~\bibnamefont{Jin}},
  \bibinfo{author}{\bibfnamefont{C.}~\bibnamefont{Kallidonis}},
  \bibinfo{author}{\bibfnamefont{N.}~\bibnamefont{Karthik}},
  \bibinfo{author}{\bibfnamefont{S.}~\bibnamefont{Mukherjee}},
  \bibinfo{author}{\bibfnamefont{P.}~\bibnamefont{Petreczky}},
  \bibinfo{author}{\bibfnamefont{C.}~\bibnamefont{Shugert}}, \bibnamefont{and}
  \bibinfo{author}{\bibfnamefont{S.}~\bibnamefont{Syritsyn}},
  \bibinfo{journal}{Phys. Rev. D} \textbf{\bibinfo{volume}{100}},
  \bibinfo{pages}{034516} (\bibinfo{year}{2019}), \eprint{1905.06349}.

\bibitem[{\citenamefont{Pochinsky}(2008--present)}]{qlua}
\bibinfo{author}{\bibfnamefont{A.}~\bibnamefont{Pochinsky}},
  \emph{\bibinfo{title}{{Qlua lattice software suite}}},
  \bibinfo{howpublished}{\url{https://usqcd.lns.mit.edu/qlua}}
  (\bibinfo{year}{2008--present}).

\bibitem[{\citenamefont{Clark et~al.}(2010)\citenamefont{Clark, Babich, Barros,
  Brower, and Rebbi}}]{Clark:2009wm}
\bibinfo{author}{\bibfnamefont{M.~A.} \bibnamefont{Clark}},
  \bibinfo{author}{\bibfnamefont{R.}~\bibnamefont{Babich}},
  \bibinfo{author}{\bibfnamefont{K.}~\bibnamefont{Barros}},
  \bibinfo{author}{\bibfnamefont{R.~C.} \bibnamefont{Brower}},
  \bibnamefont{and} \bibinfo{author}{\bibfnamefont{C.}~\bibnamefont{Rebbi}},
  \bibinfo{journal}{Comput. Phys. Commun.} \textbf{\bibinfo{volume}{181}},
  \bibinfo{pages}{1517} (\bibinfo{year}{2010}), \eprint{0911.3191}.

\bibitem[{\citenamefont{Babich et~al.}(2011)\citenamefont{Babich, Clark, Joo,
  Shi, Brower, and Gottlieb}}]{Babich:2011np}
\bibinfo{author}{\bibfnamefont{R.}~\bibnamefont{Babich}},
  \bibinfo{author}{\bibfnamefont{M.~A.} \bibnamefont{Clark}},
  \bibinfo{author}{\bibfnamefont{B.}~\bibnamefont{Joo}},
  \bibinfo{author}{\bibfnamefont{G.}~\bibnamefont{Shi}},
  \bibinfo{author}{\bibfnamefont{R.~C.} \bibnamefont{Brower}},
  \bibnamefont{and} \bibinfo{author}{\bibfnamefont{S.}~\bibnamefont{Gottlieb}},
  in \emph{\bibinfo{booktitle}{{SC11 International Conference for High
  Performance Computing, Networking, Storage and Analysis}}}
  (\bibinfo{year}{2011}), \eprint{1109.2935}.

\bibitem[{\citenamefont{Shintani et~al.}(2015)\citenamefont{Shintani, Arthur,
  Blum, Izubuchi, Jung, and Lehner}}]{Shintani:2014vja}
\bibinfo{author}{\bibfnamefont{E.}~\bibnamefont{Shintani}},
  \bibinfo{author}{\bibfnamefont{R.}~\bibnamefont{Arthur}},
  \bibinfo{author}{\bibfnamefont{T.}~\bibnamefont{Blum}},
  \bibinfo{author}{\bibfnamefont{T.}~\bibnamefont{Izubuchi}},
  \bibinfo{author}{\bibfnamefont{C.}~\bibnamefont{Jung}}, \bibnamefont{and}
  \bibinfo{author}{\bibfnamefont{C.}~\bibnamefont{Lehner}},
  \bibinfo{journal}{Phys. Rev. D} \textbf{\bibinfo{volume}{91}},
  \bibinfo{pages}{114511} (\bibinfo{year}{2015}), \eprint{1402.0244}.

\bibitem[{\citenamefont{Bhattacharya
  et~al.}(2015{\natexlab{b}})\citenamefont{Bhattacharya, Cirigliano, Gupta,
  Lin, and Yoon}}]{Bhattacharya:2015esa}
\bibinfo{author}{\bibfnamefont{T.}~\bibnamefont{Bhattacharya}},
  \bibinfo{author}{\bibfnamefont{V.}~\bibnamefont{Cirigliano}},
  \bibinfo{author}{\bibfnamefont{R.}~\bibnamefont{Gupta}},
  \bibinfo{author}{\bibfnamefont{H.-W.} \bibnamefont{Lin}}, \bibnamefont{and}
  \bibinfo{author}{\bibfnamefont{B.}~\bibnamefont{Yoon}},
  \bibinfo{journal}{Phys. Rev. Lett.} \textbf{\bibinfo{volume}{115}},
  \bibinfo{pages}{212002} (\bibinfo{year}{2015}{\natexlab{b}}),
  \eprint{1506.04196}.

\bibitem[{\citenamefont{Gracey}(2022)}]{Gracey:2022vqr}
\bibinfo{author}{\bibfnamefont{J.~A.} \bibnamefont{Gracey}},
  \bibinfo{journal}{Phys. Rev. D} \textbf{\bibinfo{volume}{106}},
  \bibinfo{pages}{085008} (\bibinfo{year}{2022}), \eprint{2208.14527}.

\bibitem[{\citenamefont{Chetyrkin}(1997)}]{Chetyrkin:1997dh}
\bibinfo{author}{\bibfnamefont{K.~G.} \bibnamefont{Chetyrkin}},
  \bibinfo{journal}{Phys. Lett. B} \textbf{\bibinfo{volume}{404}},
  \bibinfo{pages}{161} (\bibinfo{year}{1997}), \eprint{hep-ph/9703278}.

\bibitem[{\citenamefont{Aoki
  et~al.}(2022)}]{FlavourLatticeAveragingGroupFLAG:2021npn}
\bibinfo{author}{\bibfnamefont{Y.}~\bibnamefont{Aoki}} \bibnamefont{et~al.}
  (\bibinfo{collaboration}{Flavour Lattice Averaging Group (FLAG)}),
  \bibinfo{journal}{Eur. Phys. J. C} \textbf{\bibinfo{volume}{82}},
  \bibinfo{pages}{869} (\bibinfo{year}{2022}), \eprint{2111.09849}.

\bibitem[{\citenamefont{Park et~al.}(2022)\citenamefont{Park, Gupta, Yoon,
  Mondal, Bhattacharya, Jang, Jo\'o, and Winter}}]{Park:2021ypf}
\bibinfo{author}{\bibfnamefont{S.}~\bibnamefont{Park}},
  \bibinfo{author}{\bibfnamefont{R.}~\bibnamefont{Gupta}},
  \bibinfo{author}{\bibfnamefont{B.}~\bibnamefont{Yoon}},
  \bibinfo{author}{\bibfnamefont{S.}~\bibnamefont{Mondal}},
  \bibinfo{author}{\bibfnamefont{T.}~\bibnamefont{Bhattacharya}},
  \bibinfo{author}{\bibfnamefont{Y.-C.} \bibnamefont{Jang}},
  \bibinfo{author}{\bibfnamefont{B.}~\bibnamefont{Jo\'o}}, \bibnamefont{and}
  \bibinfo{author}{\bibfnamefont{F.}~\bibnamefont{Winter}}
  (\bibinfo{collaboration}{Nucleon Matrix Elements (NME)}),
  \bibinfo{journal}{Phys. Rev. D} \textbf{\bibinfo{volume}{105}},
  \bibinfo{pages}{054505} (\bibinfo{year}{2022}), \eprint{2103.05599}.

\bibitem[{\citenamefont{Abramczyk et~al.}(2020)\citenamefont{Abramczyk, Blum,
  Izubuchi, Jung, Lin, Lytle, Ohta, and Shintani}}]{Abramczyk:2019fnf}
\bibinfo{author}{\bibfnamefont{M.}~\bibnamefont{Abramczyk}},
  \bibinfo{author}{\bibfnamefont{T.}~\bibnamefont{Blum}},
  \bibinfo{author}{\bibfnamefont{T.}~\bibnamefont{Izubuchi}},
  \bibinfo{author}{\bibfnamefont{C.}~\bibnamefont{Jung}},
  \bibinfo{author}{\bibfnamefont{M.}~\bibnamefont{Lin}},
  \bibinfo{author}{\bibfnamefont{A.}~\bibnamefont{Lytle}},
  \bibinfo{author}{\bibfnamefont{S.}~\bibnamefont{Ohta}}, \bibnamefont{and}
  \bibinfo{author}{\bibfnamefont{E.}~\bibnamefont{Shintani}},
  \bibinfo{journal}{Phys. Rev. D} \textbf{\bibinfo{volume}{101}},
  \bibinfo{pages}{034510} (\bibinfo{year}{2020}), \eprint{1911.03524}.

\bibitem[{\citenamefont{Harris et~al.}(2019)\citenamefont{Harris, von Hippel,
  Junnarkar, Meyer, Ottnad, Wilhelm, Wittig, and Wrang}}]{Harris:2019bih}
\bibinfo{author}{\bibfnamefont{T.}~\bibnamefont{Harris}},
  \bibinfo{author}{\bibfnamefont{G.}~\bibnamefont{von Hippel}},
  \bibinfo{author}{\bibfnamefont{P.}~\bibnamefont{Junnarkar}},
  \bibinfo{author}{\bibfnamefont{H.~B.} \bibnamefont{Meyer}},
  \bibinfo{author}{\bibfnamefont{K.}~\bibnamefont{Ottnad}},
  \bibinfo{author}{\bibfnamefont{J.}~\bibnamefont{Wilhelm}},
  \bibinfo{author}{\bibfnamefont{H.}~\bibnamefont{Wittig}}, \bibnamefont{and}
  \bibinfo{author}{\bibfnamefont{L.}~\bibnamefont{Wrang}},
  \bibinfo{journal}{Phys. Rev. D} \textbf{\bibinfo{volume}{100}},
  \bibinfo{pages}{034513} (\bibinfo{year}{2019}), \eprint{1905.01291}.

\bibitem[{\citenamefont{Djukanovic et~al.}(2019)\citenamefont{Djukanovic,
  Meyer, Ottnad, von Hippel, Wilhelm, and Wittig}}]{Djukanovic:2019gvi}
\bibinfo{author}{\bibfnamefont{D.}~\bibnamefont{Djukanovic}},
  \bibinfo{author}{\bibfnamefont{H.}~\bibnamefont{Meyer}},
  \bibinfo{author}{\bibfnamefont{K.}~\bibnamefont{Ottnad}},
  \bibinfo{author}{\bibfnamefont{G.}~\bibnamefont{von Hippel}},
  \bibinfo{author}{\bibfnamefont{J.}~\bibnamefont{Wilhelm}}, \bibnamefont{and}
  \bibinfo{author}{\bibfnamefont{H.}~\bibnamefont{Wittig}},
  \bibinfo{journal}{PoS} \textbf{\bibinfo{volume}{LATTICE2019}},
  \bibinfo{pages}{158} (\bibinfo{year}{2019}), \eprint{1911.01177}.

\bibitem[{\citenamefont{Hasan et~al.}(2019)\citenamefont{Hasan, Green, Meinel,
  Engelhardt, Krieg, Negele, Pochinsky, and Syritsyn}}]{Hasan:2019noy}
\bibinfo{author}{\bibfnamefont{N.}~\bibnamefont{Hasan}},
  \bibinfo{author}{\bibfnamefont{J.}~\bibnamefont{Green}},
  \bibinfo{author}{\bibfnamefont{S.}~\bibnamefont{Meinel}},
  \bibinfo{author}{\bibfnamefont{M.}~\bibnamefont{Engelhardt}},
  \bibinfo{author}{\bibfnamefont{S.}~\bibnamefont{Krieg}},
  \bibinfo{author}{\bibfnamefont{J.}~\bibnamefont{Negele}},
  \bibinfo{author}{\bibfnamefont{A.}~\bibnamefont{Pochinsky}},
  \bibnamefont{and} \bibinfo{author}{\bibfnamefont{S.}~\bibnamefont{Syritsyn}},
  \bibinfo{journal}{Phys. Rev. D} \textbf{\bibinfo{volume}{99}},
  \bibinfo{pages}{114505} (\bibinfo{year}{2019}), \eprint{1903.06487}.

\bibitem[{\citenamefont{Yamanaka et~al.}(2018)\citenamefont{Yamanaka,
  Hashimoto, Kaneko, and Ohki}}]{Yamanaka:2018uud}
\bibinfo{author}{\bibfnamefont{N.}~\bibnamefont{Yamanaka}},
  \bibinfo{author}{\bibfnamefont{S.}~\bibnamefont{Hashimoto}},
  \bibinfo{author}{\bibfnamefont{T.}~\bibnamefont{Kaneko}}, \bibnamefont{and}
  \bibinfo{author}{\bibfnamefont{H.}~\bibnamefont{Ohki}}
  (\bibinfo{collaboration}{JLQCD}), \bibinfo{journal}{Phys. Rev. D}
  \textbf{\bibinfo{volume}{98}}, \bibinfo{pages}{054516}
  (\bibinfo{year}{2018}), \eprint{1805.10507}.

\bibitem[{\citenamefont{Green et~al.}(2012)\citenamefont{Green, Negele,
  Pochinsky, Syritsyn, Engelhardt, and Krieg}}]{Green:2012ej}
\bibinfo{author}{\bibfnamefont{J.~R.} \bibnamefont{Green}},
  \bibinfo{author}{\bibfnamefont{J.~W.} \bibnamefont{Negele}},
  \bibinfo{author}{\bibfnamefont{A.~V.} \bibnamefont{Pochinsky}},
  \bibinfo{author}{\bibfnamefont{S.~N.} \bibnamefont{Syritsyn}},
  \bibinfo{author}{\bibfnamefont{M.}~\bibnamefont{Engelhardt}},
  \bibnamefont{and} \bibinfo{author}{\bibfnamefont{S.}~\bibnamefont{Krieg}},
  \bibinfo{journal}{Phys. Rev. D} \textbf{\bibinfo{volume}{86}},
  \bibinfo{pages}{114509} (\bibinfo{year}{2012}), \eprint{1206.4527}.

\bibitem[{\citenamefont{Aoki et~al.}(2010)\citenamefont{Aoki, Blum, Lin, Ohta,
  Sasaki, Tweedie, Zanotti, and Yamazaki}}]{Aoki:2010xg}
\bibinfo{author}{\bibfnamefont{Y.}~\bibnamefont{Aoki}},
  \bibinfo{author}{\bibfnamefont{T.}~\bibnamefont{Blum}},
  \bibinfo{author}{\bibfnamefont{H.-W.} \bibnamefont{Lin}},
  \bibinfo{author}{\bibfnamefont{S.}~\bibnamefont{Ohta}},
  \bibinfo{author}{\bibfnamefont{S.}~\bibnamefont{Sasaki}},
  \bibinfo{author}{\bibfnamefont{R.}~\bibnamefont{Tweedie}},
  \bibinfo{author}{\bibfnamefont{J.}~\bibnamefont{Zanotti}}, \bibnamefont{and}
  \bibinfo{author}{\bibfnamefont{T.}~\bibnamefont{Yamazaki}},
  \bibinfo{journal}{Phys. Rev. D} \textbf{\bibinfo{volume}{82}},
  \bibinfo{pages}{014501} (\bibinfo{year}{2010}), \eprint{1003.3387}.

\bibitem[{\citenamefont{Fan et~al.}(2020)\citenamefont{Fan, Gao, Li, Lin,
  Karthik, Mukherjee, Petreczky, Syritsyn, Yang, and Zhang}}]{Fan:2020nzz}
\bibinfo{author}{\bibfnamefont{Z.}~\bibnamefont{Fan}},
  \bibinfo{author}{\bibfnamefont{X.}~\bibnamefont{Gao}},
  \bibinfo{author}{\bibfnamefont{R.}~\bibnamefont{Li}},
  \bibinfo{author}{\bibfnamefont{H.-W.} \bibnamefont{Lin}},
  \bibinfo{author}{\bibfnamefont{N.}~\bibnamefont{Karthik}},
  \bibinfo{author}{\bibfnamefont{S.}~\bibnamefont{Mukherjee}},
  \bibinfo{author}{\bibfnamefont{P.}~\bibnamefont{Petreczky}},
  \bibinfo{author}{\bibfnamefont{S.}~\bibnamefont{Syritsyn}},
  \bibinfo{author}{\bibfnamefont{Y.-B.} \bibnamefont{Yang}}, \bibnamefont{and}
  \bibinfo{author}{\bibfnamefont{R.}~\bibnamefont{Zhang}},
  \bibinfo{journal}{Phys. Rev. D} \textbf{\bibinfo{volume}{102}},
  \bibinfo{pages}{074504} (\bibinfo{year}{2020}), \eprint{2005.12015}.

\bibitem[{\citenamefont{Jo\'o et~al.}(2019)\citenamefont{Jo\'o, Karpie,
  Orginos, Radyushkin, Richards, and Zafeiropoulos}}]{Joo:2019jct}
\bibinfo{author}{\bibfnamefont{B.}~\bibnamefont{Jo\'o}},
  \bibinfo{author}{\bibfnamefont{J.}~\bibnamefont{Karpie}},
  \bibinfo{author}{\bibfnamefont{K.}~\bibnamefont{Orginos}},
  \bibinfo{author}{\bibfnamefont{A.}~\bibnamefont{Radyushkin}},
  \bibinfo{author}{\bibfnamefont{D.}~\bibnamefont{Richards}}, \bibnamefont{and}
  \bibinfo{author}{\bibfnamefont{S.}~\bibnamefont{Zafeiropoulos}},
  \bibinfo{journal}{JHEP} \textbf{\bibinfo{volume}{12}}, \bibinfo{pages}{081}
  (\bibinfo{year}{2019}), \eprint{1908.09771}.

\bibitem[{\citenamefont{Jo\'o et~al.}(2020)\citenamefont{Jo\'o, Karpie,
  Orginos, Radyushkin, Richards, and Zafeiropoulos}}]{Joo:2020spy}
\bibinfo{author}{\bibfnamefont{B.}~\bibnamefont{Jo\'o}},
  \bibinfo{author}{\bibfnamefont{J.}~\bibnamefont{Karpie}},
  \bibinfo{author}{\bibfnamefont{K.}~\bibnamefont{Orginos}},
  \bibinfo{author}{\bibfnamefont{A.~V.} \bibnamefont{Radyushkin}},
  \bibinfo{author}{\bibfnamefont{D.~G.} \bibnamefont{Richards}},
  \bibnamefont{and}
  \bibinfo{author}{\bibfnamefont{S.}~\bibnamefont{Zafeiropoulos}},
  \bibinfo{journal}{Phys. Rev. Lett.} \textbf{\bibinfo{volume}{125}},
  \bibinfo{pages}{232003} (\bibinfo{year}{2020}), \eprint{2004.01687}.

\bibitem[{\citenamefont{Bhat et~al.}(2021)\citenamefont{Bhat, Cichy,
  Constantinou, and Scapellato}}]{Bhat:2020ktg}
\bibinfo{author}{\bibfnamefont{M.}~\bibnamefont{Bhat}},
  \bibinfo{author}{\bibfnamefont{K.}~\bibnamefont{Cichy}},
  \bibinfo{author}{\bibfnamefont{M.}~\bibnamefont{Constantinou}},
  \bibnamefont{and}
  \bibinfo{author}{\bibfnamefont{A.}~\bibnamefont{Scapellato}},
  \bibinfo{journal}{Phys. Rev. D} \textbf{\bibinfo{volume}{103}},
  \bibinfo{pages}{034510} (\bibinfo{year}{2021}), \eprint{2005.02102}.

\bibitem[{\citenamefont{Karpie et~al.}(2021)\citenamefont{Karpie, Orginos,
  Radyushkin, and Zafeiropoulos}}]{Karpie:2021pap}
\bibinfo{author}{\bibfnamefont{J.}~\bibnamefont{Karpie}},
  \bibinfo{author}{\bibfnamefont{K.}~\bibnamefont{Orginos}},
  \bibinfo{author}{\bibfnamefont{A.}~\bibnamefont{Radyushkin}},
  \bibnamefont{and}
  \bibinfo{author}{\bibfnamefont{S.}~\bibnamefont{Zafeiropoulos}}
  (\bibinfo{collaboration}{HadStruc}), \bibinfo{journal}{JHEP}
  \textbf{\bibinfo{volume}{11}}, \bibinfo{pages}{024} (\bibinfo{year}{2021}),
  \eprint{2105.13313}.

\bibitem[{\citenamefont{Egerer et~al.}(2021)\citenamefont{Egerer, Edwards,
  Kallidonis, Orginos, Radyushkin, Richards, Romero, and
  Zafeiropoulos}}]{Egerer:2021ymv}
\bibinfo{author}{\bibfnamefont{C.}~\bibnamefont{Egerer}},
  \bibinfo{author}{\bibfnamefont{R.~G.} \bibnamefont{Edwards}},
  \bibinfo{author}{\bibfnamefont{C.}~\bibnamefont{Kallidonis}},
  \bibinfo{author}{\bibfnamefont{K.}~\bibnamefont{Orginos}},
  \bibinfo{author}{\bibfnamefont{A.~V.} \bibnamefont{Radyushkin}},
  \bibinfo{author}{\bibfnamefont{D.~G.} \bibnamefont{Richards}},
  \bibinfo{author}{\bibfnamefont{E.}~\bibnamefont{Romero}}, \bibnamefont{and}
  \bibinfo{author}{\bibfnamefont{S.}~\bibnamefont{Zafeiropoulos}}
  (\bibinfo{collaboration}{HadStruc}), \bibinfo{journal}{JHEP}
  \textbf{\bibinfo{volume}{11}}, \bibinfo{pages}{148} (\bibinfo{year}{2021}),
  \eprint{2107.05199}.

\bibitem[{\citenamefont{Bhat et~al.}(2022)\citenamefont{Bhat, Chomicki, Cichy,
  Constantinou, Green, and Scapellato}}]{Bhat:2022zrw}
\bibinfo{author}{\bibfnamefont{M.}~\bibnamefont{Bhat}},
  \bibinfo{author}{\bibfnamefont{W.}~\bibnamefont{Chomicki}},
  \bibinfo{author}{\bibfnamefont{K.}~\bibnamefont{Cichy}},
  \bibinfo{author}{\bibfnamefont{M.}~\bibnamefont{Constantinou}},
  \bibinfo{author}{\bibfnamefont{J.~R.} \bibnamefont{Green}}, \bibnamefont{and}
  \bibinfo{author}{\bibfnamefont{A.}~\bibnamefont{Scapellato}},
  \bibinfo{journal}{Phys. Rev. D} \textbf{\bibinfo{volume}{106}},
  \bibinfo{pages}{054504} (\bibinfo{year}{2022}), \eprint{2205.07585}.

\bibitem[{\citenamefont{Karpie et~al.}(2018)\citenamefont{Karpie, Orginos, and
  Zafeiropoulos}}]{Karpie:2018zaz}
\bibinfo{author}{\bibfnamefont{J.}~\bibnamefont{Karpie}},
  \bibinfo{author}{\bibfnamefont{K.}~\bibnamefont{Orginos}}, \bibnamefont{and}
  \bibinfo{author}{\bibfnamefont{S.}~\bibnamefont{Zafeiropoulos}},
  \bibinfo{journal}{JHEP} \textbf{\bibinfo{volume}{11}}, \bibinfo{pages}{178}
  (\bibinfo{year}{2018}), \eprint{1807.10933}.

\bibitem[{\citenamefont{Bl\"umlein et~al.}(2021)\citenamefont{Bl\"umlein,
  Marquard, Schneider, and Sch\"onwald}}]{Blumlein:2021enk}
\bibinfo{author}{\bibfnamefont{J.}~\bibnamefont{Bl\"umlein}},
  \bibinfo{author}{\bibfnamefont{P.}~\bibnamefont{Marquard}},
  \bibinfo{author}{\bibfnamefont{C.}~\bibnamefont{Schneider}},
  \bibnamefont{and}
  \bibinfo{author}{\bibfnamefont{K.}~\bibnamefont{Sch\"onwald}},
  \bibinfo{journal}{Nucl. Phys. B} \textbf{\bibinfo{volume}{971}},
  \bibinfo{pages}{115542} (\bibinfo{year}{2021}), \eprint{2107.06267}.

\bibitem[{\citenamefont{Petreczky and Weber}(2022)}]{Petreczky:2020tky}
\bibinfo{author}{\bibfnamefont{P.}~\bibnamefont{Petreczky}} \bibnamefont{and}
  \bibinfo{author}{\bibfnamefont{J.~H.} \bibnamefont{Weber}},
  \bibinfo{journal}{Eur. Phys. J. C} \textbf{\bibinfo{volume}{82}},
  \bibinfo{pages}{64} (\bibinfo{year}{2022}), \eprint{2012.06193}.

\bibitem[{\citenamefont{Gao et~al.}(2020)\citenamefont{Gao, Jin, Kallidonis,
  Karthik, Mukherjee, Petreczky, Shugert, Syritsyn, and Zhao}}]{Gao:2020ito}
\bibinfo{author}{\bibfnamefont{X.}~\bibnamefont{Gao}},
  \bibinfo{author}{\bibfnamefont{L.}~\bibnamefont{Jin}},
  \bibinfo{author}{\bibfnamefont{C.}~\bibnamefont{Kallidonis}},
  \bibinfo{author}{\bibfnamefont{N.}~\bibnamefont{Karthik}},
  \bibinfo{author}{\bibfnamefont{S.}~\bibnamefont{Mukherjee}},
  \bibinfo{author}{\bibfnamefont{P.}~\bibnamefont{Petreczky}},
  \bibinfo{author}{\bibfnamefont{C.}~\bibnamefont{Shugert}},
  \bibinfo{author}{\bibfnamefont{S.}~\bibnamefont{Syritsyn}}, \bibnamefont{and}
  \bibinfo{author}{\bibfnamefont{Y.}~\bibnamefont{Zhao}},
  \bibinfo{journal}{Phys. Rev. D} \textbf{\bibinfo{volume}{102}},
  \bibinfo{pages}{094513} (\bibinfo{year}{2020}), \eprint{2007.06590}.

\bibitem[{\citenamefont{Vogelsang}(1998)}]{Vogelsang:1997ak}
\bibinfo{author}{\bibfnamefont{W.}~\bibnamefont{Vogelsang}},
  \bibinfo{journal}{Phys. Rev. D} \textbf{\bibinfo{volume}{57}},
  \bibinfo{pages}{1886} (\bibinfo{year}{1998}), \eprint{hep-ph/9706511}.

\bibitem[{\citenamefont{Ji et~al.}(2022)\citenamefont{Ji, Yao, and
  Zhang}}]{Ji:2022thb}
\bibinfo{author}{\bibfnamefont{Y.}~\bibnamefont{Ji}},
  \bibinfo{author}{\bibfnamefont{F.}~\bibnamefont{Yao}}, \bibnamefont{and}
  \bibinfo{author}{\bibfnamefont{J.-H.} \bibnamefont{Zhang}}
  (\bibinfo{year}{2022}), \eprint{2212.14415}.

\bibitem[{\citenamefont{Karpie et~al.}(2019)\citenamefont{Karpie, Orginos,
  Rothkopf, and Zafeiropoulos}}]{Karpie:2019eiq}
\bibinfo{author}{\bibfnamefont{J.}~\bibnamefont{Karpie}},
  \bibinfo{author}{\bibfnamefont{K.}~\bibnamefont{Orginos}},
  \bibinfo{author}{\bibfnamefont{A.}~\bibnamefont{Rothkopf}}, \bibnamefont{and}
  \bibinfo{author}{\bibfnamefont{S.}~\bibnamefont{Zafeiropoulos}},
  \bibinfo{journal}{JHEP} \textbf{\bibinfo{volume}{04}}, \bibinfo{pages}{057}
  (\bibinfo{year}{2019}), \eprint{1901.05408}.

\bibitem[{\citenamefont{Del~Debbio et~al.}(2021)\citenamefont{Del~Debbio,
  Giani, Karpie, Orginos, Radyushkin, and Zafeiropoulos}}]{DelDebbio:2020rgv}
\bibinfo{author}{\bibfnamefont{L.}~\bibnamefont{Del~Debbio}},
  \bibinfo{author}{\bibfnamefont{T.}~\bibnamefont{Giani}},
  \bibinfo{author}{\bibfnamefont{J.}~\bibnamefont{Karpie}},
  \bibinfo{author}{\bibfnamefont{K.}~\bibnamefont{Orginos}},
  \bibinfo{author}{\bibfnamefont{A.}~\bibnamefont{Radyushkin}},
  \bibnamefont{and}
  \bibinfo{author}{\bibfnamefont{S.}~\bibnamefont{Zafeiropoulos}},
  \bibinfo{journal}{JHEP} \textbf{\bibinfo{volume}{02}}, \bibinfo{pages}{138}
  (\bibinfo{year}{2021}), \eprint{2010.03996}.

\bibitem[{\citenamefont{Bazavov et~al.}(2014)}]{HotQCD:2014kol}
\bibinfo{author}{\bibfnamefont{A.}~\bibnamefont{Bazavov}} \bibnamefont{et~al.}
  (\bibinfo{collaboration}{HotQCD}), \bibinfo{journal}{Phys. Rev. D}
  \textbf{\bibinfo{volume}{90}}, \bibinfo{pages}{094503}
  (\bibinfo{year}{2014}), \eprint{1407.6387}.

\bibitem[{\citenamefont{Bazavov et~al.}(2016)\citenamefont{Bazavov, Brambilla,
  Ding, Petreczky, Schadler, Vairo, and Weber}}]{Bazavov:2016uvm}
\bibinfo{author}{\bibfnamefont{A.}~\bibnamefont{Bazavov}},
  \bibinfo{author}{\bibfnamefont{N.}~\bibnamefont{Brambilla}},
  \bibinfo{author}{\bibfnamefont{H.~T.} \bibnamefont{Ding}},
  \bibinfo{author}{\bibfnamefont{P.}~\bibnamefont{Petreczky}},
  \bibinfo{author}{\bibfnamefont{H.~P.} \bibnamefont{Schadler}},
  \bibinfo{author}{\bibfnamefont{A.}~\bibnamefont{Vairo}}, \bibnamefont{and}
  \bibinfo{author}{\bibfnamefont{J.~H.} \bibnamefont{Weber}},
  \bibinfo{journal}{Phys. Rev. D} \textbf{\bibinfo{volume}{93}},
  \bibinfo{pages}{114502} (\bibinfo{year}{2016}), \eprint{1603.06637}.

\bibitem[{\citenamefont{Bazavov
  et~al.}(2018{\natexlab{a}})\citenamefont{Bazavov, Petreczky, and
  Weber}}]{Bazavov:2017dsy}
\bibinfo{author}{\bibfnamefont{A.}~\bibnamefont{Bazavov}},
  \bibinfo{author}{\bibfnamefont{P.}~\bibnamefont{Petreczky}},
  \bibnamefont{and} \bibinfo{author}{\bibfnamefont{J.~H.} \bibnamefont{Weber}},
  \bibinfo{journal}{Phys. Rev. D} \textbf{\bibinfo{volume}{97}},
  \bibinfo{pages}{014510} (\bibinfo{year}{2018}{\natexlab{a}}),
  \eprint{1710.05024}.

\bibitem[{\citenamefont{Bazavov
  et~al.}(2018{\natexlab{b}})\citenamefont{Bazavov, Brambilla, Petreczky,
  Vairo, and Weber}}]{Bazavov:2018wmo}
\bibinfo{author}{\bibfnamefont{A.}~\bibnamefont{Bazavov}},
  \bibinfo{author}{\bibfnamefont{N.}~\bibnamefont{Brambilla}},
  \bibinfo{author}{\bibfnamefont{P.}~\bibnamefont{Petreczky}},
  \bibinfo{author}{\bibfnamefont{A.}~\bibnamefont{Vairo}}, \bibnamefont{and}
  \bibinfo{author}{\bibfnamefont{J.~H.} \bibnamefont{Weber}}
  (\bibinfo{collaboration}{TUMQCD}), \bibinfo{journal}{Phys. Rev. D}
  \textbf{\bibinfo{volume}{98}}, \bibinfo{pages}{054511}
  (\bibinfo{year}{2018}{\natexlab{b}}), \eprint{1804.10600}.

\bibitem[{\citenamefont{Petreczky et~al.}(2022)\citenamefont{Petreczky,
  Steinbei\ss{}er, and Weber}}]{Petreczky:2021mef}
\bibinfo{author}{\bibfnamefont{P.}~\bibnamefont{Petreczky}},
  \bibinfo{author}{\bibfnamefont{S.}~\bibnamefont{Steinbei\ss{}er}},
  \bibnamefont{and} \bibinfo{author}{\bibfnamefont{J.~H.} \bibnamefont{Weber}},
  \bibinfo{journal}{PoS} \textbf{\bibinfo{volume}{LATTICE2021}},
  \bibinfo{pages}{471} (\bibinfo{year}{2022}), \eprint{2112.00788}.

\bibitem[{\citenamefont{Gao et~al.}(2021)\citenamefont{Gao, Lee, Mukherjee,
  Shugert, and Zhao}}]{Gao:2021hxl}
\bibinfo{author}{\bibfnamefont{X.}~\bibnamefont{Gao}},
  \bibinfo{author}{\bibfnamefont{K.}~\bibnamefont{Lee}},
  \bibinfo{author}{\bibfnamefont{S.}~\bibnamefont{Mukherjee}},
  \bibinfo{author}{\bibfnamefont{C.}~\bibnamefont{Shugert}}, \bibnamefont{and}
  \bibinfo{author}{\bibfnamefont{Y.}~\bibnamefont{Zhao}},
  \bibinfo{journal}{Phys. Rev. D} \textbf{\bibinfo{volume}{103}},
  \bibinfo{pages}{094504} (\bibinfo{year}{2021}), \eprint{2102.01101}.

\bibitem[{\citenamefont{Xiong et~al.}(2014)\citenamefont{Xiong, Ji, Zhang, and
  Zhao}}]{Xiong:2013bka}
\bibinfo{author}{\bibfnamefont{X.}~\bibnamefont{Xiong}},
  \bibinfo{author}{\bibfnamefont{X.}~\bibnamefont{Ji}},
  \bibinfo{author}{\bibfnamefont{J.-H.} \bibnamefont{Zhang}}, \bibnamefont{and}
  \bibinfo{author}{\bibfnamefont{Y.}~\bibnamefont{Zhao}},
  \bibinfo{journal}{Phys. Rev. D} \textbf{\bibinfo{volume}{90}},
  \bibinfo{pages}{014051} (\bibinfo{year}{2014}), \eprint{1310.7471}.

\bibitem[{\citenamefont{Lepage}(2022{\natexlab{a}})}]{lsqfit:11.5.1}
\bibinfo{author}{\bibfnamefont{P.}~\bibnamefont{Lepage}},
  \emph{\bibinfo{title}{gplepage/lsqfit: lsqfit version 12.0.3}}
  (\bibinfo{year}{2022}{\natexlab{a}}),
  \bibinfo{note}{\url{https://github.com/gplepage/lsqfit}}.

\bibitem[{\citenamefont{Lepage}(2022{\natexlab{b}})}]{gvar:11.2}
\bibinfo{author}{\bibfnamefont{P.}~\bibnamefont{Lepage}},
  \emph{\bibinfo{title}{gplepage/gvar: gvar version 11.9.5}}
  (\bibinfo{year}{2022}{\natexlab{b}}),
  \bibinfo{note}{\url{https://github.com/gplepage/gvar}}.

\bibitem[{\citenamefont{Hunter}(2007)}]{Hunter:2007}
\bibinfo{author}{\bibfnamefont{J.~D.} \bibnamefont{Hunter}},
  \bibinfo{journal}{Comput. Sci. Eng.} \textbf{\bibinfo{volume}{9}},
  \bibinfo{pages}{90} (\bibinfo{year}{2007}).

\bibitem[{\citenamefont{Bhattacharya et~al.}(2014)\citenamefont{Bhattacharya,
  Cohen, Gupta, Joseph, Lin, and Yoon}}]{Bhattacharya:2013ehc}
\bibinfo{author}{\bibfnamefont{T.}~\bibnamefont{Bhattacharya}},
  \bibinfo{author}{\bibfnamefont{S.~D.} \bibnamefont{Cohen}},
  \bibinfo{author}{\bibfnamefont{R.}~\bibnamefont{Gupta}},
  \bibinfo{author}{\bibfnamefont{A.}~\bibnamefont{Joseph}},
  \bibinfo{author}{\bibfnamefont{H.-W.} \bibnamefont{Lin}}, \bibnamefont{and}
  \bibinfo{author}{\bibfnamefont{B.}~\bibnamefont{Yoon}},
  \bibinfo{journal}{Phys. Rev. D} \textbf{\bibinfo{volume}{89}},
  \bibinfo{pages}{094502} (\bibinfo{year}{2014}), \eprint{1306.5435}.

\bibitem[{\citenamefont{Gracey}(2003)}]{Gracey:2003yr}
\bibinfo{author}{\bibfnamefont{J.~A.} \bibnamefont{Gracey}},
  \bibinfo{journal}{Nucl. Phys. B} \textbf{\bibinfo{volume}{662}},
  \bibinfo{pages}{247} (\bibinfo{year}{2003}), \eprint{hep-ph/0304113}.

\end{thebibliography}

\end{document}